\documentclass[final,3p,times]{elsarticle}

\usepackage{graphicx} 
\graphicspath{{Pictures/}} 
\usepackage{subfig}  
\usepackage[export]{adjustbox}
\usepackage{amsmath,amsfonts,amssymb,amsthm} 
\usepackage{hyperref} 
\usepackage{enumitem}
\setlist{nolistsep} 
\usepackage{booktabs}
\usepackage{xcolor}
\definecolor{MyRed}{RGB}{194,21,21} 
\definecolor{TUMGruen}{RGB}{162,173,0} 
\definecolor{TUMBlau}{RGB}{0,101,189}
\definecolor{utorange}{RGB}{191, 87, 0}
\usepackage{pifont}

\usepackage{placeins}

\journal{Elsevier}

\begin{document}

\begin{frontmatter}



  \title{Bridging Scales: a Hybrid Model to Simulate Vascular Tumor Growth and
    Treatment Response}



  \affiliation[cern]{organization={CERN},
    city={Geneva},
    country={Switzerland}}
  \affiliation[tum]{
    organization={School for Computation, Information, and Technology,
        Technical Universtity of Munich},
    country={Germany}}
  \affiliation[uta-oden]{
    organization={Oden Institute for Computational Engineering and Sciences,
        The Universtity of Texas at Austin},
    country={United States of America}}
  \affiliation[uta-computing]{
    organization={Texas Advanced Computing Center,
        The Universtity of Texas at Austin},
    country={United States of America}}

  \author[cern,tum]{Tobias Duswald}
  \ead{tobias.duswald@tum.de}
  \author[uta-oden,uta-computing]{Ernesto A.B.F. Lima}
  \author[uta-oden]{J. Tinsley Oden}
  \author[tum]{Barbara Wohlmuth}

  \begin{abstract}
    \textit{Cancer} is a disease driven by random DNA mutations and the interaction
    of many complex phenomena. To improve the understanding and ultimately find more
    effective treatments, researchers leverage computer simulations mimicking the
    tumor growth \textit{in silico}. The challenge here is to account for the many
    phenomena influencing the disease progression and treatment protocols.
    This work introduces a computational model to simulate vascular
    tumor growth and the response to drug treatments in 3D. It consists of
    two agent-based models for the tumor cells and the vasculature. Moreover,
    partial differential equations govern the diffusive dynamics of the nutrients,
    the vascular endothelial growth factor, and two cancer drugs.
    The model focuses explicitly on breast cancer cells over-expressing
    HER2 receptors and a treatment combining standard chemotherapy (Doxorubicin) and
    monoclonal antibodies with anti-angiogenic properties (Trastuzumab). However,
    large parts of the model generalize to other scenarios.
    We show that the model qualitatively captures the effects of the combination
    therapy by comparing our simulation results with previously published
    pre-clinical data. Furthermore, we demonstrate the scalability of the model and
    the associated C++ code by simulating a vascular tumor occupying a volume of
    400$mm^3$ using a total of 92.5 million agents.
  \end{abstract}


\end{frontmatter}



\section{Introduction}

According to the WHO \cite{Sung2021}, cancer is one of the deadliest diseases
worldwide and was responsible for one out of six fatalities in 2020.
In the same year, officials registered 18 million new cases, roughly matching
the entire population of the Netherlands. The sheer number of people
suffering from cancer and the accompanying protracted fight against the disease
drew many scientists into cancer research. Experimentalists and theoreticians
alike strive to foster the understanding of tumor growth, disease progression,
and different treatment protocols. United in their goal to battle
cancer and improve the life quality of patients, experimentalists gather
quantitative data on cancerous systems, while at the same time, theoreticians
explore mathematical models for the disease, i.e., attempting to predict
its evolution and reaction to treatment.

Mathematical cancer models usually belong to one of the following three
categories:
(1)~models based on ordinary differential equations (ODEs),
(2)~models based on partial differential equations (PDEs),
and (3)~models based on discrete cell representations, which we refer to as
agent-based models (ABMs).
In 2016, Oden and co-workers \cite{Oden2016} reviewed these
approaches and embedded them into the wider context of the
predictive, computational sciences, and the associated data-generating
experiments. Earlier work from Byrne~\cite{Byrne2010} and
Beerenwinkel~\cite{Beerenwinkel2015} documented the progression of the field and
offer a great introduction to the topic.

Each of these modeling approaches involves strengths and weaknesses.
For instance, ODE models are comparatively cheap to compute but fail to
resolve spatial structures. PDEs incorporate spatial information but are
significantly more expensive to implement computationally.
Further, they employ homogenized tumor
properties (i.e., from tumor cells to cell densities), which may benefit more
extensive tumor simulations but limits their ability to resolve effects on a
cellular scale. This scale is best described with ABMs resolving
the individual tumor cells and allowing a natural way to
include cellular information. Unfortunately, the computational
costs can quickly get out of hand. In 2019, Metzcar and
coworkers~\cite{Metzcar2019} reviewed ABMs and their application in theoretical
cancer research. Their work offers an excellent overview of the state of the art
and the wide range of models leveraged by scientists.

Mathematical tumor models tend to consider simplified scenarios and, ABMs in
particular, often focus on small simulations because of the computational
costs. While simple models should generally be preferred \cite{Oden2017},
cancer thrives from complex interactions. To better understand them, the
complexity must, of course, be captured by the mathematical models.
In the experimental literature, researchers working on \textit{in vitro} drug screening
have long realized that, for instance, 3D cell cultures and tumor spheroids
better match \textit{in vivo} studies than flat, 2D cultures
\cite{BrassardJollive2020} and that the complex
tumor microenvironment has a strong influence on the tumor development
\cite{Balkwill2012,Wang2017,Anderson2020,Whiteside2008}.
Thus, it is of great interest to replicate the system's complexity
\textit{in silico} to test and improve the current understanding with
computational models \cite{Oden2003}.
However, the complexity poses new challenges as it requires
significant software development effort upfront before
a specific problem in a complex environment can be studied.

In this work, we present a novel hybrid model (ABMs + PDEs) simulating
vascular tumor growth and the response to therapy combining
Doxorubicin and Trastuzumab in 3D.
Our model consists of three major components:
(1)~a set of PDEs governing the diffusive dynamics of the nutrients,
the vascular endothelial growth factor (VEGF), and cancer drug compounds,
(2)~an off-lattice, center-based ABM with spherical agents for the tumor cells
governed by a cell cycle,
and (3)~another off-lattice, center-based ABM with cylindrical
agents describing the vasculature and sprouting angiogenesis.
The PDEs and ABMs are coupled in both ways, i.e., the continua influence the
agents and vice versa.
The vasculature supplies nutrients and treatment drugs, which are consumed by
the tumor cells; in contrast, tumor cells secrete VEGF triggering
vascular growth via sprouting angiogenesis.
Moreover, the tumor cells interact via two-particle forces.
To overcome the previously outlined limitations, we base our implementation on
the highly-efficient ABM simulation platform
BioDynaMo~\cite{Breitwieser2021, Breitwieser2023}, which enables our C++
application code to scale seamlessly from single-core machines to modern compute
nodes with hundreds of threads.
The entire source code of the project is available, together
with a Docker container and bash scripts to reproduce the findings.
\footnote{
  Available after final publication on GitHub:
  \href{https://github.com/TobiasDuswald/angiogenesis}{TobiasDuswald/angiogenesis};
  \href{https://github.com/TobiasDuswald/bdm-angiogenesis-reproducer}{TobiasDuswald/bdm-angiogenesis-reproducer}.
  Currently upon request.
}

While adopting ideas from previous research, most notably
\cite{Macklin2012,Rocha2018,Lima2021,Phillips2020,Breitwieser2021},
this work
represents several significant advances. First, we extend the previous ABMs
\cite{Rocha2018,Lima2021,Phillips2020} to a treatment scenario accounting for
two different cancer drugs and move from 2D to 3D.
Second, we show a novel way to model the vasculature and angiogenesis
in a general ABM context and couple it with PDE models.
Third, we show that the model
qualitatively describes several aspects of tumor dynamics
and captures the expected characteristics of the combination therapy
as hypothesized by Jain \cite{Jain2001} and experimentally investigated
in~\cite{Sorace2016}.
Lastly, we demonstrate that the code can handle tissue-relevant sizes by
simulating a $9 \times 9 \times 9 mm^3$ large volume hosting up to 92.5 million
agents over 27 days. Computationally, this significantly exceeds
that in previous work.

We first introduce the biological mechanisms of vascular tumor growth, the
considered cancer drug treatment, and the associated preclinical study
\cite{Sorace2016}
in Section~\ref{sec:medical_background}.
We proceed by detailing the mathematical model in Section~\ref{sec:model} and
devote Section~\ref{sec:estimate_parameters} to discussing our parameter
choices. In Section~\ref{sec:Results}, we run
the fully coupled model demonstrating the model's ability to simulate
vascular tumor growth and treatment by comparing the simulation results to the
preclinical study. We scale our simulation to
tissue-relevant scales in Section~\ref{sec:large-scale}.
We critically review our approach and address shortcomings in
Section~\ref{sec:discussion}.
Additionally, \ref{appendix: data} displays the data used in
Section~\ref{sec:large-scale}, \ref{appendix:model_parameter} gives an overview
of all model parameter, and \ref{appendix:stochastic-vessels} explains our
approach for statistically mimicking the initial vasculature for the
tissue scale.

\section{Preliminaries and Model Framework}\label{sec:medical_background}

In this work, we present a hybrid model simulating the vascular growth of a
tumor and its decline under treatment.
This section establishes the biological and medical background to
understand the model's components and reviews the literature.
We begin with summarizing the most important biological concepts of vascular
tumor growth and point the reader to related mathematical literature.
In Section~\ref{sec:dox_and_tra}, we sketch the mechanism of action of the two
cancer drugs considered by our model, Doxorubicin and Trastuzumab, and
outline why a treatment combining both may excel in efficacy in contrast
to current practice \cite{Jain2001}. The preclinical study supporting this
hypothesis~\cite{Sorace2016} is
presented in Section~\ref{sec:data} and provides the most relevant data for this
study.

\subsection{Vascular Tumor Growth and Mathematical Models}

Cancer is a disease evolving on a cellular scale; on the most fundamental level,
seemingly random mutations of the cell DNA occur during the regular cell cycle.
These DNA changes trigger abnormal behavior, mainly affecting cell
proliferation and mobility. Typically, cancerous cells replicate quicker than
healthy cells enabling them to locally out-compete the normal cells for
resources. However, increased proliferation and mobility are only two among
many phenomena that differentiate tumor cells from regular cells. In a seminal
series of papers \cite{Hanahan2000, Hanahan2011, Hanahan2022}, Hanahan and
Weinberg identified the \textit{hallmarks of cancer}, i.e., specific properties
that either tumor cells or populations thereof show in contrast to normal
tissue due to the altered DNA. They describe ten hallmarks and four further
candidates as of 2022 \cite[Fig.~1]{Hanahan2022}. While these hallmarks
characterize the tumor cells on small scales, Nia et al. \cite{Nia2020} linked
the hallmarks to macroscopic properties, which they called the
\textit{physical traits of cancer}. These traits encompass stress, pressure,
stiffness, and the complexity of the tumor microenvironment.
The hallmarks and physical traits of cancer form a solid basis for the
theoretical investigation of cancerous systems using mathematical tools
\cite{Magi2017}.

Among the ten hallmarks, \textit{inducing and accessing vasculature} is
particularly important for the present study.
If a local population of tumor cells grows, a commonly observed pattern is that
it drains the locally available energy resources, e.g., oxygen and glucose, and
creates a deadly, hypoxic environment for all cell types.
The tumor cells enter a hypoxic state and secrete signaling
substances such as VEGF to attract new vasculature, an observation usually
attributed to Folkman \cite{Folkman1971}. The existing vasculature reacts by
forming new sprouts that grow towards the hypoxic region to supply oxygen and
rescue the dying cells. This process, called sprouting
angiogenesis, forms a central component of this study. We note that there are
alternative mechanisms to increase the tumor's vascular density. However,
sprouting angiogenesis is usually dominant, and
consequently, we focus on it in the present work
(see~\cite[1.2.1]{Vilanova2015}).

As explained in the introduction,
cancer dynamics are typically modeled with ODEs, PDEs, ABMs, or combinations
thereof. An excellent summary of ODE methods can be found in Benzekry's
work~\cite{Benzekry2014}.
PDE based models leverage diffusive terms~\cite{Jiang2014, Lipkova2019} or a
phase field description~\cite{Xu2017, Xu2020, Fritz2023}.
More recently, models involving fractional diffusion dynamics have been
considered \cite{Fritz2021_subdiffusion}; i.e., diffusion processes that deviate
from the traditional Flick's law. ABMs have recently been reviewed by
Metzcar~\cite{Metzcar2019}. It is common practice to combine the three
approaches, e.g., using an ABM with cell internals modeled with an ODE system
while diffusing substances are modelled with PDEs \cite{Rocha2018,Rahman2017}.

Similar modeling paradigms have also been used to model the phenomena of
(sprouting)
angiogenesis. Villanova \cite{Vilanova2015} presents an introduction to
the topic in his Ph.D. thesis by summarizing different modeling approaches and
explaining the biological background. In his research
\cite{Vilanova2013,Vilanova2017}, he combines a discrete model for the tip cells
with a phase field model following them and classifying regions as
being vasculature or not, an approach similar to \cite{Milde2008}.
Fritz and co-workers \cite{Fritz2021Modelling} describe a complex,
coupled PDE model with a network growth algorithm considering the
vasculature's statistical features.
For more general reviews of angiogenesis models, we refer the reader to
\cite{Heck2015,Apeldoorn2022}, but for the present work, purely agent-based
angiogenesis models are at the center of attention.
Arguably one of the most important works in this regard has been carried out
by Bentley et al. \cite{Bentley2008,Bentley2009}.
They describe an initial blood vessel by points
located on a cylinder connected via mechanical springs and reacting to external
substances. Each point resembles an agent acting independently, forming sprouts
and predecessors to vessels.
Perfahl and co-workers \cite{Perfahl2017} modelled the vasculature as a chain
of spherical agents connected via springs showing similarities to our approach.
In contrast, Phillips et al. \cite{Phillips2020} model angiogenesis in a 2D
setting resolving the individual cells of the vessels modelled as tip and stalk
cells. Their cancer model shares significant features with ours, but the
angiogenesis module is conceptually different. Furthermore, we use the evolving
vasculature to model the supply of Doxorubicin and Trastuzumab, two drugs
discussed in the next section.

\subsection{Doxorubicin and Trastuzumab}\label{sec:dox_and_tra}

We consider a treatment protocol involving the two well-known
cancer drugs:
Doxorubicin (DOX) and Trastuzumab (TRA).
The U.S. Food and Drug Administration approved these drugs in 1974 and 1998,
respectively, and they routinely find use in clinical applications.
DOX is an \textit{anthracycline} frequently used in chemotherapy,
popular because of its high efficacy in fighting many different types of
cancer. In typical treatment scenarios, DOX is injected into the patient's
veins, from where it spreads through the body and, ultimately, begins
interacting with the cells. Effectively, DOX interrupts the DNA duplication by a
process referred to as \textit{intercalation} \cite{Tewey1984, Box2007}.
Once cells fail to duplicate their DNA, they trigger safety mechanisms, often
leading to the cell's death\cite{Cutts1996, Cutts2005}.
For more information on DOX and its effects on cells, we refer the reader to
\cite{Weiss1992, Carvalho2009, Rivankar2014, Sritharan2021} and references
therein.

TRA is a \textit{monoclonal antibody} and, thus, is more specific in its
therapeutic action than DOX.
In general, a monoclonal antibody is an antibody that only
binds to a specific molecular structure (e.g., a protein). After binding, the
antibody induces an immune reaction targeting its binding partner, which
may depend on the monoclonal antibody and the binding partner.
Historically, monoclonal antibodies had much success in cancer therapy
\cite{Scott2012}.
TRA specifically binds to the so-called
\textit{human epidermal growth factor receptor type-2} (HER2) located at the
surface of some tumor cells. HER2 is often
over-expressed in dangerous breast cancer variations (20-30\%).
The associated pathways lead to increased
proliferation and, thus, tumor formation.
When TRA binds to HER2, it inhibits
proliferation and reduces survival. Moreover, there is evidence that TRA shows
anti-angiogenic properties; i.e., it stops the formation of new blood vessels
and prunes and regularizes the exiting tumor vasculature
\cite{Petit1997, Petit2001}.
For an in-depth literature review, see \cite{Hudis2007}.

While both drugs have proven effective in fighting cancer, they may also
have severe side effects. For instance, DOX has been linked to
cardiotoxicity, neurological disturbances, and many other maladies
(see references in \cite[Section 3]{Carvalho2009}).
TRA is less harmful, but side effects still
occur \cite[Toxicity]{Hudis2007}. \textit{Combination therapy} strives to
combine different drugs into one therapy strategy such that the drugs
enhance each other's anti-tumor tumor properties while minimizing their
toxicity, i.e., damage to the patient. In 2001, Jain \cite{Jain2001}
suggested a new paradigm for combining anti-angiogenic therapies with regular
tumor treatment. He argued that anti-angiogenic drugs could be used to
regularize the tumor vasculature, allowing it to deliver other anti-cancer drugs
more effectively. For the case at hand, TRA would regularize the
vasculature improving its supply properties. Afterward, lower doses of DOX may
be sufficient to eradicate the tumor cell population.
In the present work, we provide a computational model designed to illustrate
this effect and to compare it to preclinical data introduced in the next
section.

\subsection{\textit{In vivo} Experiments for Combination Therapy}\label{sec:data}

\begin{figure}
  \centering
  \subfloat[Group 1]{
    {\includegraphics[width=.31\textwidth,valign=c]{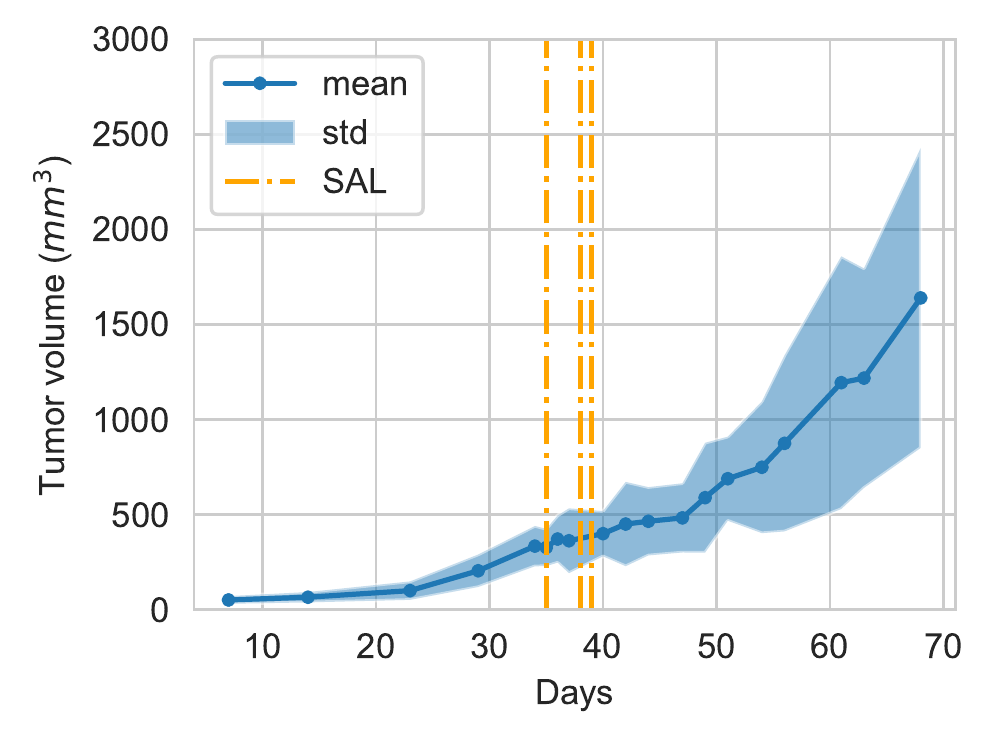}}
  }
  \ \
  \subfloat[Group 2]{
    {\includegraphics[width=.31\textwidth,valign=c]{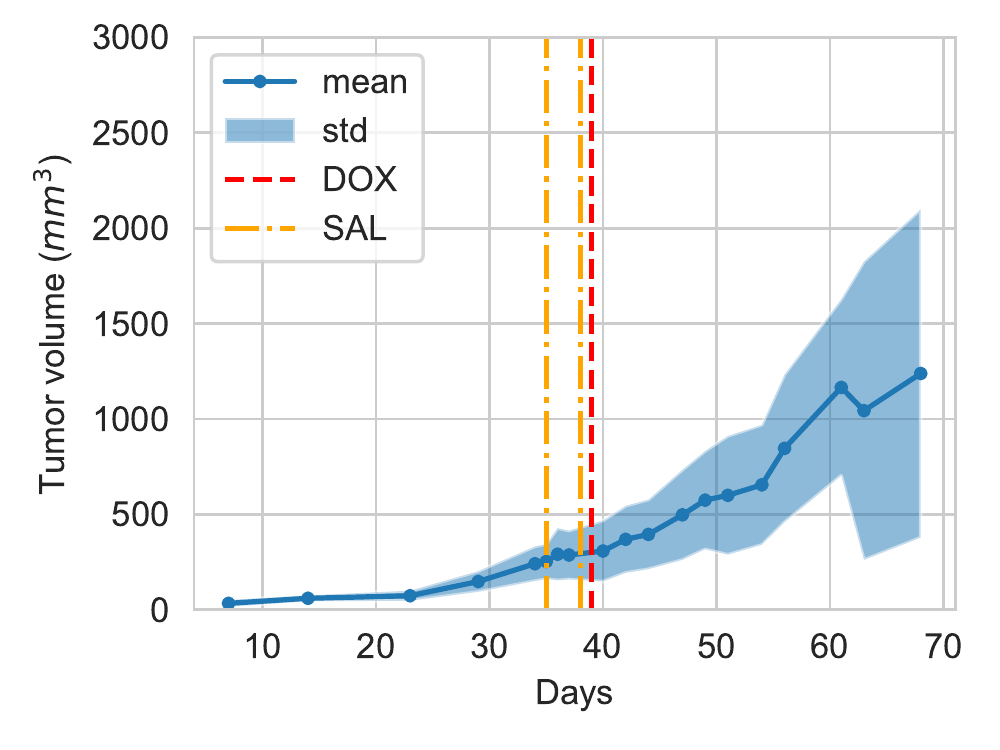}}
  }
  \ \
  \subfloat[Group 3]{
    {\includegraphics[width=.31\textwidth,valign=c]{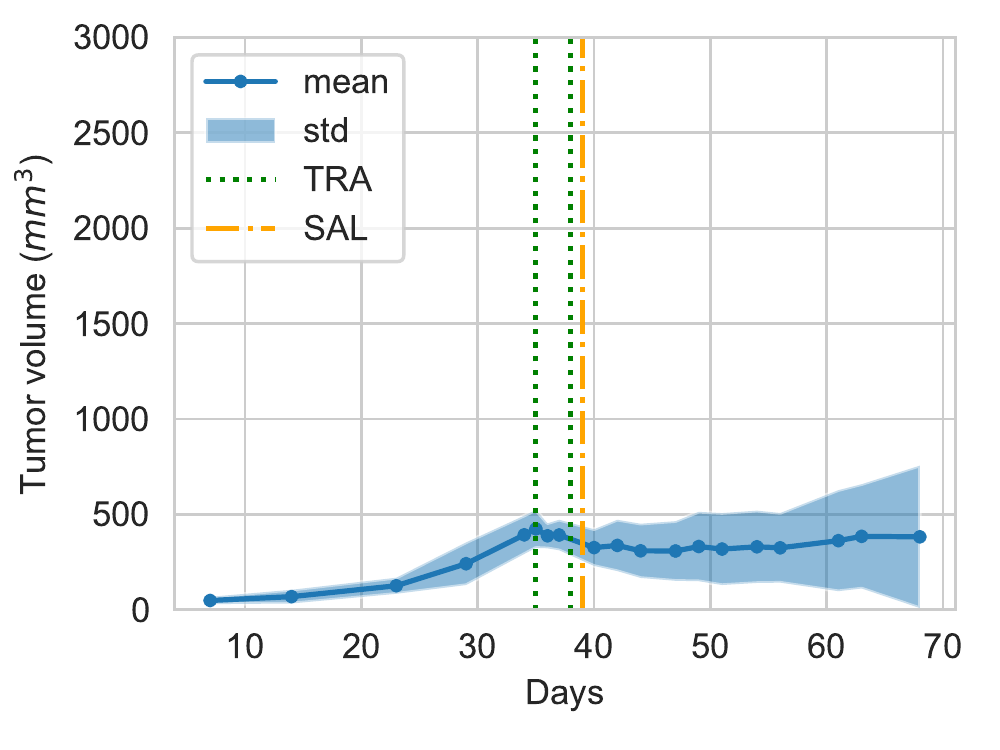}}
  } \\
  \subfloat[Group 4]{
    {\includegraphics[width=.31\textwidth,valign=c]{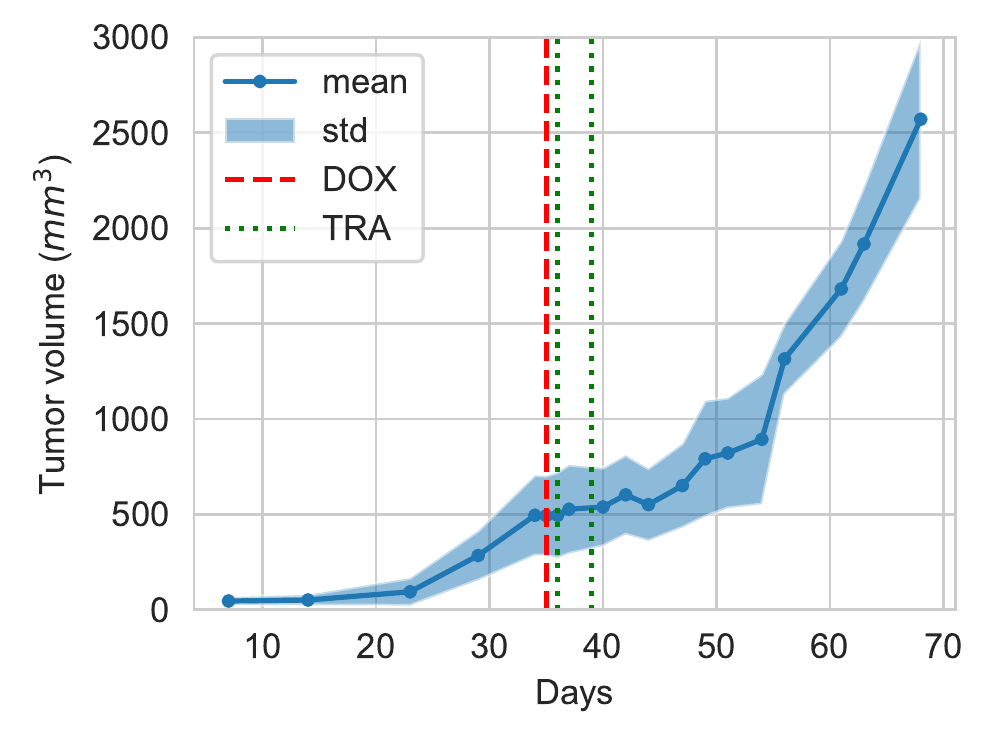}}
  }
  \ \
  \subfloat[Group 5]{
    {\includegraphics[width=.31\textwidth,valign=c]{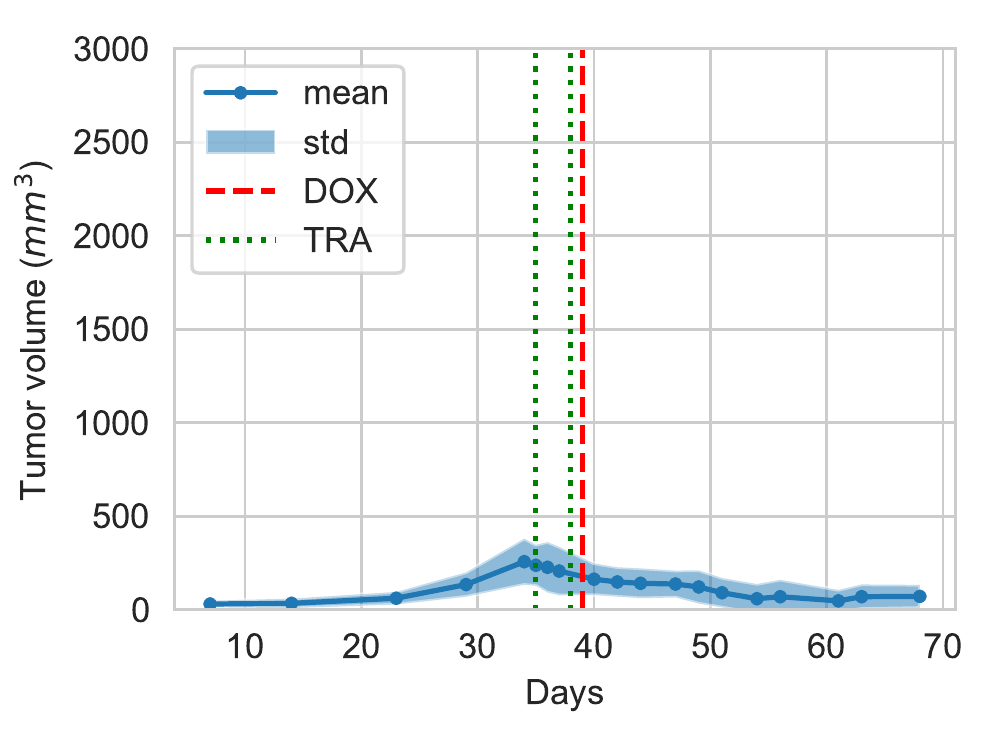}}
  }
  \ \
  \subfloat[Group 6]{
    {\includegraphics[width=.31\textwidth,valign=c]{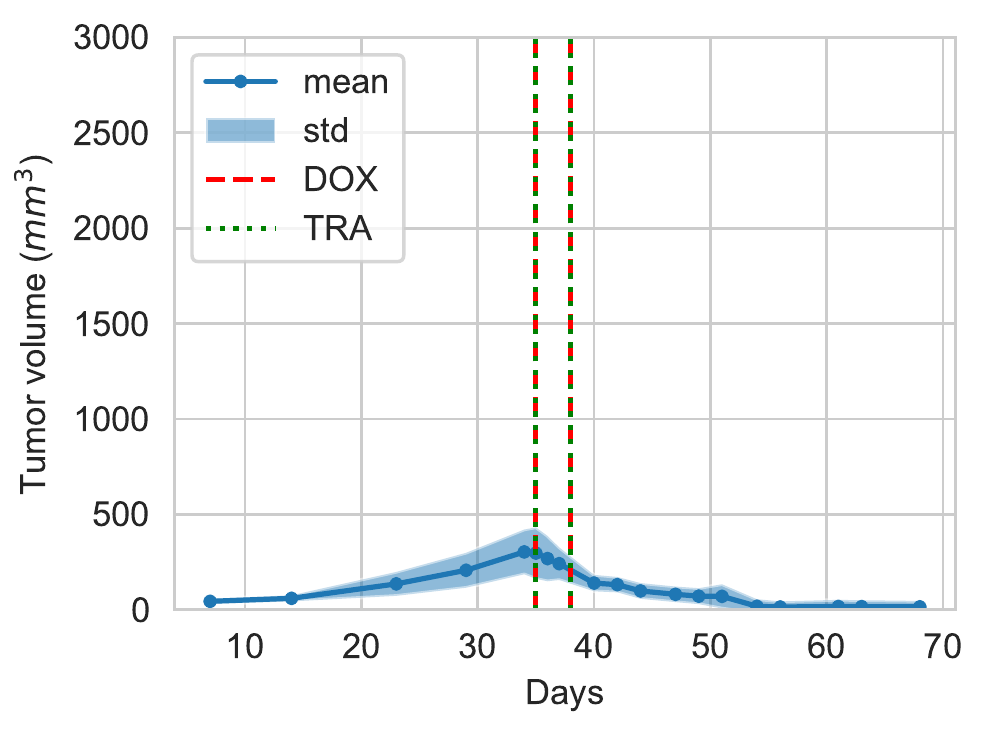}}
  }
  \caption{
    Mean and standard deviation of the tumor volume, measured over 70 days, of six
    different treatment groups: (a) group 1, (b) group 2, (c) group 3, (d) group 4,
    (e) group 5, and (f) group 6. The vertical lines indicate the day when each
    treatment was delivered (Doxorubicin (DOX), Trastuzumab (TRA),
    Saline (SAL)).
    Data taken from \cite{Sorace2016} and \cite[Tab.~1 and 5]{Lima2022}.
  }
  \label{fig:growth-data}
\end{figure}

Sorace et al. \cite{Sorace2016} tested Jain's hypothesis \cite{Jain2001} in a
pre-clinical \textit{in vivo} study. They injected HER2+ breast cancer cells
(BT474, ATCC) into the murine subjects and observed the tumor evolution over
70 days. They split the 42 murine subjects into six different treatment
groups:
\begin{itemize}
  \item Group 1: control group, treated with saline,
  \item Group 2: treated with DOX only,
  \item Group 3: treated with TRA only,
  \item Group 4: first treated with DOX, subsequently with TRA,
  \item Group 5: first treated with TRA, subsequently treated with DOX,
  \item Group 6: simultaneously treated with DOX and TRA.
\end{itemize}
All 42 animals
remained untreated for 35 days and showed similar disease progression. Once the
treatment started, Sorace and coworkers observed significant differences in
tumor volume over time between the groups. The observations are displayed in
Fig.~\ref{fig:growth-data}, which shows that the tumor volume grows exponentially
before the treatment begins. Furthermore, the treatments of groups 2
and 4 are observed to be ineffective.
For group 3, we observe stagnation, and for groups 5 and 6
a significant decline in tumor volume.
The data of these experiments were published in \cite[Tab.~1 and 5]{Lima2022}
together with a calibrated ODE model.
We merged the pre-treatment stages of the six groups into a separate
dataset given in Tab.~\ref{tab:tumor-evolution-data-merged}
in~\ref{appendix: data}.
These data, specifically the pre-treatment stage and the groups 2, 3, 4, and 5,
play a fundamental role in assessing the quality of our hybrid model later in
the results and discussion sections. We now shift our attention to the
core of this work: the hybrid model.

\section{The Hybrid Model}\label{sec:model}

Our model incorporates the evolution of the tumor mass, its nutrient and blood
supply, and the effects of the therapy in one comprehensive hybrid model.
The tumor mass is described with an agent-based model composed
of individual, spherical tumor cells independently progressing in their cell
cycle but dependent on the concentration of external substances.
The cells interact via two-particle forces.
The blood vessels are modeled with individual, cylindrical agents managed
in a tree-like structure, i.e., each agent has precisely one predecessor and either
one or two successors.
The developing vasculature
delivers the nutrients and drug compounds to the cells but also reacts to the
local VEGF gradient.
Four substances are modeled as scalar
fields: nutrients, VEGF, DOX, and TRA.
All four substances obey
reaction-diffusion equations and are coupled to the ABM via source and
sink terms proportional to regularized $\delta$-distributions marking the
agents' locations. Hence, the coupling between PDEs and ABMs is bi-directional.
We implemented the model in C++ based on the highly efficient BioDynaMo
framework \cite{Breitwieser2021, Breitwieser2023}.

In the following subsections, different parts of the model and their
interactions are described.
We begin with the tumor cells, their cell cycle,
and their interaction forces in Section \ref{sec:tumor_cell}.
We continue with the blood
vessels and explain the rules governing the dynamics of angiogenesis in
Section \ref{sec:blood_vessel}. The equations governing the scalar fields are
detailed in Section \ref{sec:continuum_models}. After discussing the separate
components of the model, we couple them in Section \ref{sec:coupling}.

\subsection{Tumor cell}
\label{sec:tumor_cell}

Our model describes the tumor on the cellular scale; i.e., each tumor cell
is explicitly modeled as a spherical agent with stochastic behaviors. In the
simulation, a tumor cell is a C++ object with various attributes. Focusing on
the most relevant examples, a tumor cell is characterized by
(1) a unique ID,
(2) its position $\vec{x}$ in the three dimensional space,
(3) its nuclear, physical, and action radii ($r_n,r_p,r_a$),
(4) its cell state $s$, and
(5) an internal clock tracking the time since the last state transition
$\Delta t_s$ to model the time-dependent phases of the cell cycle. All but the
cell state $s$ are real, possibly vector-valued numbers.

The cell state $s$ is a categorical
variable taking values $s \in \{ Q, SG2, G1, H, D \}$. $Q$ is the quiescent
state in which the cell is idle and no special events occur. $SG2$ and $G1$
denote the proliferative cell states. In $SG2$, the cell duplicates its inner
components and prepares for cell division. The volume-preserving cell
division marks the transition from $SG2$ to $G1$. In $G1$, the cells grow until
they reach their natural size. The states $H$ and $D$ denote the hypoxic and
dead states, respectively.

The different cell states form the core of the tumor growth model. The
transitions between them depend on the values of the four continua - the
nutrients, VEGF, DOX, and TRA. We denote them as $u_n$, $u_v$,
$u_d$, and $u_t$, respectively. The second basic component of the tumor model is
the force model consisting of
the repulsive and adhesive forces governing the cell-cell interaction.
In what follows, we describe the stochastic model underlying the
state transitions and detail the forces and their algorithmic computation
afterward. The model of the cell cycle and the forces are, in part, based on
previous work \cite{Macklin2012,Rocha2018,Phillips2020,Lima2021}.

\subsubsection{Cell cycle}

\begin{figure}
  \centering
  \includegraphics[width=0.7\textwidth]{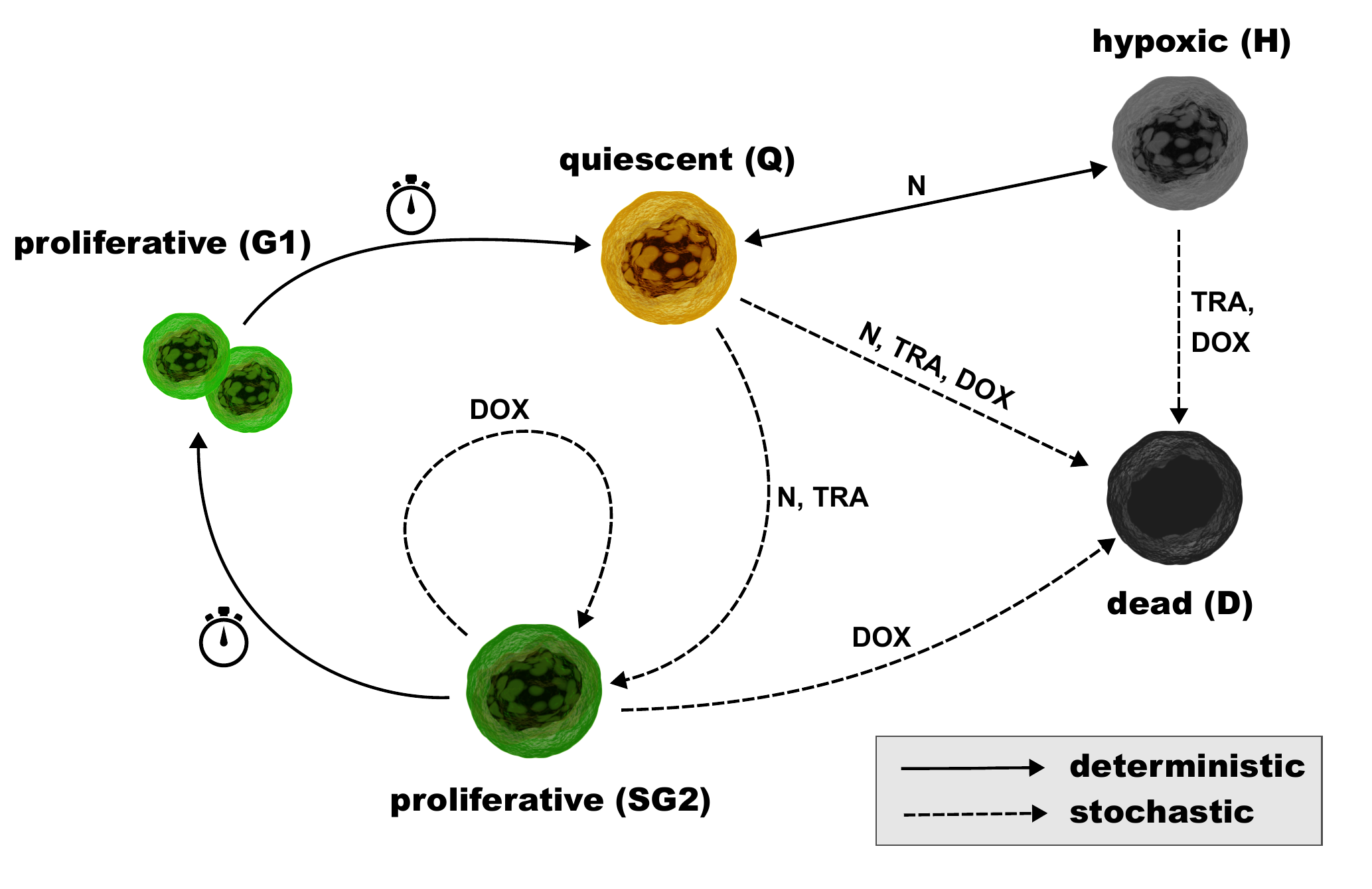}
  \caption{The cell cycle for the tumor cells. Deterministic and stochastic
    transitions are indicated by solid and dashed lines, respectively. The arrows
    indicate the direction of the transition. Transitions depending on the
    concentration of the nutrients (N), DOX, or TRA are labeled accordingly.
    Transitions solely depending on an internal clock are label with a
    stopwatch.\\
    \begin{scriptsize}
      Cell representation: \textit{Cancer cell} from the \textit{DataBase Center
        for Life Science (DBCLS)} distributed under \textit{Creative Commons
        Attribution 4.0 International license} (modified).
      Stopwatch: \textit{Stop Watch} from \textit{SimpleIcon} distributed
      under \textit{Creative Commons Attribution 3.0 Unported}.
    \end{scriptsize}
  }
  \label{fig:cell_cycle}
\end{figure}

The progression of a tumor cell in its cell cycle depends on the local
substance concentrations but not on the surrounding cells.
The state transitions are governed by stochastic as well as
deterministic rules. Given the five states $(Q, SG2, G1, H, D)$, our cell cycle
allows transitions $Q \rightarrow SG2 / H / D$, $SG2 \rightarrow G1 / D$,
$G1 \rightarrow Q$, and $H \rightarrow Q / D$. In its entirety, the cell cycle
is best represented graphically as depicted in Fig.~\ref{fig:cell_cycle}.

We first focus on the three deterministic transitions: $SG2 \rightarrow G1$,
$G1 \rightarrow Q$, and $Q \rightarrow H$. The first two transitions simply
require the cell to spend a given time in the cell state assuming that the time
for the cells to duplicate their internals and their growth is fixed.
For $SG2 \rightarrow G1$ and $G1 \rightarrow Q$, the times are $T_{SG2}$ and
$T_{G1}$, respectively. The transition from $Q$ to $H$ depends on the nutrient
concentration, i.e., if the concentration falls below the hypoxic
threshold, $u_n^H$, the cells transition from $Q$ to $H$; if the concentration
raises above $u_n^H$, the cells transition from $H$ to $Q$. The deterministic
transitions are indicated by solid lines in Fig.~\ref{fig:cell_cycle}.
We use the
Iverson brackets $[ \cdot ]$ to denote an if-statement: if the condition inside
the brackets evaluates to true, the cell moves from one state to the other.
With this notation, we describe the deterministic transitions as
\begin{align}
  SG2 \rightarrow G1  : & \ \ [ \Delta t_s \geq T_{SG2} ] \ ,    \\
  G1 \rightarrow Q    : & \ \ [ \Delta t_s \geq T_{G1} ] \ ,     \\
  Q \rightarrow H     : & \ \ [  u_n < u_n^H ] \ ,  \ \text{and} \\
  H \rightarrow Q     : & \ \ [  u_n \geq u_n^H ] \ .
\end{align}

To characterize the stochastic transitions, we first introduce two functions
$\varsigma$ and $\varrho$ \cite{Lima2021,Macklin2012} appearing repeatedly in the transition
probabilities describing a smoothed Heaviside function and linear increase,
respectively. The functions are given by the following equations:
\begin{align}
  \varsigma (x;a,b,\bar{x}) & =
  1 -
  \exp \left(
  -\left(
    a +
    \frac{1}{1+ \exp(2 \cdot b
        \cdot (x - \bar{x}))}
    \right)\Delta t
  \right) \ , \ \text{and}      \\
  \varrho (x; c, \bar{x})   & =
  1 -
  \exp \left(
  - \max
  \left(
    c \cdot \frac{x-\bar{x}}{1-\bar{x}} , 0
    \right) \Delta t
  \right) \ .
\end{align}
The dependent variable $x$ and its parameters are
separated by a semi-colon. The construction of the functions implicitly assumes
bounded values $x, \bar{x} \in [0,1] \subset \mathbb{R}$.
Moreover, $\Delta t$ denotes the simulation time step.
For $\varsigma$, the parameter $a$ offsets the function along the y-axis, the parameter
$b$ models the sharpness of the transition, and the parameter $\bar{x}$
defines the transition point.
For $\varrho$, the parameter $a$ describes the slope, and $\bar{x}$ defines the
starting point of the linear increase.

The stochastic transitions in the cell cycle are indicated by the dashed
lines in Fig.~\ref{fig:cell_cycle}. We extend the work of
\cite{Macklin2012,Rocha2018,Phillips2020,Lima2021} to account for the
concentration of TRA and DOX. The transition probability for a tumor
cell at position $\vec{x}$ from $Q \rightarrow SG2$ is modeled as
\begin{equation}\label{eq:Q_to_SG2}
  P_{Q \rightarrow SG2} =
  \varrho \left(
  u_n(\vec{x});
  c_{Q \rightarrow SG2}, u_n^{Q \rightarrow SG2}
  \right) \cdot
  \exp \left(
  -\lambda_{Q \rightarrow SG2}
  u_t(\vec{x})
  \right) \ ,
\end{equation}
where we introduce three parameters characterizing the $Q \rightarrow SG2$
transition indicated by a sub- or superscript. Note that
$\lambda_{Q \rightarrow SG2} \geq 0$. The exponential suppression
is introduced because
TRA leads to cell cycle arrest \cite{Albanell2003}.
Introducing more parameters, we express the remaining stochastic transitions as
\begin{align}
  P_{Q \rightarrow D}     & =
  \varsigma \left(\
  u_n(\vec{x});
  a_{Q \rightarrow d}, b_{Q \rightarrow d}, u_n^{Q \rightarrow d}
  \right) \cdot
  \left(
  1
  + \xi_d^{Q \rightarrow D} u_d
  + \xi_t^{Q \rightarrow D} u_t
  + \xi_{dt}^{Q \rightarrow D} u_d u_t
  \right) ,                   \\
  P_{SG2 \rightarrow SG2} & =
  \varrho \left(
  u_d(\vec{x});
  c_{SG2 \rightarrow SG2}, u_d^{SG2 \rightarrow SG2}
  \right) \ ,                 \\
  P_{SG2 \rightarrow D}   & =
  \varrho \left(
  u_n(\vec{x});
  c_{SG2 \rightarrow D}, u_n^{SG2 \rightarrow D}
  \right) \ ,                 \\
  P_{H \rightarrow D}     & =
  \left(
  r_{H \rightarrow D} \cdot \Delta t
  \right) \cdot
  \left(
  1 + \xi_d^{H \rightarrow D} u_d
  + \xi_t^{H \rightarrow D} u_t
  + \xi_{dt}t^{H \rightarrow D} u_d u_t
  \right) \ . \label{eq:P_H_D}
\end{align}
Here, linear and cross terms are added to parametrize the treatment effect.
These terms
appear in the $Q \rightarrow D$ and $H \rightarrow D$ transitions. We
further introduce the $SG2 \rightarrow SG2$ transition triggering a reset of
the internal clock. This transition models DOX's ability to interfere with the
DNA duplication process via intercalation \cite{Tewey1984, Box2007}. If the DNA
duplication process fails, cells may die which we capture with the added
$SG2 \rightarrow D$ transition. Theoretically, the probabilities for
$Q \rightarrow D$ and  $H \rightarrow D$ may exceed 1 for certain parameter
choices; practically however, this does not harm the implementation. One may
formally rewrite the transitions as $min(max(0,\bullet ),1)$ to ensure a proper
probability interpretation.
We note that the cell cycle is parametrized by a total of 20 parameters
(see Tab.~\ref{tab:ParameterCellCycle} in \ref{appendix:model_parameter}
for all parameter values).

\subsubsection{Cell-cell forces}

The cell-cell forces are taken from Rocha et al. \cite{Rocha2018}.
In summary, the cells
are represented as spheres with three radii (action, regular, and nuclear).
The cell-to-cell force is a two-particle force depending on the distance
$d$ between two cells.
It has adhesive and repulsive components that depend on the overlap of the
different radii.

To compute the displacement of cell $(i)$, all contributions from all
other cells are summed; e.g.,
\begin{equation}
  \vec{F}_i = \sum_{j \neq i} \vec{F}_{ij} \ .
\end{equation}
We note that the force is zero if $d \geq R_A$, where $R_A$ is the sum of the
two action radii of the cells involved in the interaction. In other words, the
force is zero if cells do not overlap. Thus, we can define an index set
\begin{equation}
  \mathcal{N}_i = \{ j \neq i \ \vert \ d(\vec{x}_i,\vec{x}_j)
  \leq 2 \cdot \max_k (r_a^{(k)}) \}
\end{equation}
containing all neighbor cells whose action radii ($r_a$) overlap with the one
of cell-$i$, and compute the force as
\begin{equation}
  \vec{F}_i = \sum_{j \in \mathcal{N}_i} \vec{F}_{ij} \ .
\end{equation}
This is significantly cheaper to compute because only iterates over a small
index set are used rather than all other cells.
Computationally, we determine the index set $\mathcal{N}_j$ through an efficient
neighbor query based on an artificial, uniform grid with a discretization length
$h = 2 \cdot \max_k (r_a^{(k)})$, i.e., twice the largest action radius
observable in the simulation at this time
\cite{Breitwieser2021, Breitwieser2023}.
After computing $\vec{F}_i$, we update the position of the cell $(i)$ as
\begin{equation}
  \vec{x}_i(t+\Delta t) = \vec{x}_i(t) + \eta \vec{F}_i \cdot \Delta t \ ,
\end{equation}
where $\eta$ is a viscosity parameter describing the linear relationship between
displacement and force. We note that we do not consider forces between the
tumor cells and vessels which we discuss next.

\subsection{Blood Vessels and Angiogenesis}
\label{sec:blood_vessel}

To model the nutrient supply, the vasculature is decomposed into small
cylindrical compartments. Each compartment is computationally represented by
a cylindrical agent. We base our implementation on BioDynaMo's neurite class
\cite{Breitwieser2021} whose equations are detailed in \cite{Zubler2009}.

More broadly viewed, the vasculature is represented as a linked, tree-like
data structure. The structure is realized by assigning three references to each
cylindrical vessel-agent containing the address of its unique predecessor
and the optional addresses of up to two successors. We refer hereafter to these
as mother and daughters. If a vessel-agent has no daughter, it is called
the terminal end or tip cell; if it has one daughter, it is part of a regular,
longer vessel segment; and if it has two daughters, we call it a branching
point.

Assuming a given, initial vasculature, the individual vessel-agents evolve
independently from each other based on locally available information.
In the language of agent-based modeling, each agent executes the same
stochastic behaviors (rules) describing how the object changes in a time step
$\Delta t$ depending on the local information.
In this work, we focus on VEGF-triggered sprouting angiogenesis.
Our stochastic rules
differentiate between tip cells, the cells embedded in a regular vessel
segment, and the branching points. The latter remain unchanged because they
can neither form any further branches nor can they extend into any direction.

Any vessel agent that is part of the normal vasculature with a single mother and
a single daughter is a candidate for branching and, therefore, can create a new
tip cell. If the agent is allowed to branch depends on three criteria.
First, we require the
concentration of VEGF at the agent's position to surpass a threshold
$u_v(\vec{x}) \geq u_v^{\text{thres}}$. Second, it has been observed that new
tip cells only form, if there are no other tip cells in its vicinity.
Hence, we require a minimum Euclidean distance $d_{\text{tip}}$ to the closest
tip cell (see discussions and references in \cite{Phillips2020, Bentley2008}).
In other words, denoting the set of all tip cells as $\mathcal{T}$,
we demand that
\begin{equation}
  \min_{k \in \mathcal{T}} (|| \vec{x} - \vec{x}_k || ) > d_{\text{tip}} \ .
\end{equation}
Thirdly, to ensure the mechanical stability of the vessel, branching points must
be separated by a minimal distance $d_{\text{branch}}$ measured along the
vessel. Denoting the curve defined by the vessel from the agent to the preceding
and succeeding branching point as $\Gamma_p$ and $\Gamma_s$, respectively,
a valid point for branching satisfies
\begin{equation}
  \int_{\Gamma_p} ds > d_{\text{branch}} \ \ \text{and} \ \
  \int_{\Gamma_s} ds > d_{\text{branch}} \ .
\end{equation}
To evaluate the tip cell distance criteria, we leverage an octree implementation
\cite{Behley2015} updated after each simulation time step. To evaluate the
distance to the branches, we iterate over the tree structures.

If all three criteria are satisfied, the agent evaluates a stochastic
branching rule: if the generated random uniform number $X \sim U(0,1)$ is
smaller than the sprouting probability $p_s = p_{s,\text{rate}} \cdot \Delta t$,
the agent creates a second daughter whose cylinder axis lies on a (random) cone
around the VEGF gradient $\nabla u_v$. We remark that we choose the parameter
$p_{s,\text{rate}}$ to be very small, e.g., it satisfies
$p_{s,\text{rate}} \cdot \Delta t \ll 1$ for reasonable choices of $\Delta t$.

Tip cells, on the contrary, are never candidates for branching, i.e., we do not
split vessels at the terminal end.
Tip cells follow the VEGF gradient $\nabla u_v$ to establish the vasculature in
undersupplied regions characterized by high VEGF concentrations.
We allow growth if the gradient at the tip cell's position surpasses a
threshold $\nabla u_v(\vec{x})$.

When growing, we elongate the tip cells in the direction of the vector
\begin{equation}
  w_1 \nabla u_v (\vec{x}) + w_2 \vec{a} + w_3 X_3 \ ,
\end{equation}
where $w_1, w_2, w_3 \in \mathbb{R}^+$ are the modeling weights, $\vec{a}$
denotes the axis of the cylinder, and $X_3 \sim U(-1,1)^3$.
The modeling weights determine the direction of the growth, i.e., increasing
$w_1$ leads
to vessels that follow the gradient more closely, increasing $w_2$ creates
inert vessels barely changing their directions, and increasing $w_3$ allows
more and more randomness in the growth.

To avoid unlimited growth after tip cell selection, a criterion to determine
when to stop the growth is needed.
Multiple criteria may be used; e.g.,
large VEGF concentrations, strong gradients, or some engineered criteria such
as the quotient of the gradient magnitude and the concentration. In practice,
stopping the growth once vessels reach high gradients proved effective.

The description of the initial vasculature is taken up in
Section~\ref{sec:estimate_parameters}.
To connect the new,
evolving vasculature to the initial structure, we first branch from a
cylinder with diameter $d_0$. The diameter of the new branch is computed as
\begin{equation}
  d_1 = \max \Big( 5 \mu m,\min \left(0.8 \cdot d_0, \ 20 \mu m \right) \Big).
\end{equation}
Furthermore, the diameter is decreased along the vessels. Typically, we
elongate the cylinders until they reach a length of $10\mu m$. We then split
them into two cylinders of length $9\mu m$ and $1 \mu m$. Denoting the initial
diameter as $d_0$, the diameter of the second agent is computed as
\begin{equation}
  d_1 = \max \Big( 5 \mu m,\min \left(0.98 \cdot d_0, \ 20 \mu m \right) \Big).
\end{equation}
These heuristic criteria allow us to connect the
microvasculature to larger initial vessels smoothly.

\begin{figure}
  \centering
  \includegraphics[width=0.5\textwidth]{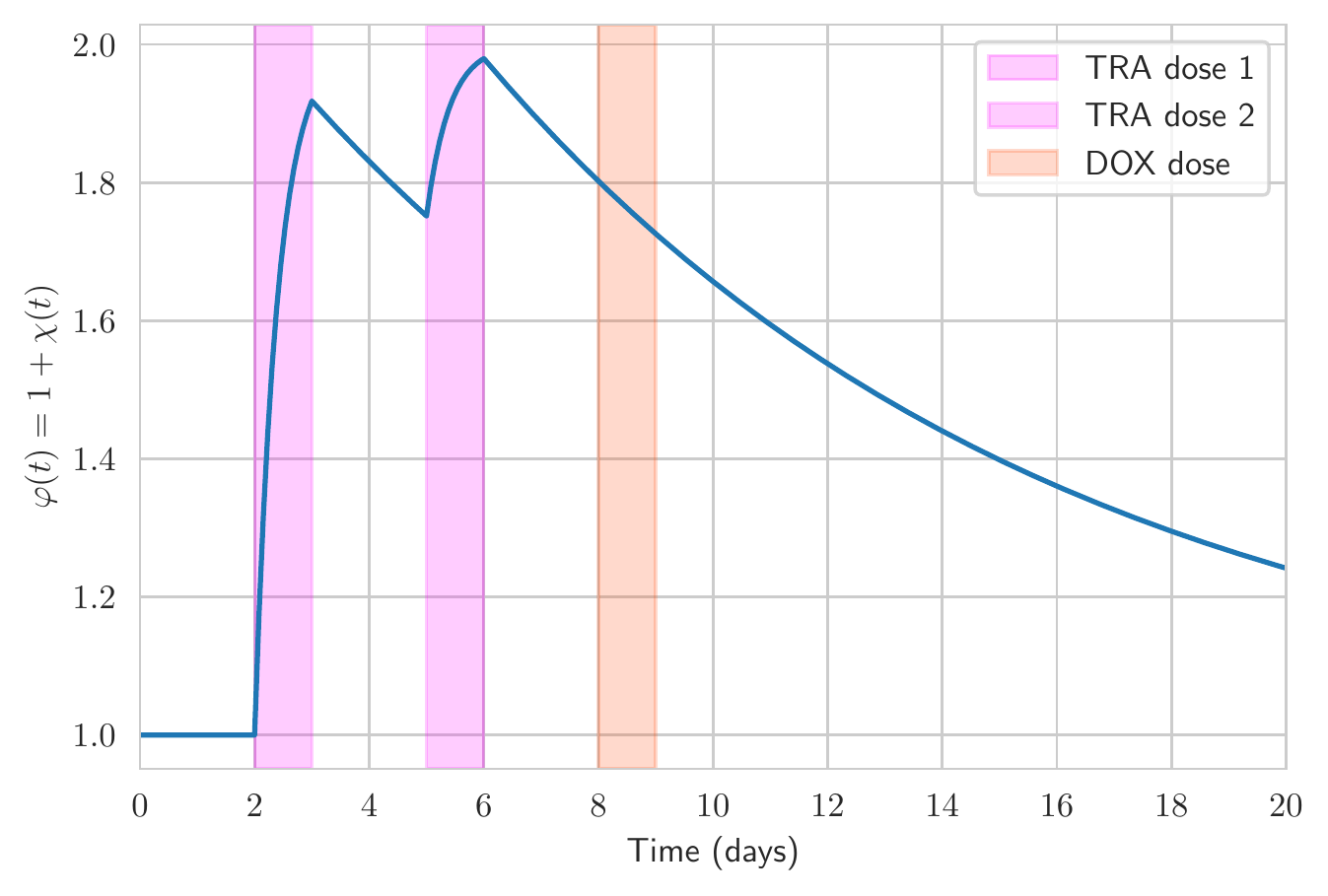}
  \caption{Exemplary evolution of the DOX supply $\varphi (t)$ during a
    treatment protocol. In this illustration, one may expect roughly 75\% more
    DOX to be delivered via the vasculature compared to no TRA treatment.
    Note that $\varphi(t)$ remains unaffected by DOX.
  }
  \label{fig:supply-over-time}
\end{figure}

Lastly, we model the effect of TRA treatment on the vessel. TRA has been shown
to regularize the vasculature and improve the supply properties
\cite{Petit1997, Petit2001}. We capture this effect by modulating the DOX supply, $\varphi (t)$,
with a time-dependent supply factor, $\chi (t)$,
\begin{equation}\label{eq:doc_supply_1}
  \varphi (t) = 1 + \chi (t) \ ,
\end{equation}
where a capacitor like ODE describes the time dependence
\begin{equation}\label{eq:doc_supply_2}
  \frac{d \chi}{d t} = \begin{cases}
    \frac{1}{\tau_\uparrow } (\chi_{\max} - \chi) \  & \
    \text{during TRA treatment,}                         \\
    \frac{-1}{\tau_\downarrow } \chi  \              & \
    \text{else.}
  \end{cases}
\end{equation}
An example of the time dependence of the supply factor is given in
Fig.~\ref{fig:supply-over-time}.

\subsection{Continuum models}
\label{sec:continuum_models}

Our model involves four continua: the nutrients, VEGF, and the drugs DOX
and TRA. Recall that we denote their concentration as $u_n$, $u_v$, $u_d$, and
$u_t$, respectively.
In brief, the four continua have the following roles.
The nutrients model the energy supply of the system. The more
nutrients, the more likely the tumor cell transition into proliferative states
and multiply. In the absence of nutrients, cells become hypoxic and eventually
die (necrosis).
VEGF is the signaling pathway allowing dying cells to trigger angiogenesis and
attract blood vessels improving the nutrient supply. TRA and DOX
disturb the regular cell cycle, prohibiting proliferation
and favoring transitions into hypoxic or dead states.

Assuming no chemical interactions between the different substances, each of the
four substances diffuse independently in the simulation domain and may
naturally decay over time.
The parameters depend on the substance under consideration. The tumor cells
and blood vessels act as source and sink terms in the continuum models and we
denote them as $c(\vec{x}; \vec{\alpha})$ and $v(\vec{x}; \vec{\beta})$,
respectively, where $\vec{\alpha}$ and $\vec{\beta}$ are parameter vectors.
Both functions are effectively a
sum of $\delta$-distributions with strictly positive coefficients, e.g.,
$c(\vec{x}; \vec{\alpha}) = \sum_j \alpha_j \delta(\vec{x}-\vec{x_j})$ and
$v(\vec{x}; \vec{\beta}) = \sum_j \beta_j \delta(\vec{x}-\vec{x_j})$
with $\alpha_j, \beta_j > 0$ (more details for the coupling in
Section~\ref{sec:coupling}). The equations governing the continuum model are
\begin{align}\label{eq:continua}
  \left(
  \frac{\partial}{\partial t} - \nabla \cdot D_n \nabla + \lambda_n
  \right) u_n & = - u_n c_{\vec{\alpha}_n} + (1-u_n) v_{\vec{\beta}_n} \
              & \text{with} \ \ \
  \vec{n} \cdot D_n\nabla u_n = 0 \ \ \mbox{on} \ \ \partial \Omega \ ,                         \\
  \left(
  \frac{\partial}{\partial t} - \nabla \cdot D_v \nabla + \lambda_v
  \right) u_v & = + (1-u_v)c_{\vec{\alpha}_v} - u_v v_{\vec{\beta}_v} \
              & \text{with} \ \ \
  \vec{n} \cdot D_v\nabla u_v = 0 \ \ \mbox{on} \ \ \partial \Omega  \ ,    \label{eq:PDE_VEGF} \\
  \left(
  \frac{\partial}{\partial t} - \nabla \cdot D_d \nabla  + \lambda_d
  \right) u_d & = - u_d c_{\vec{\alpha}_d} + \varphi(t) \ (1-u_d) v_{\vec{\beta}_d} \
              & \text{with} \ \ \
  \vec{n} \cdot D_d\nabla u_d = 0 \ \ \mbox{on} \ \ \partial \Omega  \ ,                        \\
  \left(
  \frac{\partial}{\partial t} - \nabla \cdot D_t \nabla + \lambda_t
  \right) u_t & = - u_t c_{\vec{\alpha}_t} + (1-u_t) v_{\vec{\beta}_t} \
              & \text{with} \ \ \
  \vec{n} \cdot D_t\nabla u_t = 0 \ \ \mbox{on} \ \ \partial \Omega \ .
\end{align}
Here, the diffusion and decay constants are labeled as $D_i$ and $\lambda_i$,
respectively. The terms $(u_i)$ and $(1-u_i)$ ensure that
the concentration remains bounded by zero and one.
They also indicate whether
the tumor cells and the blood vessel act as a source or a sink. For instance,
tumor cells secrete VEGF but consume nutrients, DOX, and TRA. For the blood
vessels, the opposite is true, i.e., they consume VEGF but provide nutrients,
DOX, and TRA. The factor $\varphi (t)$ was defined in
Eq.~(\ref{eq:doc_supply_1}) and (\ref{eq:doc_supply_2}).

The equations are solved with a finite difference scheme on a cube domain.
We use the forward difference in time and the
central difference in space, a scheme commonly abbreviated as FTCS scheme.
Labeling the points in the (isotropic) lattice with the triple $(i,j,k)$ and
denoting the concentration at this point at time step $n$ as $u^n_{i,j,k}$,
the discrete
stencil computation to evolve the continuum models over time is given by
\begin{equation}\label{eq:stencil-computations}
  \begin{split}
    u^{n+1}_{i,j,k} = & (1 - \lambda \Delta t) u^{n}_{i,j,k}  \\
    &+ \frac{D \Delta t }{h^2} \cdot
    \left(
    u^{n}_{i+1,j,k} + u^{n}_{i-1,j,k} + u^{n}_{i,j+1,k} + u^{n}_{i,j-1,k}
    + u^{n}_{i,j,k+1} + u^{n}_{i,j,k-1} - 6 u^{n}_{i,j,k}
    \right) \\
    &+ \Delta t (1-u_{i,j,k}^n) \cdot A_+(i,j,k)
    - \Delta t (u_{i,j,k}^n) \cdot A_-(i,j,k) \ ,
  \end{split}
\end{equation}
where $h$ is the grid size, i.e., the distance between neighboring points.
$A_+$ and $A_-$ characterize all agent source and sink terms, respectively.
For the discrete setting, the
distribution $\delta(\vec{x})$ equals one if the grid point $(i,j,k)$ is
the closest one, and is zero otherwise. In other words, if the
point $\vec{y}$ is labeled as $(l,m,n)$, the distribution
$\delta(\vec{x}-\vec{y})$ reduces to a product of Kronecker deltas
$\delta_{il}\delta_{jm}\delta_{kn}$.
For stability, the time step $\Delta t$ is bounded from above by
\begin{equation}\label{eq:stability}
  \left( \lambda + \frac{12 D}{h^2} \right) \Delta t \leq 2 \ ,
\end{equation}
which follows from a standard stability analysis of the finite difference
scheme.

\subsection{Coupling of ABM and continuum}
\label{sec:coupling}

The continuum and the agent-based model are coupled: agents
influence the evolution of the continuum via the source and sink terms, and the
continuum values determine how the tumor cells progress in the cell cycle and
drive the growth of the vasculature.
For the latter, the interactions have been detailed in
Section~\ref{sec:tumor_cell} and \ref{sec:blood_vessel}. For both agent types,
the location of its center $\vec{x}$ is identified and the closest grid value
$u_{i,j,k}$ is retrieved when we evaluate the probabilities with dependencies on
the concentration of nutrients, VEGF, DOX, or TRA.

A detailed explanation of the source and sink terms is warranted. We denote the
set of all tumor cells as $\mathcal{T}$ and the set of all blood vessel agents
as $\mathcal{V}$. For the tumor cells, an agent $i$ is identified by its center
coordinate $\vec{x}_i$. The function $c(\vec{x},\vec{\alpha_i})$ ($i=n,v,d,t$)
takes on the general form
\begin{equation}\label{eq:cell-sources}
  c(\vec{x},\vec{\alpha_i}) =
  \sum_{k \in \mathcal{T}} \alpha_{i,k} \delta(\vec{x} - \vec{x}_k) \ ,
\end{equation}
where the coefficients $\alpha_{i,k}$ are functions modeling a dependence on
the
agents attributes. For instance, the consumption of nutrients could be chosen
to be proportional to the cell size (surface or volume).
Ultimately, only the VEGF model leverages the function character of the
coefficients because only hypoxic cells secrete VEGF. Thus, the coefficients
read
\begin{equation}
  \alpha_{v,k} = \alpha_v \cdot
  \begin{cases}
    0 \ \text{if} \ s_k \neq H \\
    1 \ \text{if} \ s_k = H
  \end{cases} \ ,
\end{equation}
with the positive, scalar parameter $\alpha_v$. For all other continua
($i=n,d,t$), we reduce the coefficients to a scalar dependence
\begin{equation}
  c(\vec{x},\vec{\alpha_i}) =
  \alpha_{i} \sum_{k \in \mathcal{T}} \delta(\vec{x} - \vec{x}_i) \ ,
\end{equation}
indicating that the source and sink terms only depend on the presence of the
agent but not on any further attributes. We note that dead cells are exempt from
the interactions because they neither consume nutrients nor secrete VEGF.

For the blood vessel, we reduce an agent to a number of points along the
cylinder's center line. For a vessel-agent $i$, the number of points $m_i$ is
determined at each time step based on its length $l_i$ and the
smallest grid constant $h_j$ ($j=n,v,d,t$) as
\begin{equation}
  m_i = \max
  \left(
  3, \text{ceil}
  \left(
    \frac{2l_i}{\min_j h_j} + 1
    \right)
  \right) \ ,
\end{equation}
where $\text{ceil}:\mathbb{R} \rightarrow \mathbb{N}$ maps a real number to the
next largest integer. For convenience, we define a line-$\delta$ ($\delta_l$)
for the vessel-agent $i$ as
\begin{equation}
  \delta_{l,i}(\vec{x}) =
  \frac{2 \pi r_i l_i}{m_i}
  \sum_{k=1}^{m_i} \delta (\vec{x}-\vec{x}_{k,m_i}) \ ,
\end{equation}
where the discretization points are given by
\begin{equation}
  \vec{x}_{k,m_i} = \vec{x}_i^s
  + (k+1) \frac{\vec{x}_i^e - \vec{x}_i^s}{m_i} \ .
\end{equation}
Here, $\vec{x}_i^s$ and $\vec{x}_i^e$ denote the start and
endpoint of the cylinder's center line. Note that the line-$\delta$ scales
automatically the source and sink terms with the agent's surface and distributes
the total contribution evenly between the points. The terms are eventually given
by
\begin{equation}
  v(\vec{x},\vec{\beta_j}) =
  \sum_{i \in \mathcal{V}} \beta_{j,i} \cdot \delta_{l,i}(\vec{x}) =
  \beta_{j} \sum_{i \in \mathcal{V}} \delta_{l,i}(\vec{x}) \ ,
\end{equation}
where we again simplify $\beta_j$ to have no dependence on the agent's
attribute. However, the coefficients $\beta_d$ and $\beta_t$ involve a time
dependence; they are only non-zero during the DOX and TRA treatment.
The coupling concludes our model description. While the individual modules are
fairly simple, the number of modules and their interactions form a complex
system parameterized by many parameters. We discuss the
parameter choices and experimental setup in the next section.

\newpage
\section{Experimental Setup, Parameter Choice, and Reproducibility}
\label{sec:estimate_parameters}

\begin{figure}
  \centering
  \subfloat[view 1]{
    {\includegraphics[trim=665 135 600 180, clip, width=.47\textwidth,valign=c]{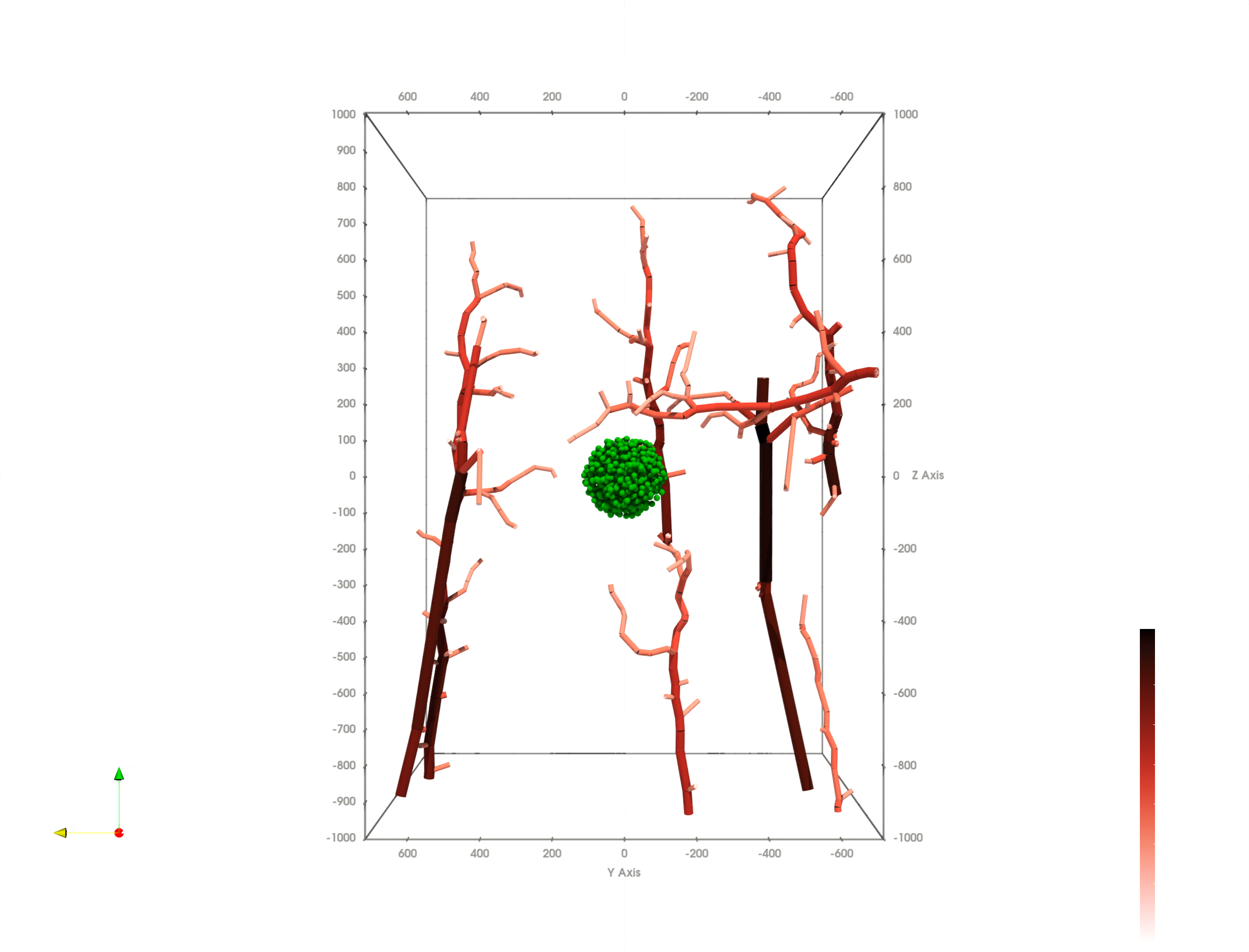}}
  }
  \subfloat[view 2]{
    {\includegraphics[trim=665 135 600 180, clip, width=.47\textwidth,valign=c]{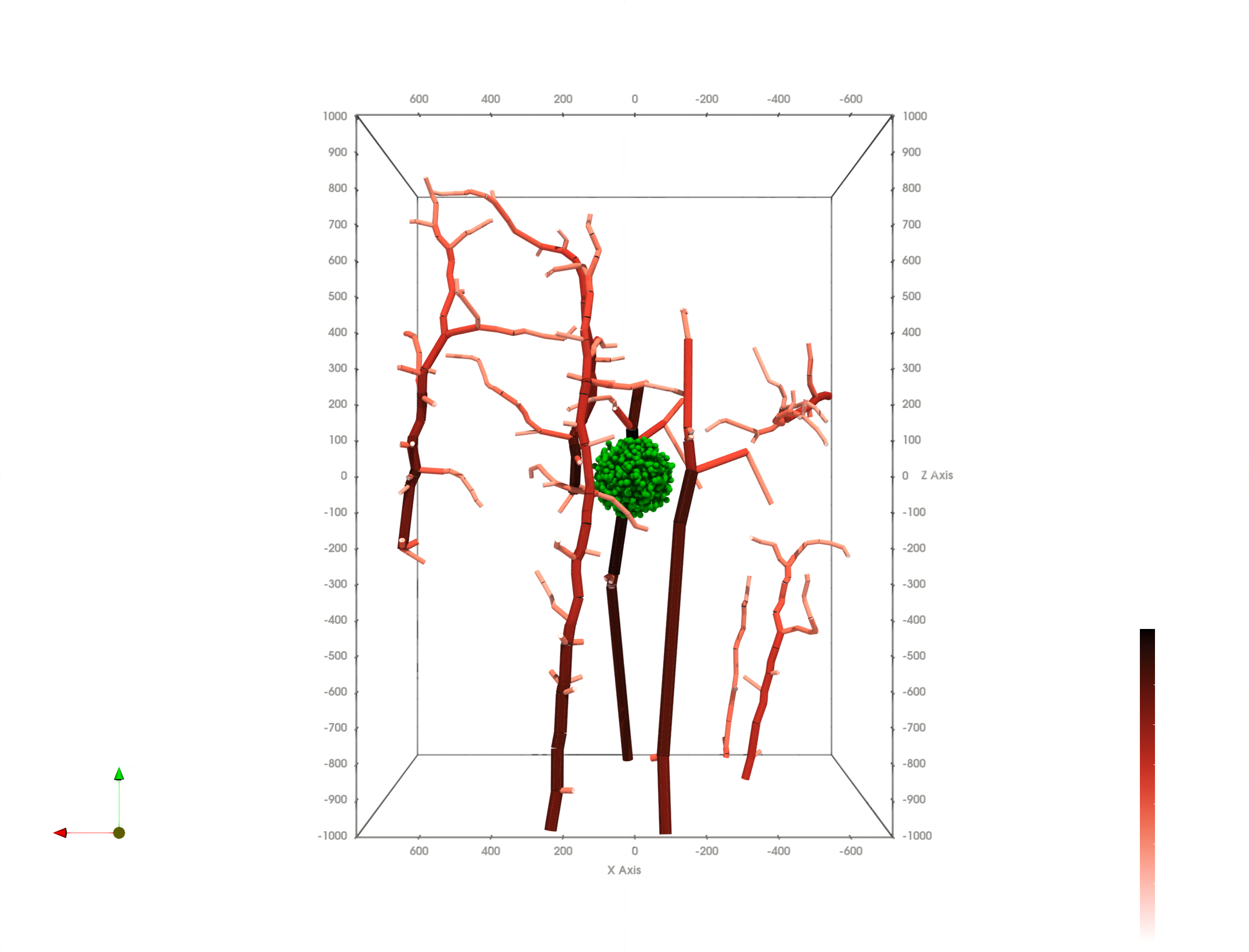}}
  }
  \caption{
    Initial configuration of the simulation. The tumor cells are shown in green
    independent of their state. The color of the vessel indicates the diameter;
    segments with larger diameters are shown in dark red, small diameters are
    shown in light red.
  }
  \label{fig:initialization}
\end{figure}

In this section, we discuss the initialization of the simulation, the choice
of parameters, and how to reproduce the findings of the present work.
We begin by detailing the initial vasculature and tumor setup.
We continue with different parameter sets. First, we discuss the parameters
for the cell geometries and the forces between them. We proceed with the
parameters related to the stochastic cell cycle and angiogenesis.
Lastly, we discuss the continuum and coupling parameters.
All parameters are given in
\ref{appendix: data}.
We conclude the section by explaining how to reproduce the computational
experiments presented in Section~\ref{sec:Results} and \ref{sec:large-scale}.

\subsection{Initial Simulation State}

At the beginning of the simulation, the tumor cells, the vessels, and the
continua are initialized to specific configurations. For the tumor cells, we
randomly place cells in a sphere centered around the origin. The radius of
the sphere is determined from the number of cells such that the resulting
spheroid is randomly but densely packed:
\begin{equation}
  r_{spheroid}^3 = \frac{N}{0.64} r_{cell}^3.
\end{equation}

In application, we use
data from
previous work~\cite{Reichold2009,Koeppl2020} to define the initial structure of
the vessels. The data covers a volume of roughly $1 \times 1 \times 2 \ mm$ and
serves as the starting point of our angiogenesis simulation.
For the continua, we enforce homogeneous Neumann boundaries
and initialize the substances to a constant value (usually 0).

The initial state of the simulation is depicted in
Fig.~\ref{fig:initialization}.
The figure shows a tumor spheroid that is surrounded by the data-realistic
vasculature. While the size of the tumor cells as well as the size and structure
of the vascular are motivated by data, the relative positioning of tumor and
vasculature is unspecified. Here, the tumor spheroid is placed in the empty
center between the vessels; however, positioning the tumor at any other place
within the simulation space is a valid choice as well.

\subsection{Cell and Force Parameters}

The tumor cells in our model describe BT-474 breast cancer cells as used in
previous ABM
studies~\cite{Rocha2018,Phillips2020,Lima2021,Macklin2012} and the
related pre-clinical study \cite{Sorace2016}. The nuclear, regular, and action
radius of the cells are chosen as $9.953 \mu m$, $5.296 \mu m$, and
$12.083 \mu m$, respectively \cite{Macklin2012}.
See Tab.~\ref{tab:ParameterCellRadii}. The parameters determining the strength
of the repulsive and adhesive force components are taken from \cite{Lima2021};
including the viscosity parameter found in the related source code.
See Tab.~\ref{tab:ParameterForces}.

\subsection{Cell Cycle and Angiogenesis Parameter}

Lima et al. \cite{Lima2021} calibrated their BT-474 breast cancer cell ABM with
experimental data. Our cell cycle shares the deterministic and some of the
stochastic transitions with their model. Here, we use their
calibrated parameters as starting point. We further obtained information for
effective parameters from exponential fits of Sorace's data \cite{Sorace2016}
to further guide our parameter choices.
These parameters describe the effective behavior of a simple exponential
surrogate, i.e., the evolution of the number of tumor cells merging the effects
of all transitions in two numbers. The cell-cycle parameters are
summarized in Tab.~\ref{tab:ParameterCellCycle}.

While adopting common ideas from earlier work (tip cells following the
gradient), the angiogenesis model is novel and, thus, parameters can hardly be
derived from the literature. Phillips and co-workers \cite{Phillips2023}
recently developed an ABM for angiogenesis and calibrated it with
data from an experiment specifically designed for this scenario. They found
that the sprouts extend with a speed of roughly $2 \frac{\mu m}{h}$, which we
adapt for this work. All other parameter choices for vessels and the
angiogenesis algorithm are summarized in Tab.~\ref{tab:ParameterAngiogenesis}.

\subsection{Continuum Parameter}

In this work, we consider four substances: the nutrients, VEGF, DOX, and TRA.
According to Eq.~(\ref{eq:continua}), they obey reaction-diffusion equations, and
their dynamics are determined by their respective diffusion
coefficients $D_x$ and the decay constants $\lambda_x$, where $x$ labels the
substances. None of these constants are directly available.

\subsubsection{Diffusion Coefficients}

To overcome this limitation,
we first adopt Lima and coworkers' \cite{Lima2021} estimate for the value of
the nutrient diffusion $D_n = 50 \frac{\mu m^2}{h}$. Intuitively, diffusion is
a passive transport phenomenon and larger, heavier objects should diffuse
more slowly. Einstein \cite{Einstein1905} showed that the diffusion coefficient
is inversely proportional to the radius of the diffusing particles, i.e.,
$D \sim r^{-1}$. Since it is non-trivial and beyond the scope of this work to
define an effective radius for complex protein structures, we work with the
simple
hypothesis that the mass of a molecule or a protein structure scales cubically
with the radius, e.g., $m \sim r^{3}$. We conclude that $D \sim m^{-1/3}$.
Thus, if the masses of two particles are related by the relationship
$m_2 = \alpha m_1$,
the respective diffusion coefficients behave as $D_2 = \alpha^{-1/3} D_1$.
The masses of all four diffusing molecules are known and, together with
Lima's estimate for $D_n$, we compute estimates for the remaining diffusion
constants. The masses, values for $\alpha$, and the diffusion constants are
given in Tab.~(\ref{tab:diffusion-scale-factors}).

\begin{table}[]
  \centering
  \begin{tabular}{@{}lcccc@{}}
    \toprule
                        & molecular mass {[}$\frac{g}{mol}${]} & $\alpha$ & $\alpha^{-1/3}$ & $D_x$ in {[}$\frac{\mu m^2}{h}${]} \\ \midrule
    Glucose (nutrients) & $180$                                & $1$      & $1$             & $50.0$                             \\
    VEGF (monomer)      & $19.3 \cdot 10^{3}$                  & $107$    & $0.21$          & $10.5$                             \\
    VEGF  (dimer)       & $38.6 \cdot 10^{3}$                  & $214$    & $0.16$          & $8.0$                              \\
    DOX                 & $543$                                & $3$      & $0.69$          & $42.5$                             \\
    TRA                 & $145 \cdot 10^{3}$                   & $806$    & $0.11$          & $5.5$                              \\ \bottomrule
  \end{tabular}
  \caption{
    The table shows 1) the molecular mass of the diffusing molecules and
    proteins, 2) the factor $\alpha$ expressing the mass in terms of the
    glucose mass, and 3) the scale factor $\alpha^{-1/3}$ for the diffusion
    coefficients.}
  \label{tab:diffusion-scale-factors}
\end{table}

\subsubsection{Decay, Source, and Sink Terms}

For the continua, the decay parameter and the tumor cell sink
terms both lead to an exponential decay. Ignoring the diffusion and sink terms,
the update rule
for the substance concentration at a given position reads
\begin{equation}
  u^{n+1} = (1 - \lambda^\prime \cdot dt) u^{n} \ .
\end{equation}
If we add the sink terms of $N$ tumor cells, we obtain the relationship
\begin{equation}
  u^{n+1} =
  \left(1 - \lambda  \cdot dt -
  \left( \sum_{i=1}^N \bar{r}_i \right) dt
  \right) u^{n}
  =
  \left(1 - \lambda  \cdot dt -
  \left( \sum_{i=1}^N \frac{r_i}{dx \cdot dy \cdot dz} \right) dt
  \right) u^{n} \ ,
\end{equation}
where we rewrite the consumed concentration $\bar{r}$ in terms of the consumed
amount $r$.
Acknowledging that $r_i$ is independent of the tumor cell
under consideration, we use a homogenization approach to rewrite the previous
equation in terms of the tumor cell density $\rho$:
\begin{equation}
  u^{n+1} =
  \left(1 - \lambda  \cdot dt -
  \frac{r N}{dx \cdot dy \cdot dz} dt
  \right) u^{n} =
  \left(1 - \lambda  \cdot dt -
  r \cdot \rho(x,y,z) dt
  \right) u^{n} \ .
\end{equation}
Comparing the previous equations yields
\begin{equation}\label{eq:diffusion-comparison}
  \lambda^\prime = \lambda + N \bar{r} = \lambda + r \rho(x,y,z) \ ,
\end{equation}
displaying that our model effectively shows a global decay of the substances as
well as a local decay depending on the density distribution of the tumor cells.
Furthermore, (\ref{eq:diffusion-comparison}) allows us to compare
the parameters $\lambda$ and $r$ with data and simple exponential surrogates
$\lambda^\prime$ attributing effects
to tumor cell presence or not.

DOX engages in different chemical reactions whose effect we model with the
decay constant and the sink terms. According to the FDA \cite{FDA2013},
DOX shows a terminal half-life of 20 to 48 hours.
Data for TRA suggest a dose-dependent half-life of 1.7 to 28 days
\cite{Boekhout2011,BGP2022}. These data imply the decay constants of
$\lambda_d^\prime \in [ 14.4 \cdot 10^{-3}, 34.7 \cdot 10^{-3}] h^{-1}$ and
$\lambda_t^\prime \in [ 1.0 \cdot 10^{-3}, 17.0 \cdot 10^{-3}] h^{-1}$,
respectively.
However, the data stems from generic experiments and is not necessarily
representative. Lima and coworkers \cite{Lima2022} recently calibrated their ODE
model to fit the same data that we consider here and obtained the decay
constants $\lambda_b^\prime = 12.0 \cdot 10^{-3} h^{-1}$ and
$\lambda_t^\prime = 20.0 \cdot 10^{-3} h^{-1}$. For DOX, their estimates are
slightly below the data range, and for TRA slightly above.

We realize that DOX is unspecific in its nature while TRA only interacts with
the HER2 receptor of the tumor cells.
We further consider the initial vasculature and the
not explicitly modeled regular tissue to be in equilibrium. Consequently, the
excess nutrients supplied via the new vasculature should primarily be consumed
via the tumor cells (see also the Warburg effect
\cite{Liberti2016, Heiden2009}).
Using these findings and further values from \cite{Lima2021},
we summarize all our parameter choices regarding the continua
in Tab.~\ref{tab:ParameterContinua} and \ref{tab:ParameterContinuumInteraction}.

\subsection{Reproducibility}\label{sec:reproducibility}

To ensure transparency and reproducibility, we share all source code and data
used in the project\footnote{available after final publication,
  currently upon request}. To reproduce the results of the following sections
(i.e., Section~\ref{sec:Results} and \ref{sec:large-scale}),
we need to fix four key components:
\begin{itemize}
  \item the version of the BioDynaMo source code,
  \item the version of the application source code including all possible
        changes,
  \item the parameters used for the simulation, and
  \item the postprocessing pipeline.
\end{itemize}
We share two repositories on GitHub: (1) the repository
\href{https://github.com/TobiasDuswald/angiogenesis}{TobiasDuswald/angiogenesis}
contains the application code, and (2) the repository
\href{https://github.com/TobiasDuswald/bdm-angiogenesis-reproducer}{TobiasDuswald/bdm-angiogenesis-reproducer}
contains the parameters, possible patches to the source code, and postprocessing
routines. We structured the latter repository such that it contains one folder
in "experiments/*" for each result shown in the main text. The respective
folders have the information to adequately initialize the code and parameters
for the simulation runs. For convenience, the initialization, build process,
execution, and post-processing of each computational experiment is
wrapped in bash scripts. Thus, to reproduce any figure in the main text,
the reader only needs to run a single bash script. Note, however, that BioDynaMo
simulations are not bit-reproducible at the time of writing.

\section{Results}\label{sec:Results}

In this section, we present the results of our computational experiments.
They are ordered such that their complexity gradually increases.
We begin with simulating the growth of tumor spheroids in the absence of
any vasculature and treatment in Section~\ref{sec:ResultsTumorSpheroid}.
Next, we investigate different vascular patterns arising from
the angiogenesis algorithm in Section~\ref{sec:ResultsAngiogenesis}.
Afterwards, we demonstrate the fully coupled model by simulating the vascular
growth and treatment. We first establish a clear and visual
understanding of the different phases of the simulation by showing a
conceptual simulation in Section~\ref{sec:ResultsVascularGrowth}.
In Section~\ref{sec:ResultsTreatment}, we then focus
on the key quantity of interest, the tumor volume, and
show how it evolves over time for different treatment scenarios.
We qualitatively compare these results to Sorace and co-workers'
observations~\cite{Sorace2016,Lima2022}. For all simulations involving
the vasculature, we consider the
initial setup described in the previous section and summarized in
Fig.~\ref{fig:initialization}.

\subsection{Tumor Spheroids and the Hypoxic Threshold}
\label{sec:ResultsTumorSpheroid}

In this subsection, we consider the growth of a tumor spheroid ignoring
vasculature and treatment protocols ($u_v = u_d = u_t = 0$).
To initialize the simulation, we
set the nutrients to a constant value and employ Dirichlet boundaries with the
same, constant value; i.e. $u_n (t=0) = 0.5$ and
$u_n (t) \vert_{\partial \Omega} = 0.5$.
In the absence of vasculature, the Dirichlet boundary conditions effectively
act as nutrient supply.
Here, we restrict ourselves to investigating the
hypoxic threshold $u_n^H$ because earlier 2D studies showed that this parameter
has significant influence on the number of proliferative cell driving the tumor
growth \cite[Fig.~6B]{Lima2021}.
The parameter marks the transition from the quiescent to the hypoxic state and,
thus, prohibits cell proliferation in regions with insufficient nutrient
availability ($u_n < \sigma_n^H$).
In other words, one may say that the hypoxic threshold defines hypoxic and
proliferative regions via a level-set function on $u_n$.
For this experiment, we place 500 tumor cells in the center of our cubical
simulation domain and simulate the growth of the tumor for 200 days with a
time step of one minute for different hypoxic thresholds
$u_n^H \in \{ 0.15, 0.13, 0.12, 0.11\}$.

\begin{figure}
  \centering
  \subfloat{
    {\includegraphics[trim=1080 480 1100 500, clip, width=.19\textwidth,valign=c]{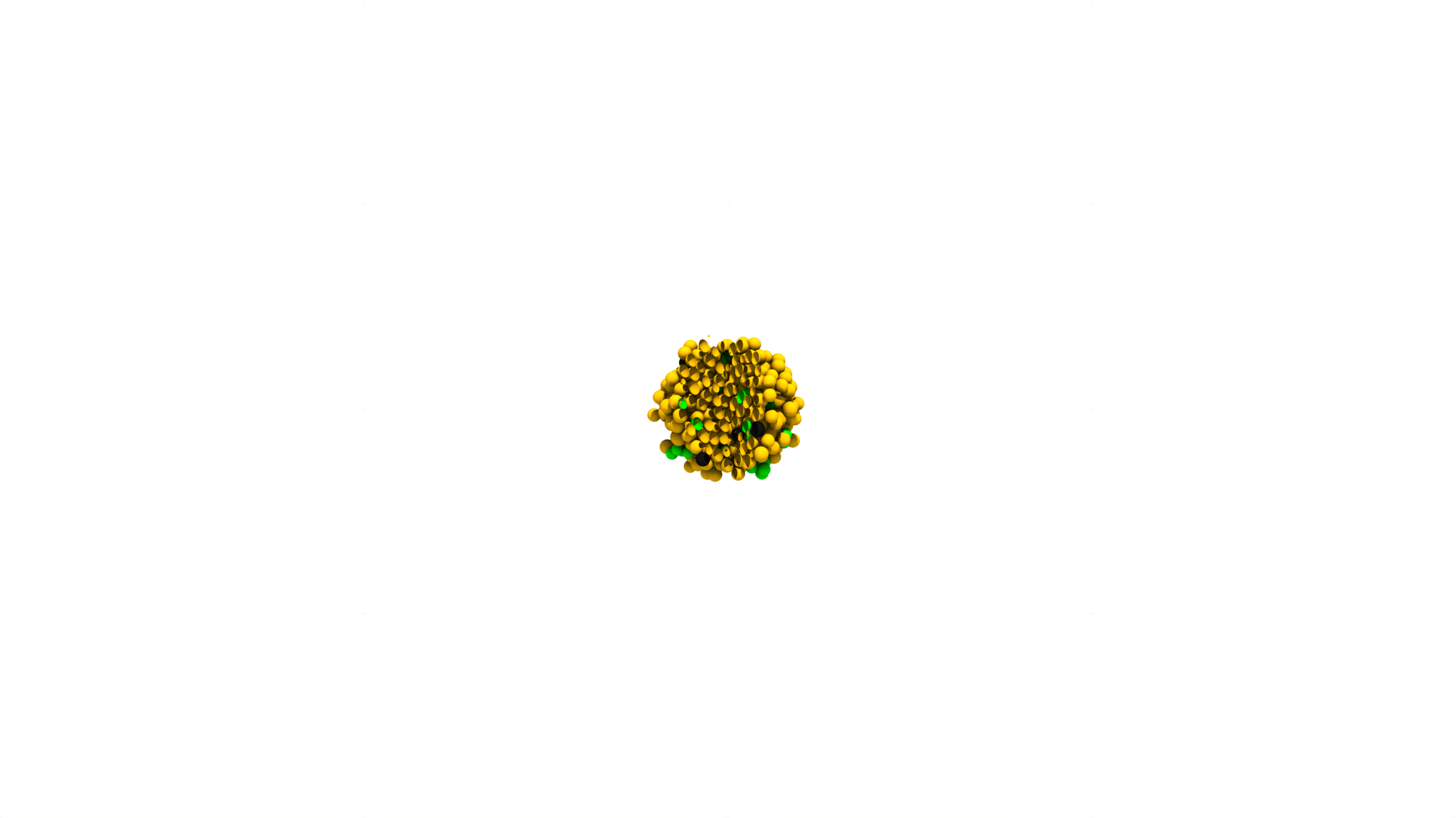}}
  }
  \subfloat{
    {\includegraphics[trim=1080 480 1100 500, clip, width=.19\textwidth,valign=c]{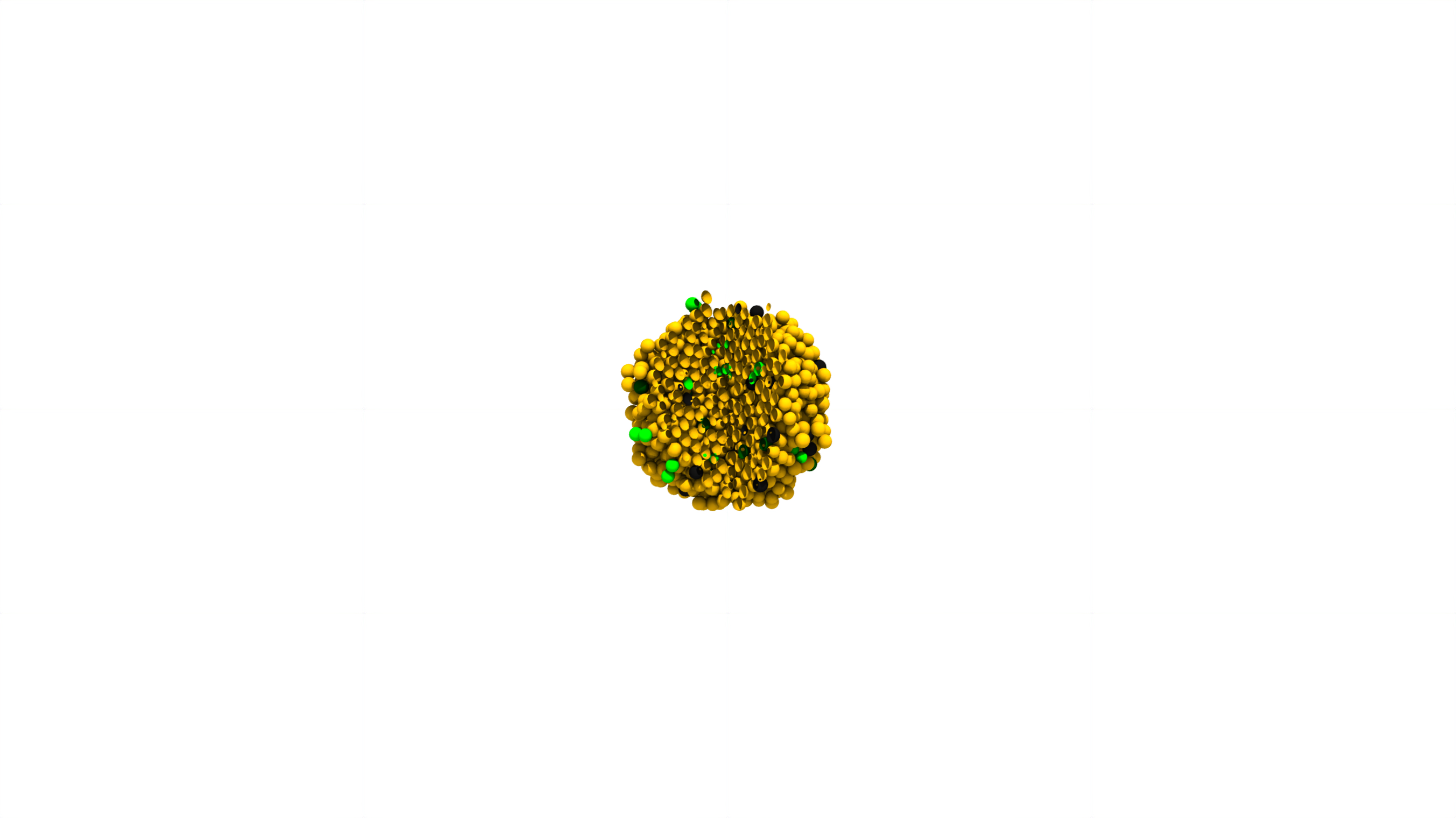}}
  }
  \subfloat{
    {\includegraphics[trim=1080 480 1100 500, clip, width=.19\textwidth,valign=c]{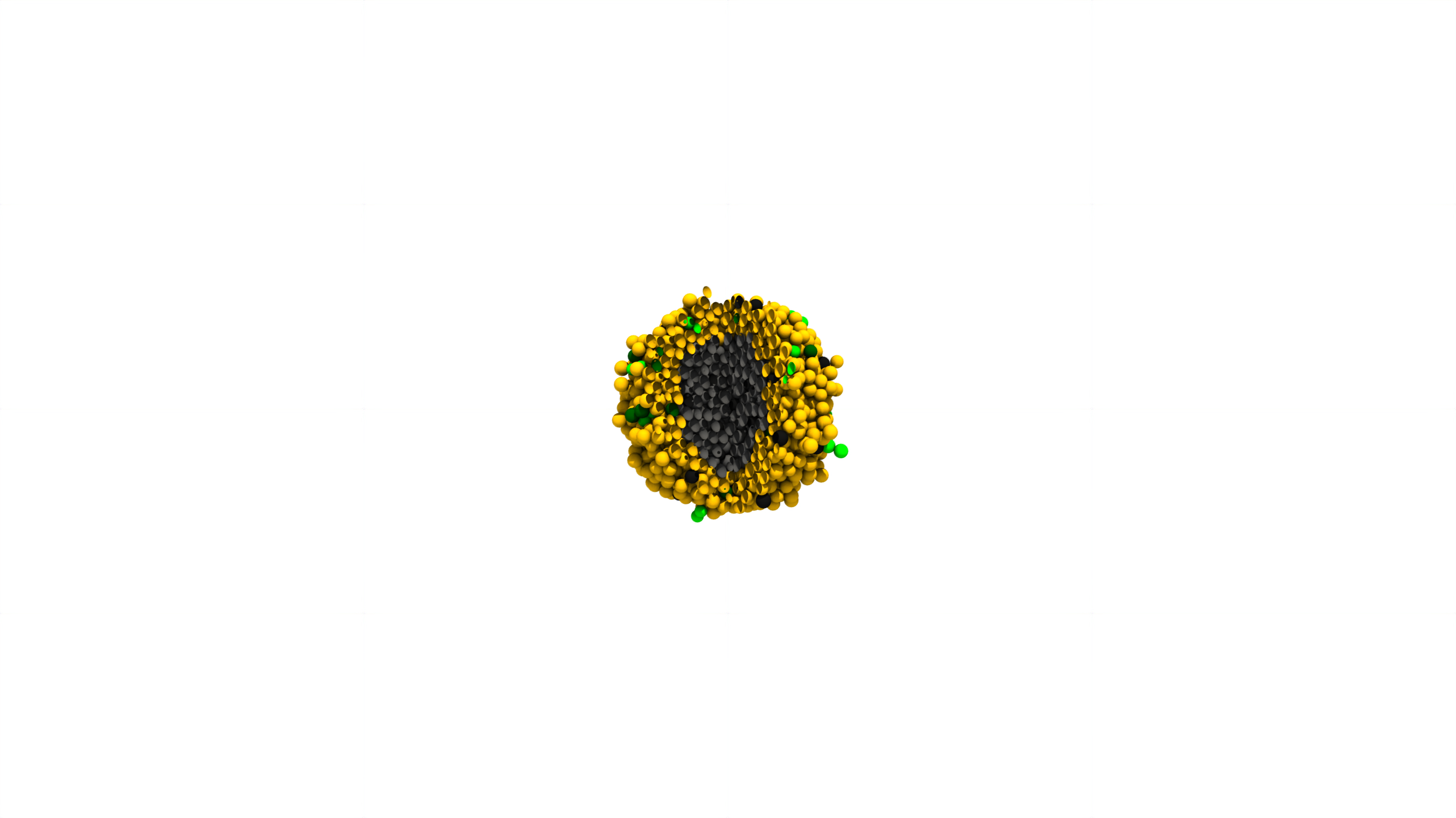}}
  }
  \subfloat{
    {\includegraphics[trim=1080 480 1100 500, clip, width=.19\textwidth,valign=c]{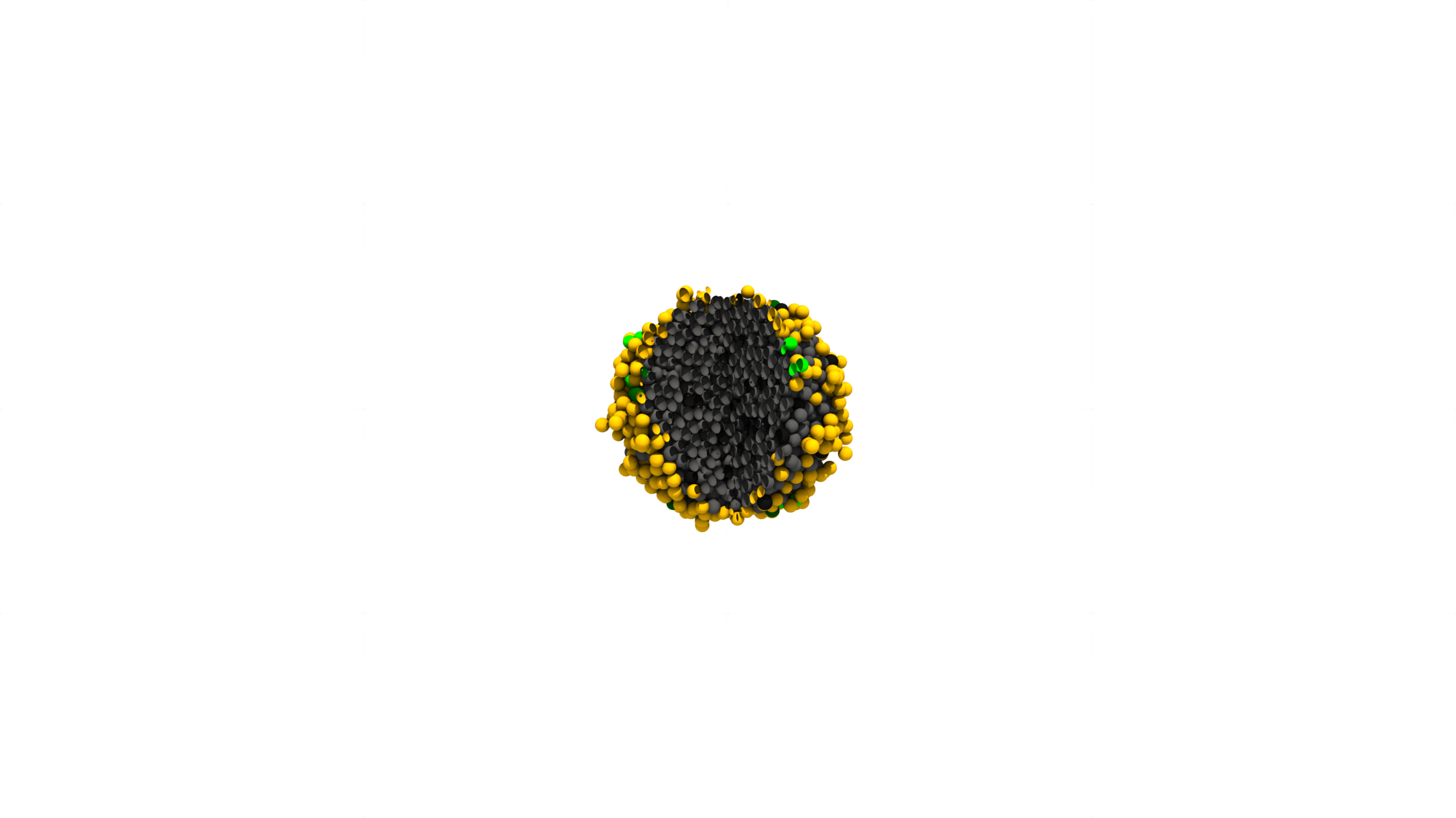}}
  }
  \subfloat{
    {\includegraphics[trim=1080 480 1100 500, clip, width=.19\textwidth,valign=c]{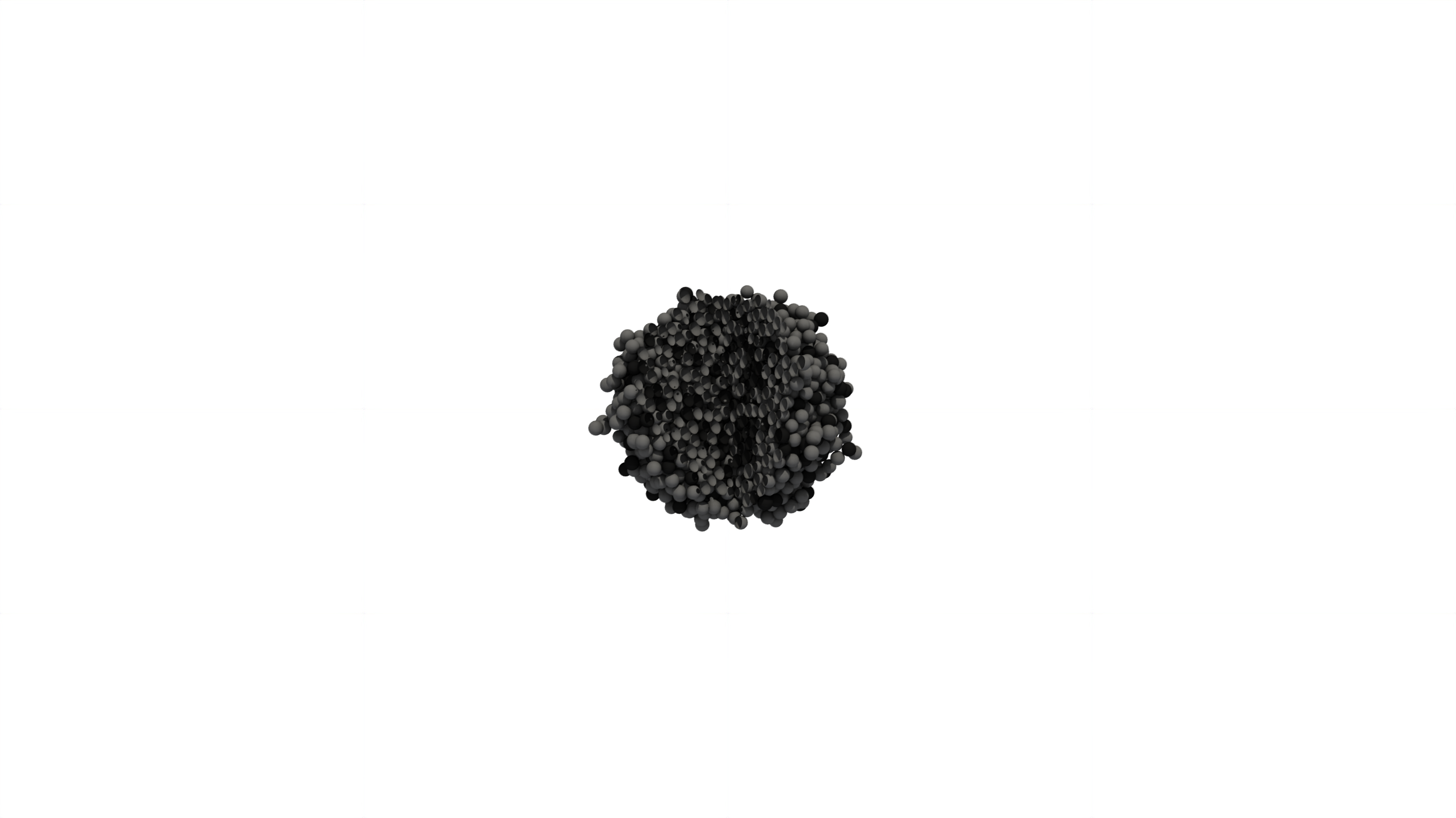}}
  }
  \caption{
    Evolution of the tumor spheroid for $u_n^H = 0.15$.
    Time progresses from left to right.
    The colors indicate the state: yellow (Q), bright
    green (G1), dark green (SG2), gray (H), and black (D).
  }
  \label{fig:spheroid-over-time}
\end{figure}

Figure~\ref{fig:spheroid-over-time} shows the spheroid over
time for the largest hypoxic threshold $u_n^H = 0.15$.
The cells are initially in proliferative states, consume nutrients, and the
spheroid grows. Over time, a hypoxic region in the center of the spheroid
begins to form.
The border between
hypoxic and proliferative regions is implicitly visualized as the sharp
transition zone between proliferative (yellow, green) and hypoxic (grey, dark
blue) tumor cells. The time-dependent border may be defined as a hypersurface
satisfying $u_n(x,t) = u_n^H$. Its dynamics together with the spacial structure
of the tumor effectively determine the evolution of the spheroid. The more
tumor cells lie outside the hypersurface, the more cells participate in the
exponential cell proliferation. Parameters such es the nutrient consumption and
the hypoxic threshold define the shape and dynamics of this surface. The
dynamics of the tumor growth effectively comes down to a race between the spatial
tumor extend and the hypersurface, i.e., if the hypersurface can spread quicker
than the tumor to eventually enclose the entire spheroid and stop the growth.
This is the case for $u_n^H = 0.15$ where the tumor eventually stops growing and
dies off, see Fig.~\ref{fig:spheroid-over-time} on the right.

\begin{figure}
  \centering
  \subfloat{
    {\includegraphics[trim=600 50 600 50, clip, width=.29\textwidth,valign=c]{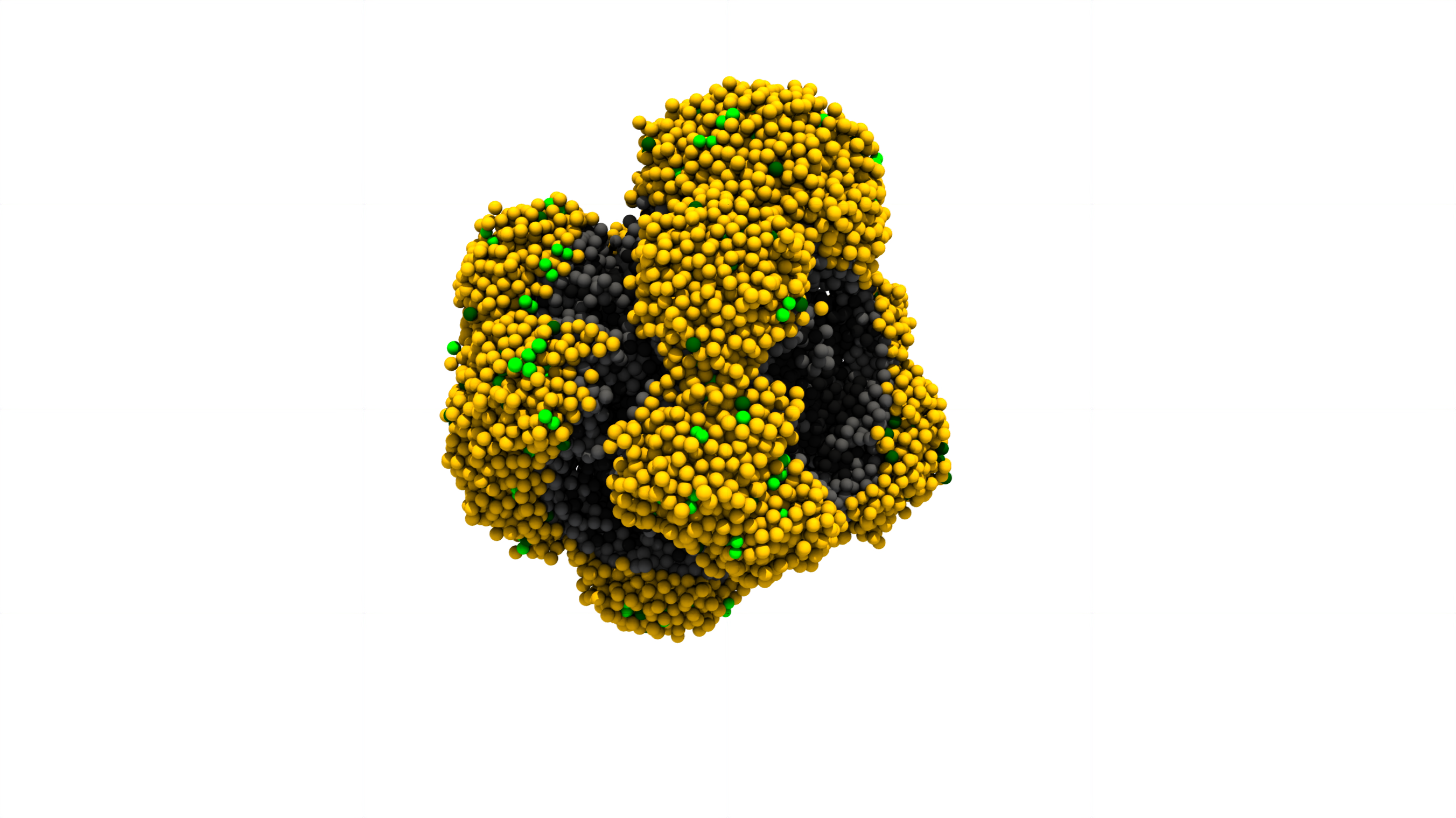}}
  }
  \subfloat{
    {\includegraphics[trim=600 50 600 50, clip, width=.29\textwidth,valign=c]{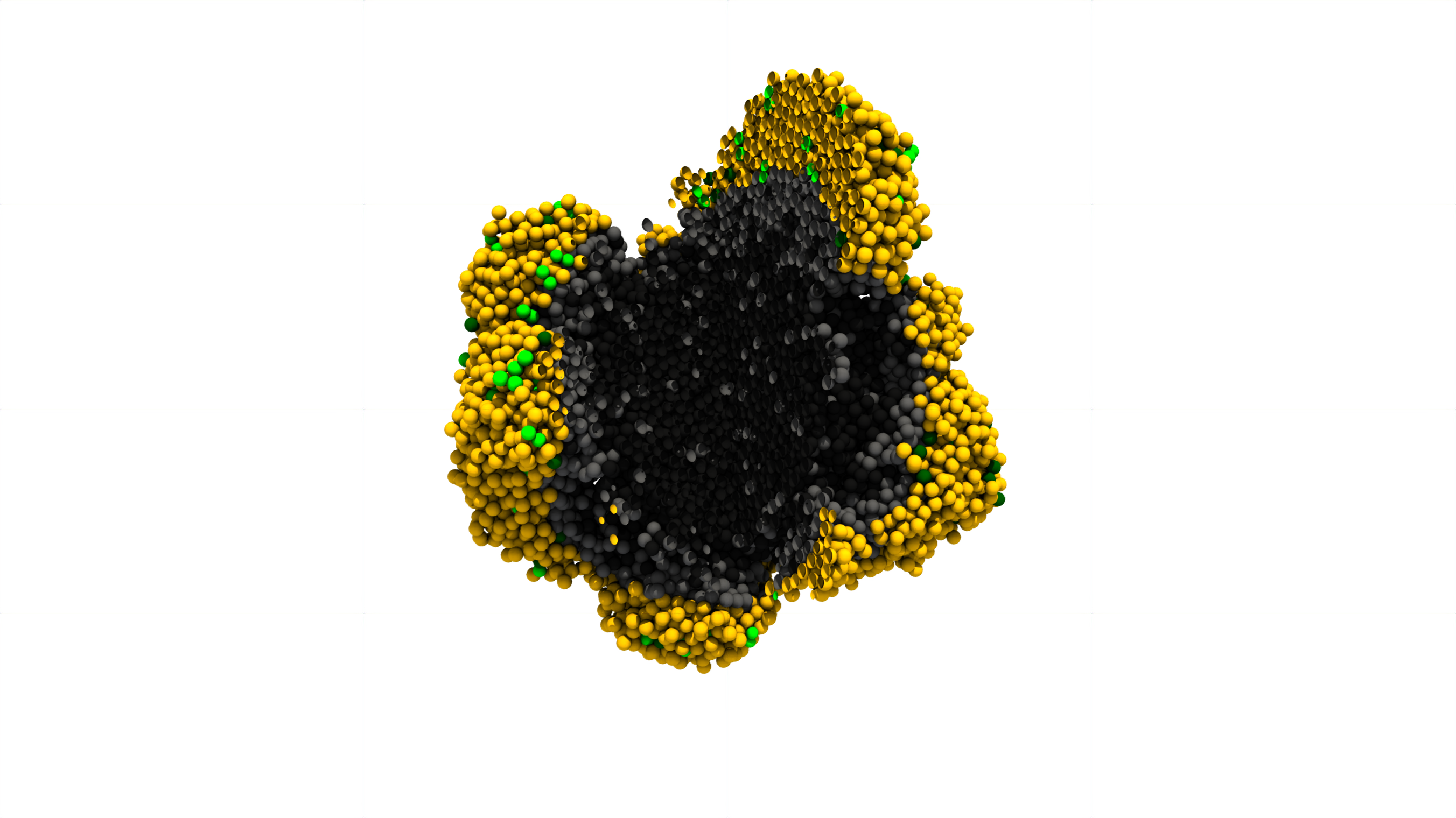}}
  }
  \subfloat{
    {\includegraphics[width=.40\textwidth,valign=c]{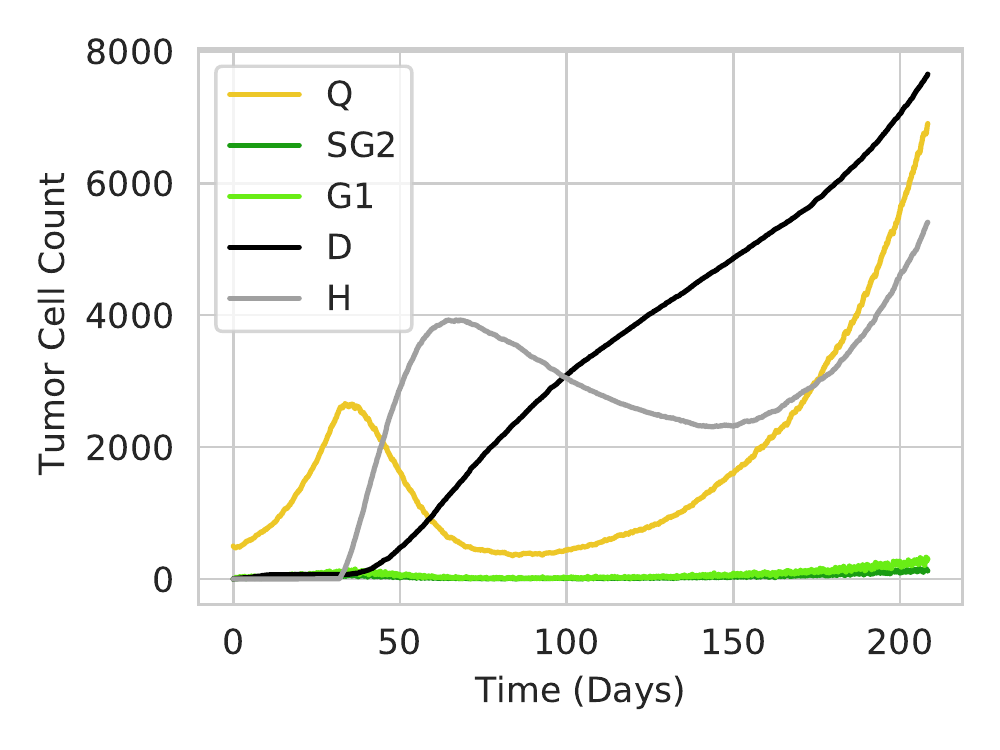}}
  } \par\noindent\rule{\textwidth}{0.1pt} \\
  \subfloat{
    {\includegraphics[trim=600 0 600 100, clip, width=.29\textwidth,valign=c]{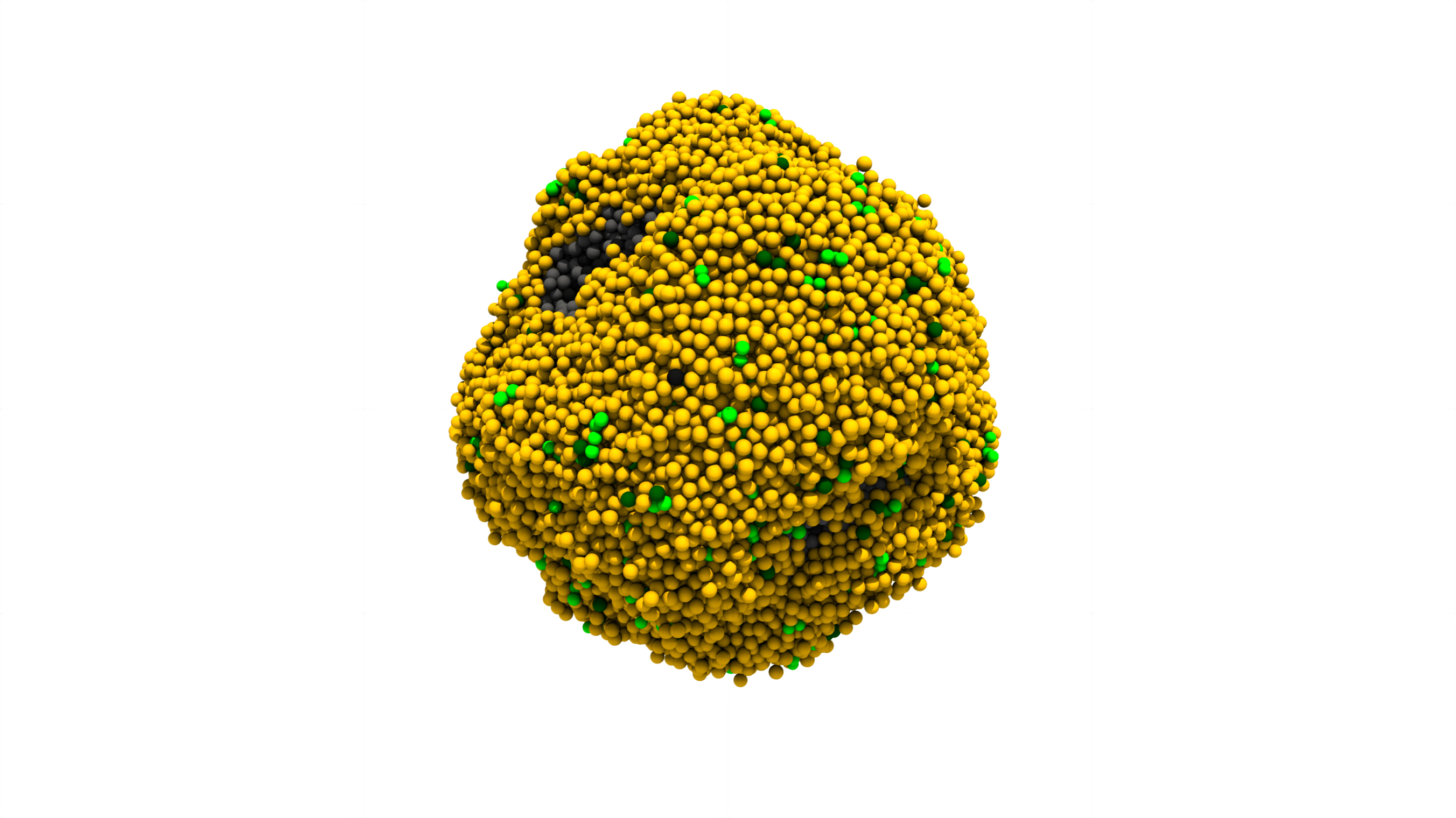}}
  }
  \subfloat{
    {\includegraphics[trim=600 0 600 100, clip, width=.29\textwidth,valign=c]{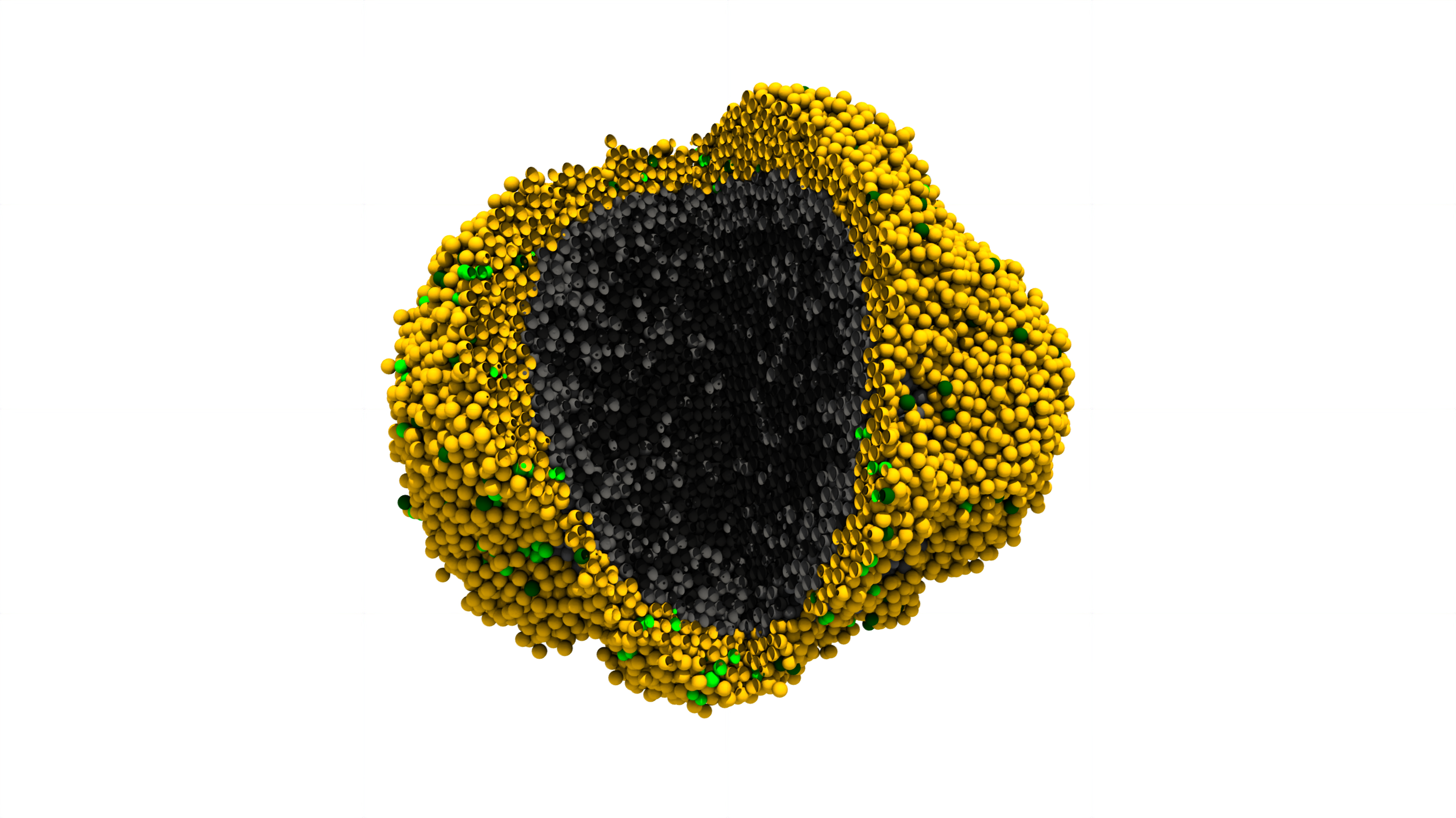}}
  }
  \subfloat{
    {\includegraphics[width=.40\textwidth,valign=c]{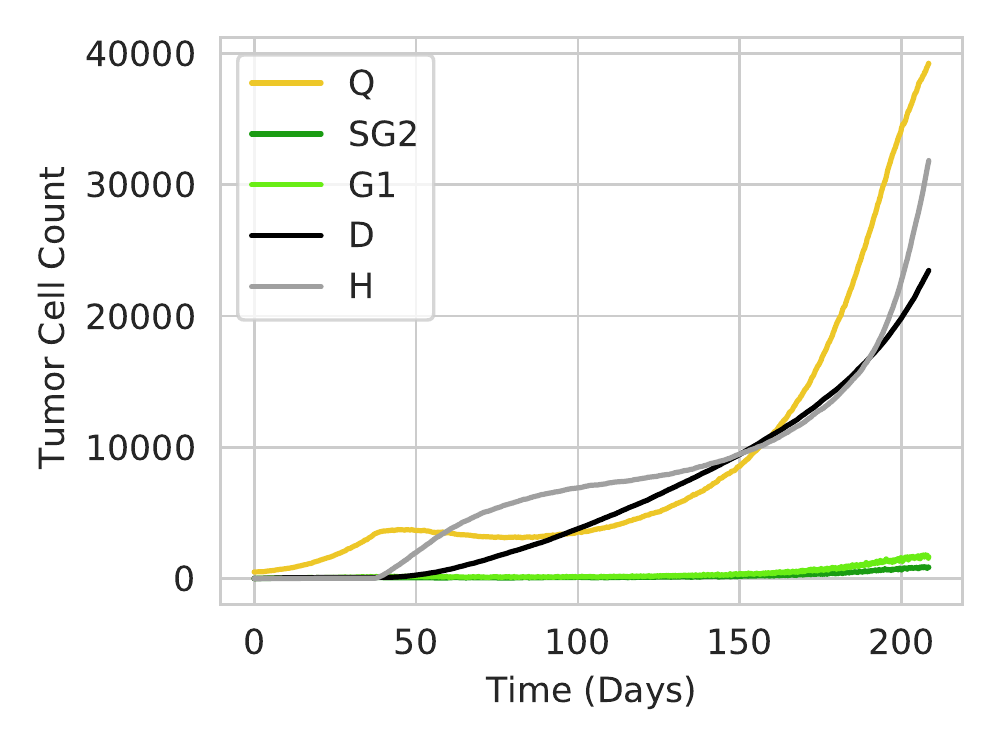}}
  } \par\noindent\rule{\textwidth}{0.1pt} \\
  \subfloat{
    {\includegraphics[trim=600 50 600 50, clip, width=.29\textwidth,valign=c]{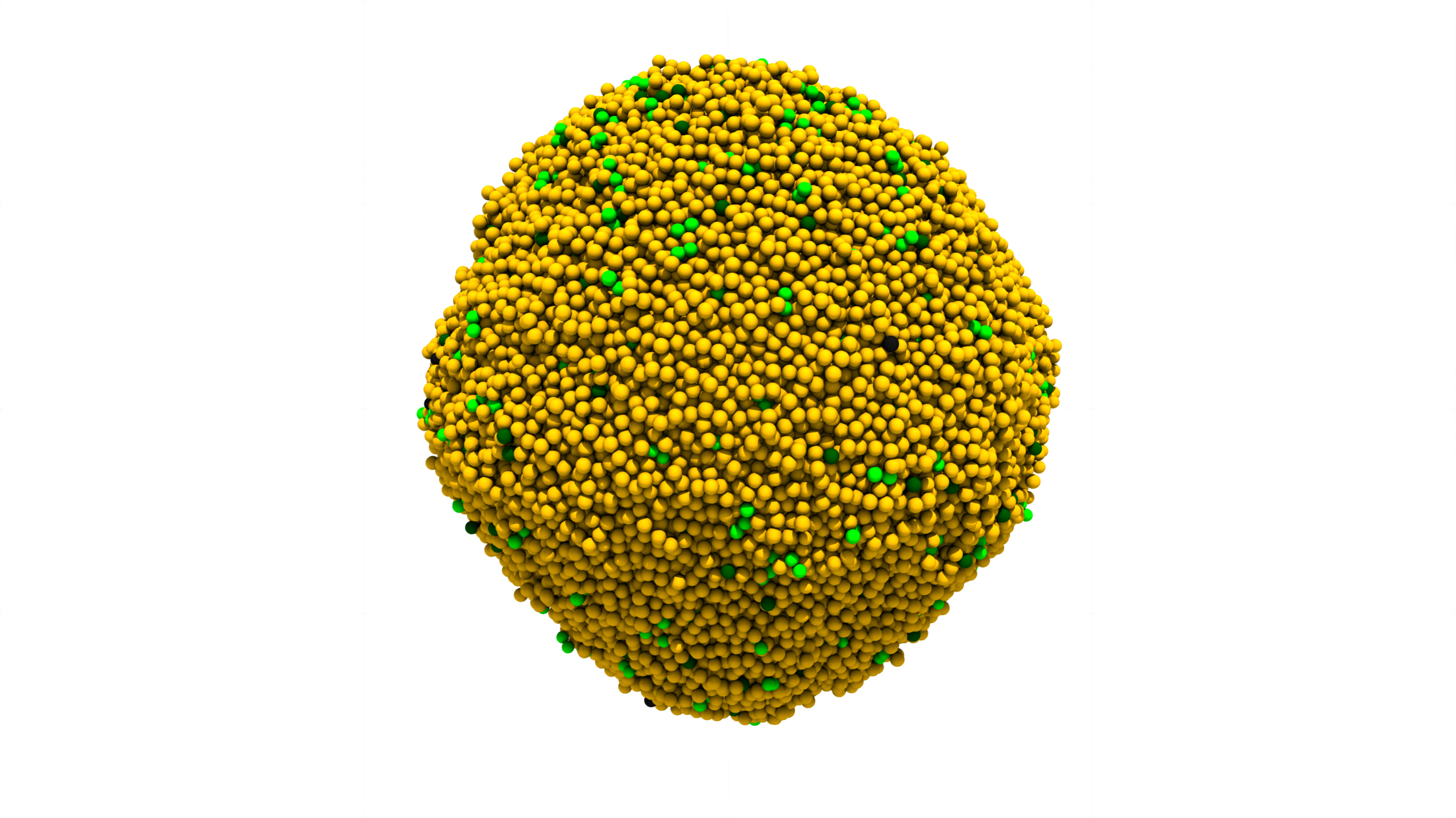}}
  }
  \subfloat{
    {\includegraphics[trim=600 50 600 50, clip, width=.29\textwidth,valign=c]{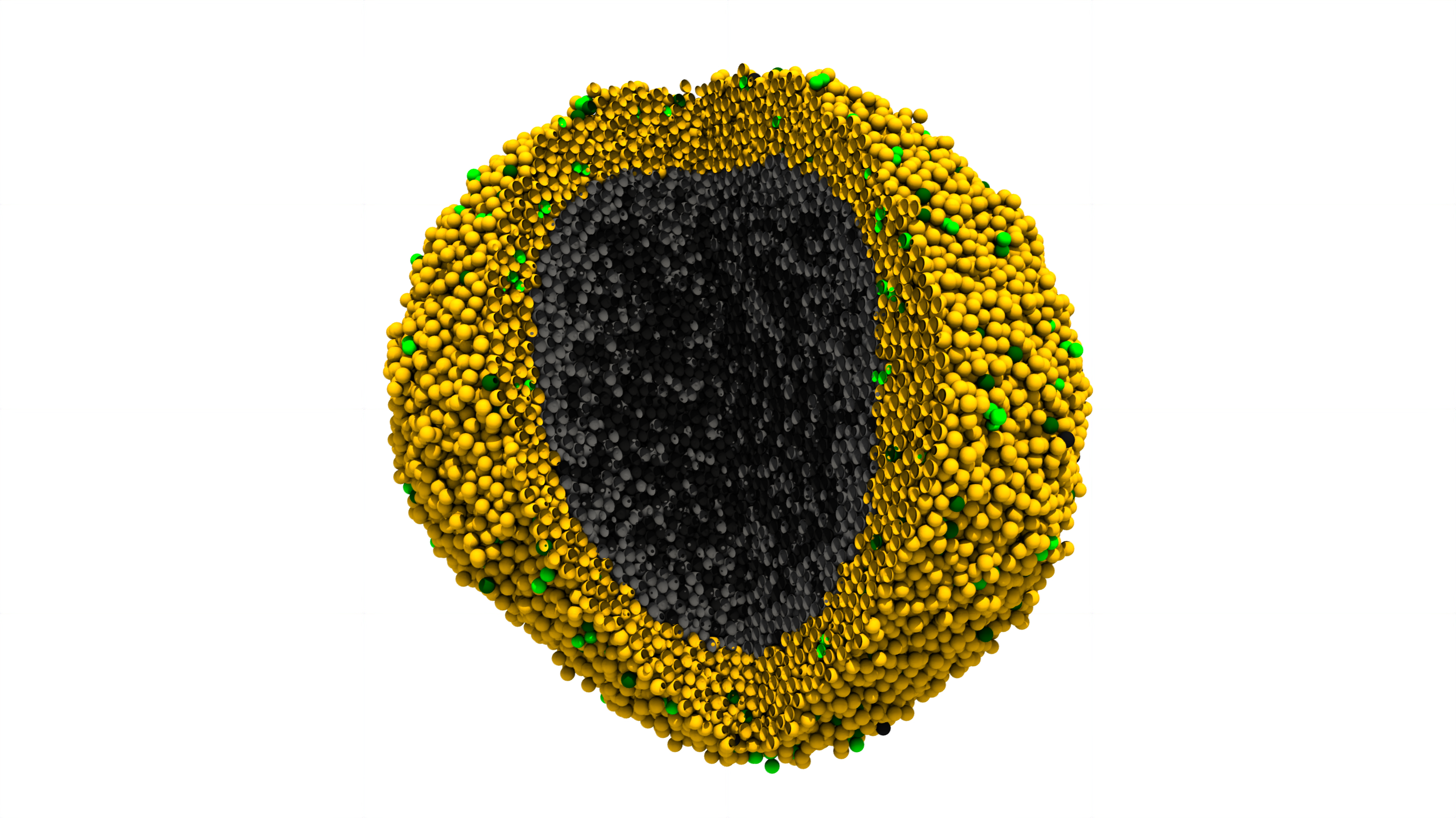}}
  }
  \subfloat{
    {\includegraphics[width=.40\textwidth,valign=c]{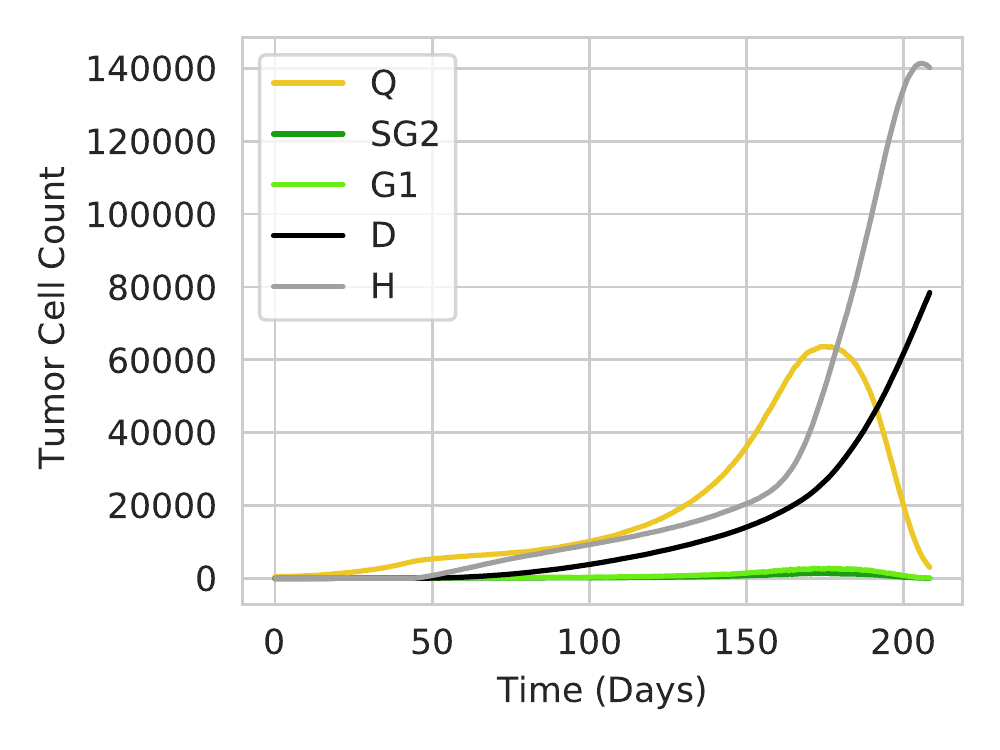}}
  }
  \caption{
    Evolution of tumor spheroids.
    The rows correspond to different hypoxic threshold
    $u_n^H \in \{ 0.13, 0.12, 0.11\}$ (top to bottom).
    The first columns shows all cells, the second colum adds a
    cut-out revealing the inner structure, and
    the last column shows the number of cells in different states over time.
    The spheroids for the hypoxic thresholds 0.13, 0.12, and 0.11 are shown
    after different simulation times, i.e., after 200, 165, and 145 days,
    respectively.
    The boundaries start to affect the simulation afterwards.
    The colors indicate the state: yellow (Q), bright
    green (G1), dark green (SG2), gray (H), and dark blue (D).
  }
  \label{fig:spheroid}
\end{figure}

With decreasing $u_n^H$, we observe vastly different growth patterns in
Fig.~\ref{fig:spheroid}.
We display the spheroids for the remaining hypoxic thresholds
($u_n^H \in \{ 0.13, 0.12, 0.11\}$) at specific time points in the simulation
(i.e., 200, 165, and 145 days).
We further
show the dynamics with a graph of the cell numbers in different states over
time. In Fig.~\ref{fig:spheroid}, each row corresponds to a hypoxic threshold.
In contrast to Fig.~\ref{fig:spheroid-over-time}, the hypersurface
cannot cover the entire spheroid and the growth never stops completely.
For instance for $u_n^H = 0.13$ (Fig.~\ref{fig:spheroid}, top row),
almost all cells transition into the
hypoxic state after roughly 50 days. However, some cells on the surface still
lie outside the hypoxic regions and continue to proliferate (similar to the
fourth spheroid in Fig.~\ref{fig:spheroid-over-time}).
These few cells eventually move on to form satellite tumors on the surface of
the original spheroid.
Further lowering the threshold to $u_n^H = 0.12$
(Fig.~\ref{fig:spheroid}, middle row), more and more cells on the
surface remain in the proliferative states and larger parts of the spheroids
are covered with proliferative cell populations.
Once we reach $u_n^H = 0.12$ (Fig.~\ref{fig:spheroid}, bottom row),
the entire surface remains proliferative
throughout the simulation and the tumor shows the characteristic proliferative
hull around a necrotic core.
Over all, the graphs clearly show that a lower hypoxic threshold leads to
stronger proliferation.
It is interesting to note that the stochasticity of the system breaks its
symmetry for $u_n^H \in \{ 0.13, 0.12 \}$. After investigating the dynamics of
the tumor growth, we now shift our attention to the development of the
vasculature via sprouting angiogenesis.

\subsection{Angiogenesis}
\label{sec:ResultsAngiogenesis}

To demonstrate the process of angiogenesis,
we simulate the system for an extended period of time (i.e., 14 days)
and observe the resulting vasculature under the influence of different
parameters.
The majority of them have straight-forward
interpretations. The tip cell distance $d_{tip}$, branching distance
$d_{branch}$, and sprouting probabilities influence the sparsity pattern of
the network.
The weights combining
the gradient information, randomness, and previous growth direction allow us to
interpolate between a random walk and smooth curves along the gradient. The
VEGF and gradient thresholds put hard limits on the signal strength that
a vessel agent has to sense to form a sprout. Here, we want to focus on the
parameter that is easily overlooked.

\begin{figure}
  \centering
  \subfloat[with VEGF consumption]{
    {\includegraphics[clip, width=.49\textwidth,valign=c]{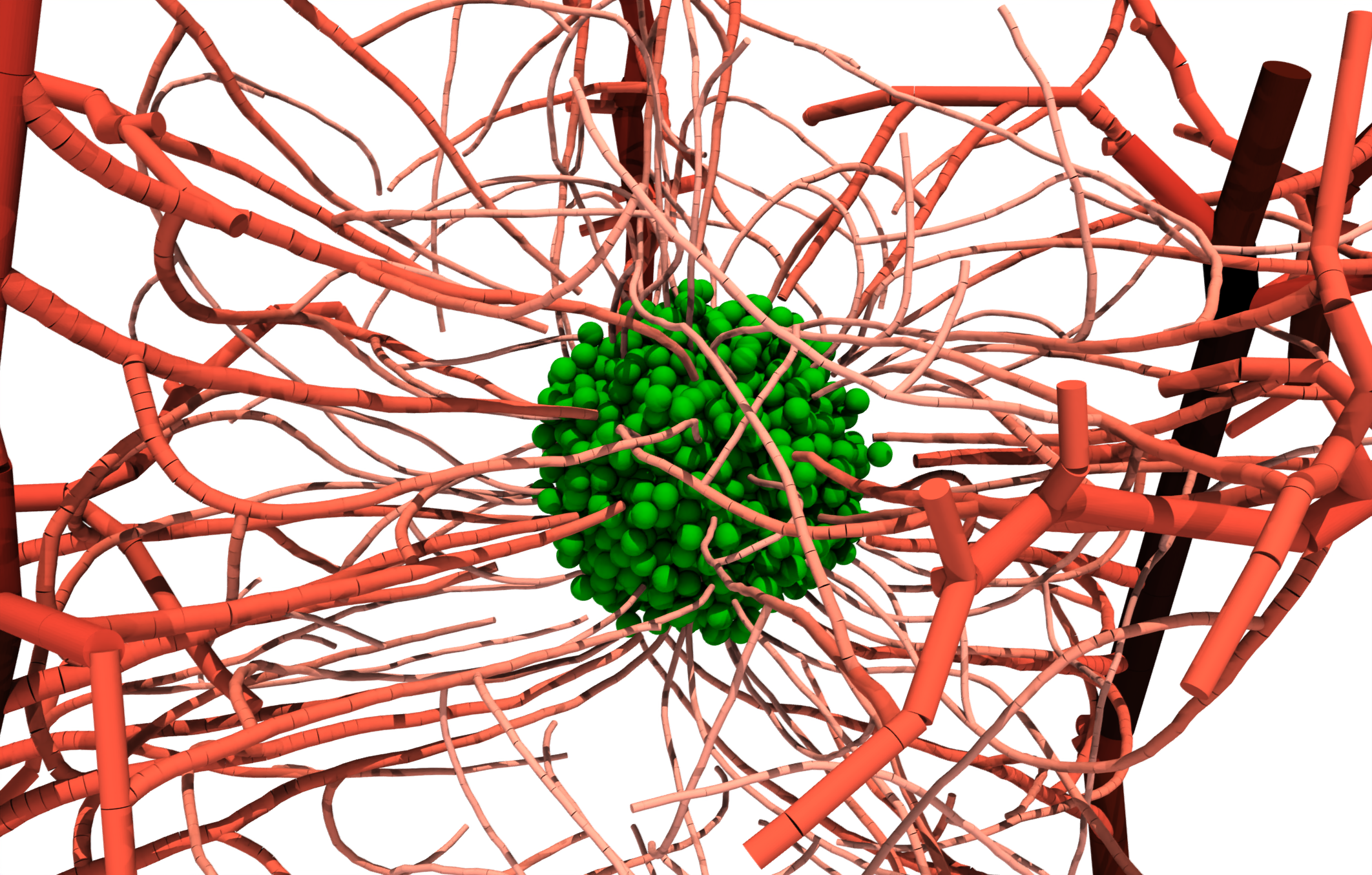}}
  }
  \subfloat[no VEGF consumption]{
    {\includegraphics[width=.49\textwidth,valign=c]{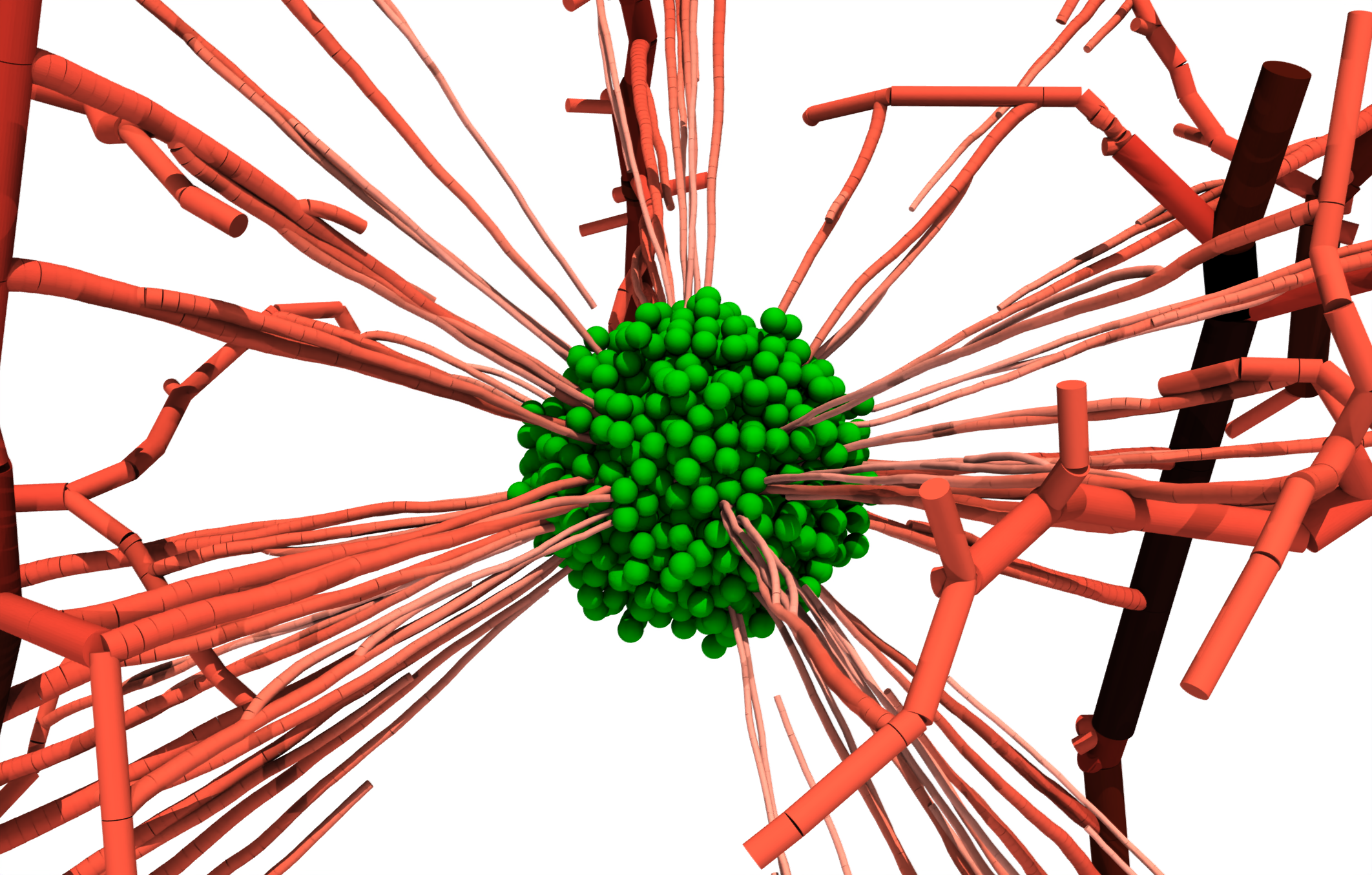}}
  }
  \caption{
    Simulated vasculature (a) with and (b) without vessels acting as sink terms.
    The tumor cells are shown in green independent of their state.
    The color of the vessel indicates the diameter;
    segments with larger diameters are shown in dark red, small diameters are
    shown in light red.
  }
  \label{fig:result-vasculature}
\end{figure}

The term coupling the vessels to the VEGF concentration is the most influential
parameter on the structure of the vasculature. While this effect is hard to
quantify, we depict the final vasculature of the simulation with and without
the coupling term in Fig.~\ref{fig:result-vasculature} (all other parameters
are kept identical). It is evident that the vasculature differs significantly
for the two scenarios. Without the coupling, the vessels simply grow towards
the tumor center in relatively straight lines. We also note that even if the
vessels branch, they follow almost the same path. In contrast, the coupling term
removes VEGF from the vessel's vicinity and, thus, it locally changes the
gradient field
encouraging new sprouts to grow away from its parent vessel, avoid
other vessels growing towards the tumor, and search for alternative paths
to revive the hypoxic regions. The network of the coupled simulation produces a
significantly more diffuse network in which vessels surround the tumor rather
than growing towards its center. In experimental work~\cite{McDonald2003},
researchers showed that
typically the blood flows on the outskirts of the tumor spheroid
indicating that the tumor-surrounding, diffuse network is more realistic than
the one ignoring the coupling.
It is worth noting that the diffuse growth process also produces
vessels that, over time, grow in other VEGF rich regions other then the tumor
core. Our model does not prune such growth; however, pruning mechanisms such
as suggested in \cite{Koeppl2020} are beyond the scope of the present work.

\subsection{Vascular Tumor Growth and Treatment}
\label{sec:ResultsVascularGrowth}

Combining the previous sections, we now aim to simulate vascular tumor growth
and treatment.
We initialize
the simulation with 1000 tumor cells in the center of the vasculature, see
Fig.~\ref{fig:initialization}. We start the simulation with no nutrients, i.e.,
all cells take the hypoxic state and secrete VEGF.
We choose to couple the vessels to the VEGF
field to achieve a more realistic, tumor-surrounding micro-vasculature.
The vasculature develops
around the tumor over time and we subsequently expect the tumor to grow into the
surrounding, nutrient-rich regions.
At day 70, we turn off the vessel growth algorithm
to avoid growth resulting from sprouts that may not hit the stopping
criteria. We let the tumor proliferate until day 102 on which we trigger the
treatment. We simulate the treatment by turning on the TRA and DOX source
terms between days 102-104 and 106-108, respectively. We remark that the time
scales are somewhat arbitrary; we choose them such that the tumor covers
the newly vascularized region before the treatment begins.

\begin{figure}
  \centering
  \subfloat[VEGF-secreting tumor cells]{
    {\includegraphics[trim=0 100 0 100, clip, width=.45\textwidth,valign=c]{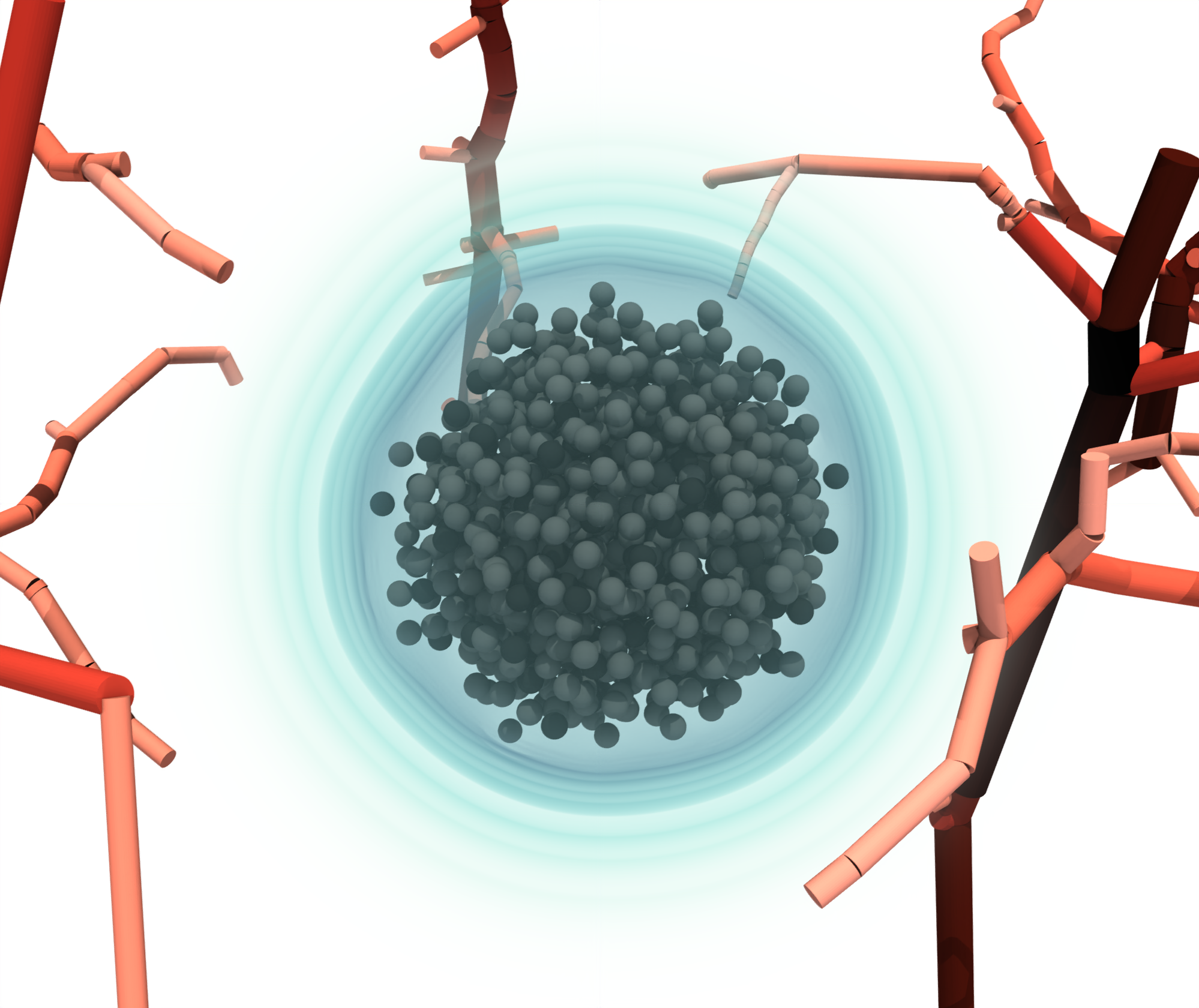}}
  }
  \qquad
  \subfloat[Early vasculature and nutrients]{
    {\includegraphics[trim=0 100 0 100, clip, width=.45\textwidth,valign=c]{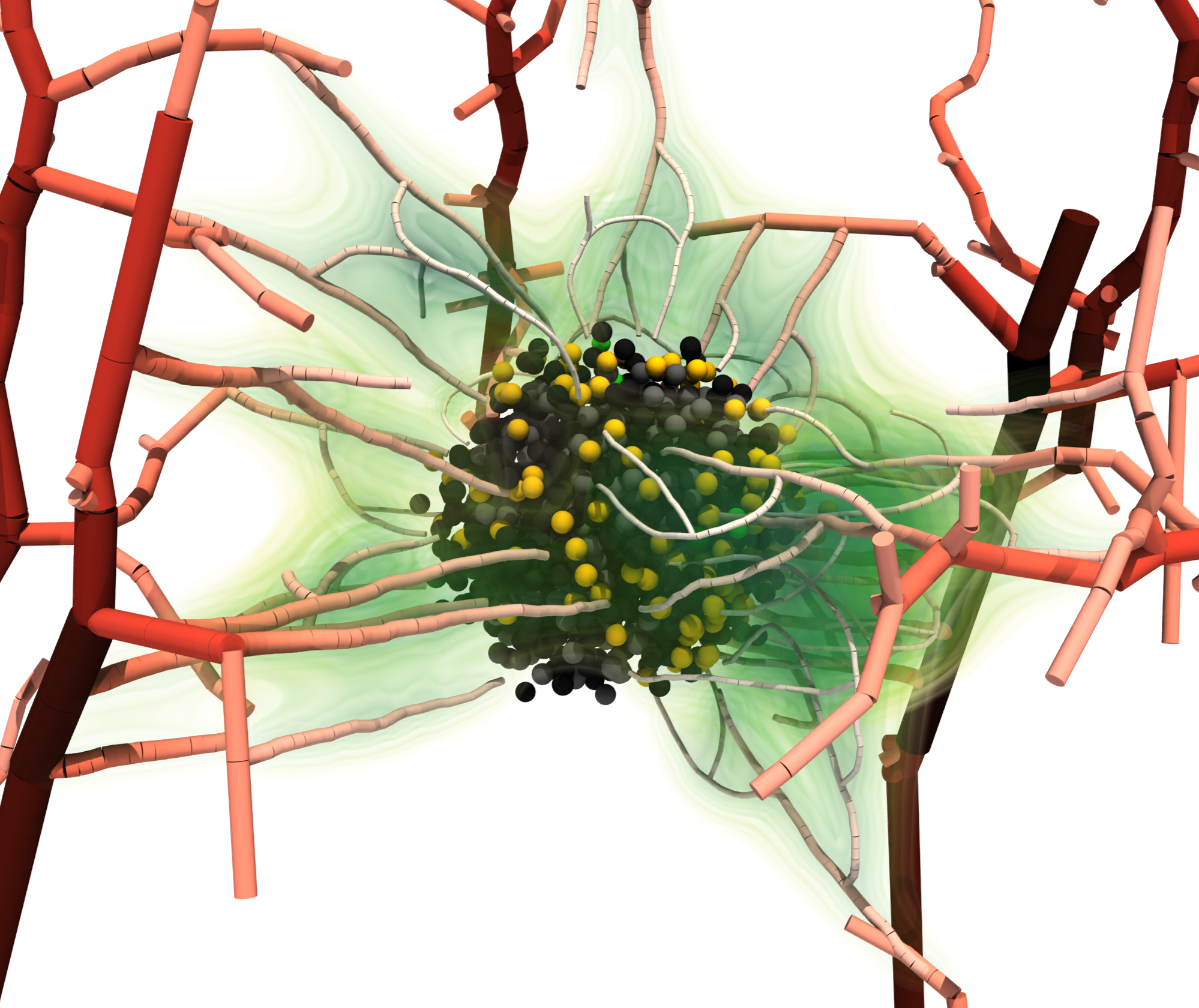}}
  } \\
  \subfloat[Final vasculature and nutrients]{
    {\includegraphics[trim=0 100 0 100, clip, width=.45\textwidth,valign=c]{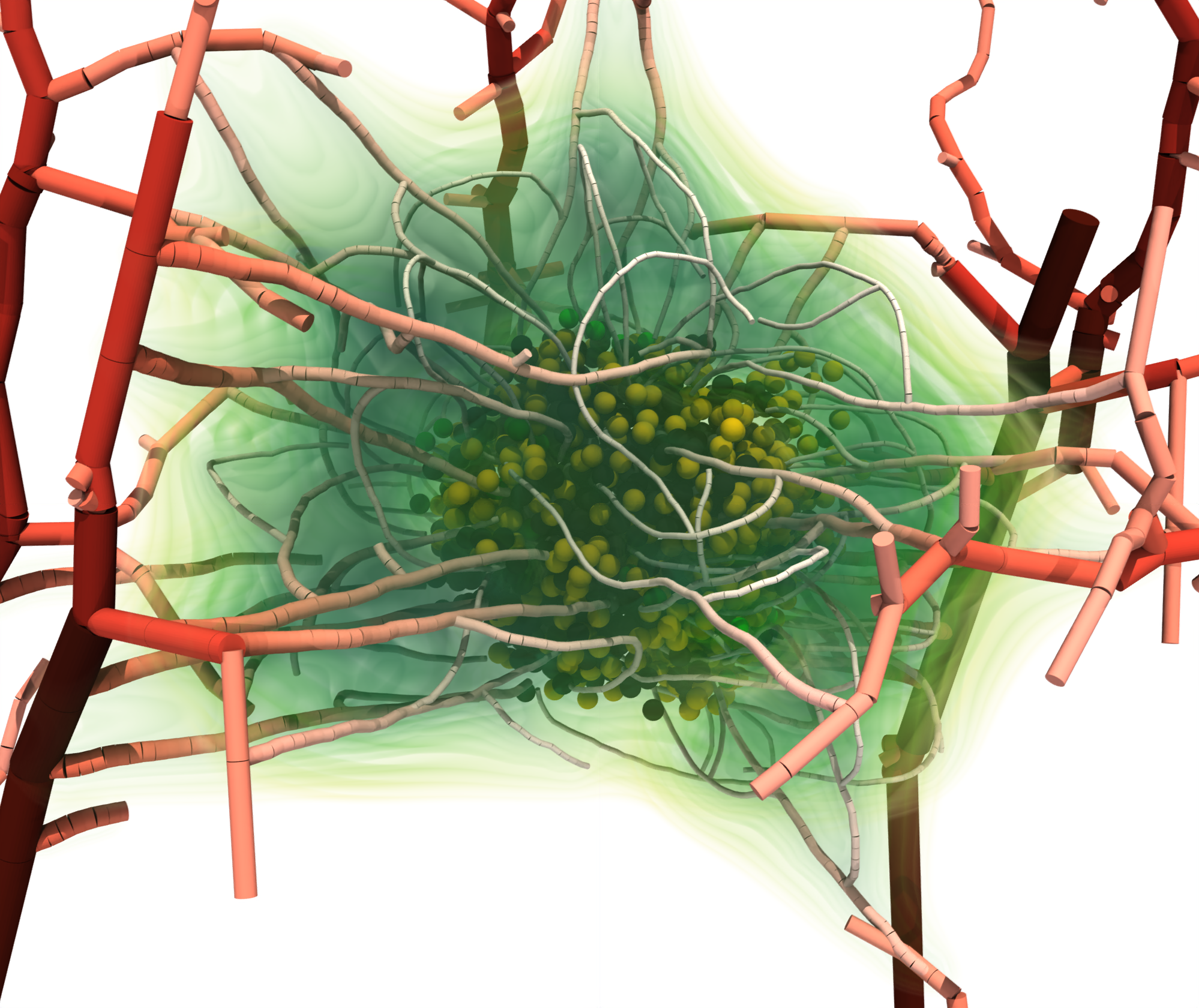}}
  }
  \qquad
  \subfloat[Tumor before treatment]{
    {\includegraphics[trim=0 100 0 100, clip, width=.45\textwidth,valign=c]{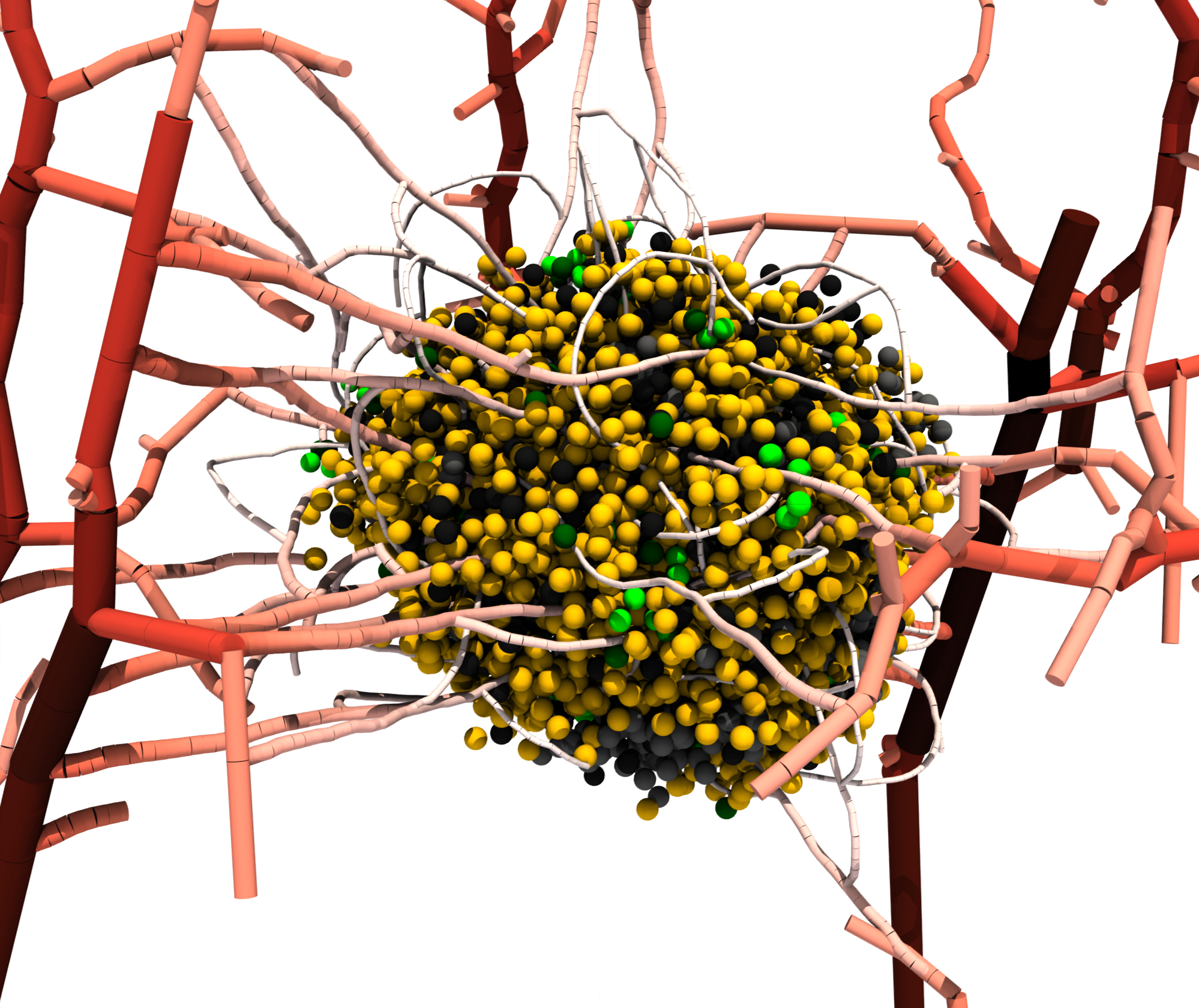}}
  } \\
  \subfloat[Treatment with TRA]{
    {\includegraphics[trim=75 220 57 220, clip, width=.6\textwidth,valign=c]{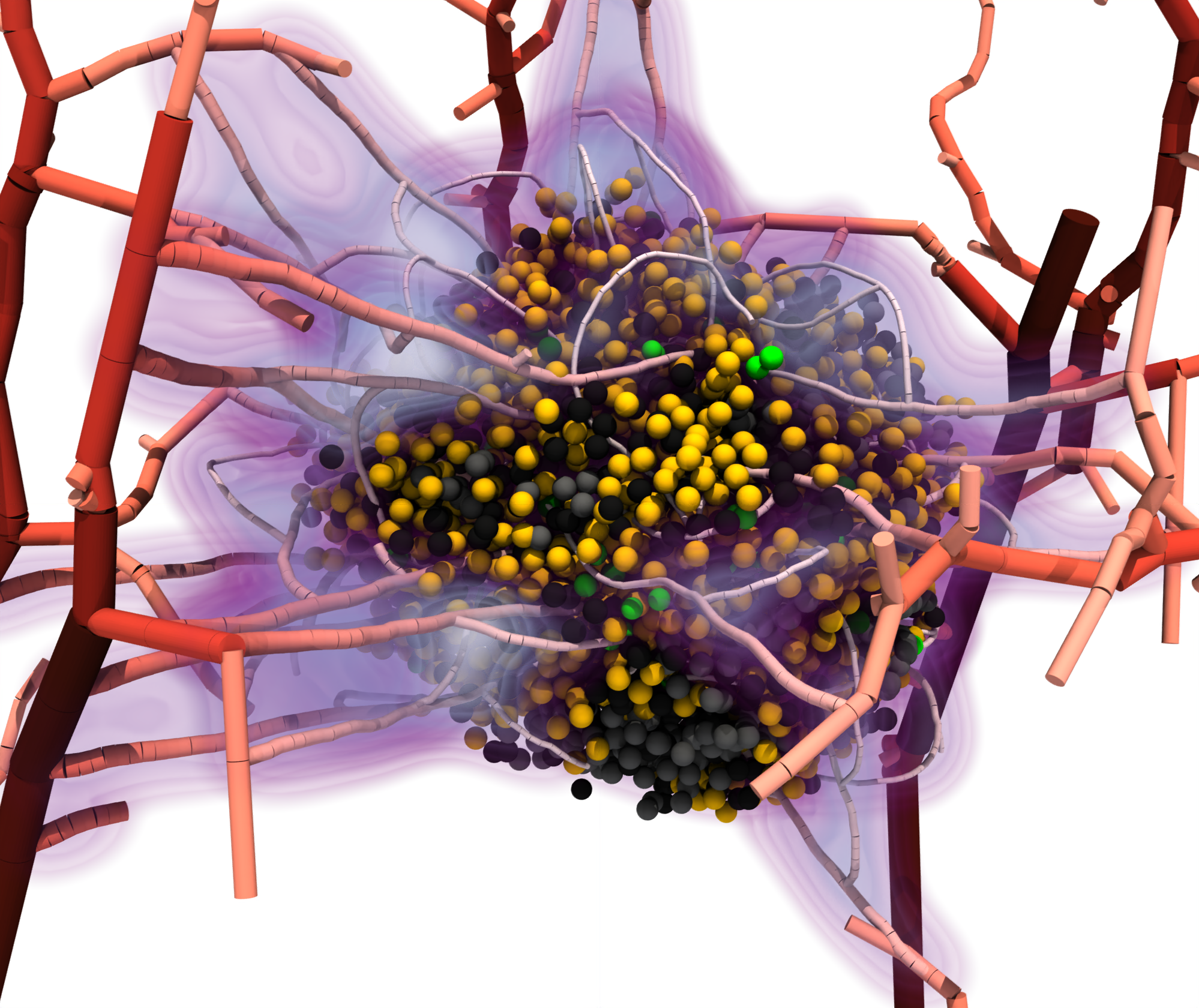}}
  }
  \caption{
    Visualization of a full simulation run with tumor cells colored by their
    cell state: Q (yellow), SG2 (dark green), G1 (light green), H (gray), and
    D (black). (a) Initial hypoxic population secreting VEGF (blue) to trigger
    angiogenesis. (b) First vessels reach the tumor surface and supply nutrients
    (green) leading to cells taking proliferative states on the surface.
    (c) Final state of the vasculature, i.e., we deactivated the vessel growth
    algorithm at this point. (d) Final tumor before treatment initialization.
    (e) Early stage of the treatment (TRA in purple).
  }
  \label{fig:result-full-concept}
\end{figure}

In Fig.~\ref{fig:result-full-concept}, we show the evolution of the simulation
over time. While we start with only 1000 tumor cells, the number increases by
a factor of 5 until the treatment begins.
Figure~\ref{fig:result-full-concept}~(a)
shows the simulation briefly after the initialization. All tumor cells are in
the hypoxic state and secrete VEGF displayed in blue. The dynamics of VEGF
are encapsulated in the PDE (Eq.~(\ref{eq:PDE_VEGF})) and, consequently, it
diffuses
from the tumor into the surrounding regions. After some time that primarily
depends on the secreted amount and the diffusion coefficient, VEGF reaches the
vasculature and, in reaction, sprouts form and move along the gradient towards
the tumor in the center. The new vasculature supplies nutrients, and once the
nutrients and vessels reach the spheroid, cells on the surface begin
transitioning into proliferative states
(Fig.~\ref{fig:result-full-concept}~(b)). Afterward, the vasculature keeps
developing until terminated at day 70
(Fig.~\ref{fig:result-full-concept}~(c)). The tumor evolves in the final
vasculature with its proliferative ring until it reaches its largest size in
Fig.~\ref{fig:result-full-concept}~(d).

It is interesting to note that the model
favors growth along the vessels because these regions provide more nutrients
and the cells are more likely to transition into the proliferative state SG2.
Thus, more tumor forms in the well-vascularized regions. Admittedly, this is not
immediately evident in Fig.~\ref{fig:result-full-concept};
however, the bottom region
in (d) is poorly vascularized and one can see that the hypoxic cells reach all
the way to the surface of the tumor spheroid (d/e). Furthermore,
considering (b), we
observe a fairly dense network on the right side of the tumor
resulting in an outgrowth to the right in (c). Overall, the model shows expected
characteristics and significant similarities to images of 3D \textit{in vitro}
studies \cite{Ehsan2014}.

After the tumor has formed, we trigger the treatment with TRA and DOX.
Figure~\ref{fig:result-full-concept}~(e) shows the tumor and
the concentration of
TRA (purple low, blue high) created by the vessel source terms.
After the treatment, tumor cells stop proliferation and begin transitioning into
the dead state inhibiting further tumor growth.
The effects of the treatment are investigated in the following section.

\subsection{Treatment Comparison}
\label{sec:ResultsTreatment}

\begin{figure}
  \centering
  \subfloat[DOX only]{
    {\includegraphics[clip, width=.4\textwidth,valign=c]{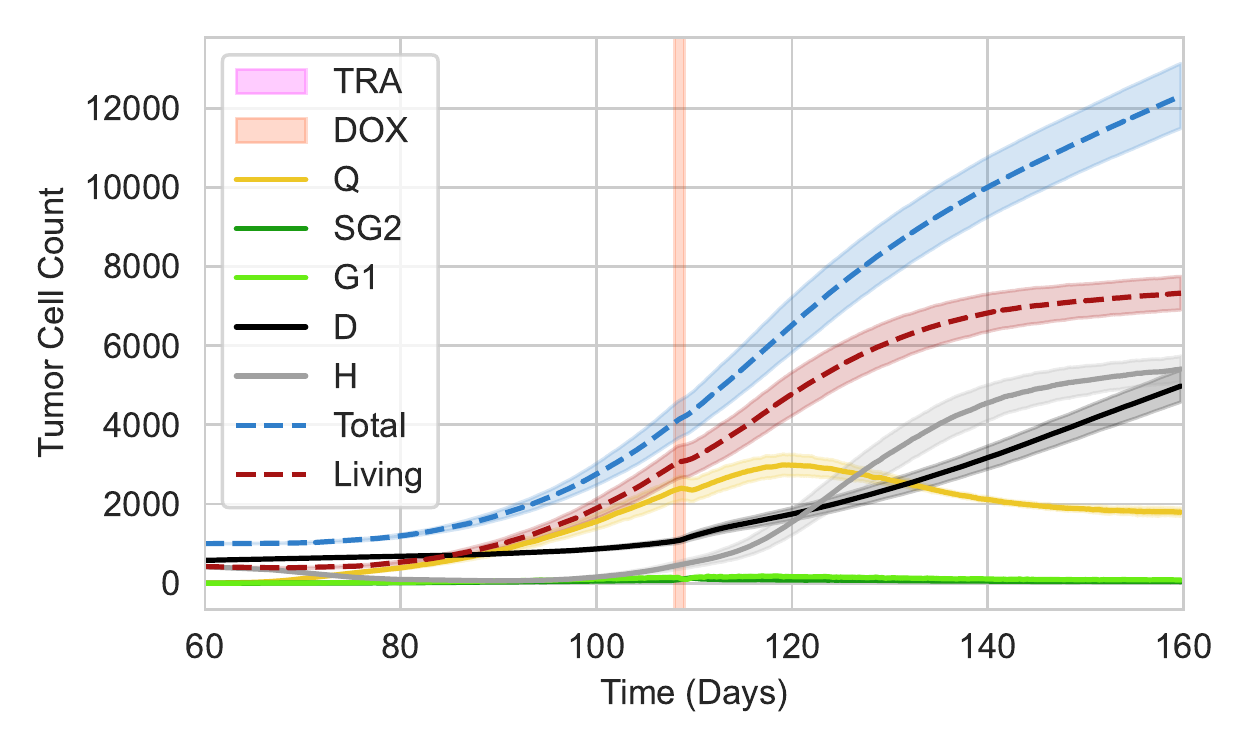}}
  } \qquad
  \subfloat[TRA-TRA]{
    {\includegraphics[clip, width=.4\textwidth,valign=c]{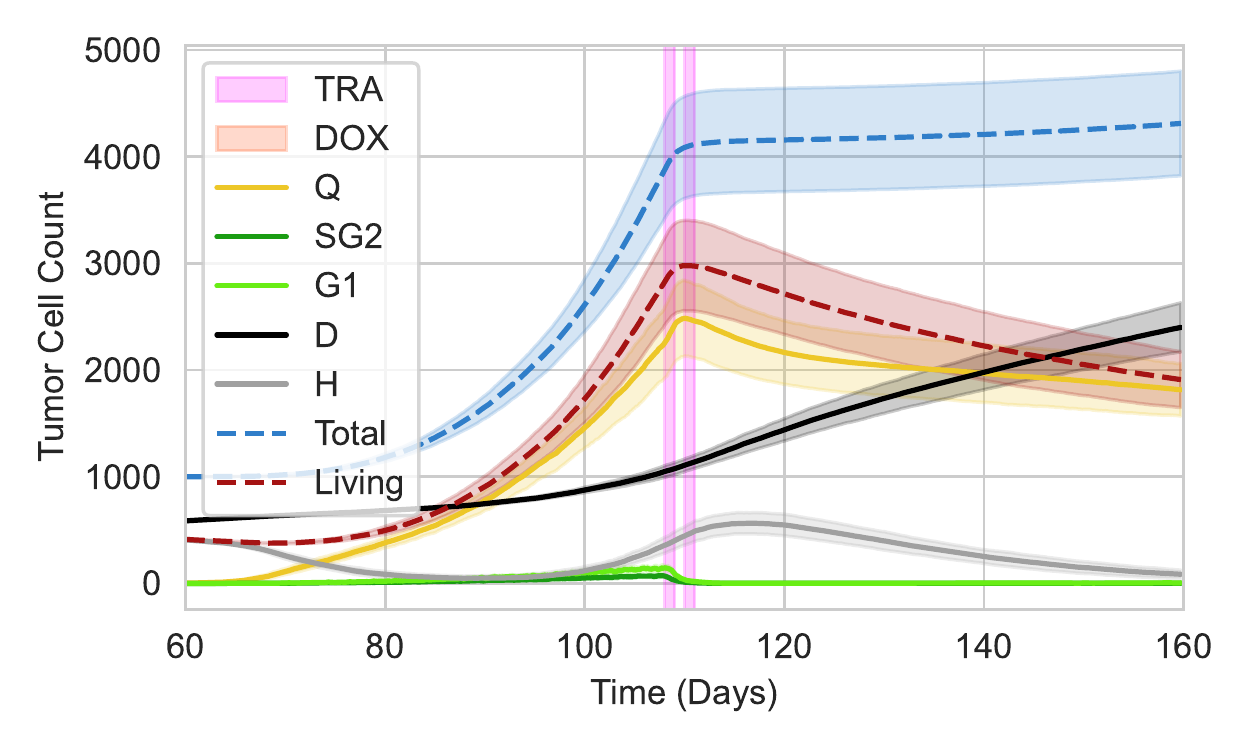}}
  } \\
  \subfloat[TRA-TRA-DOX]{
    {\includegraphics[width=.4\textwidth,valign=c]{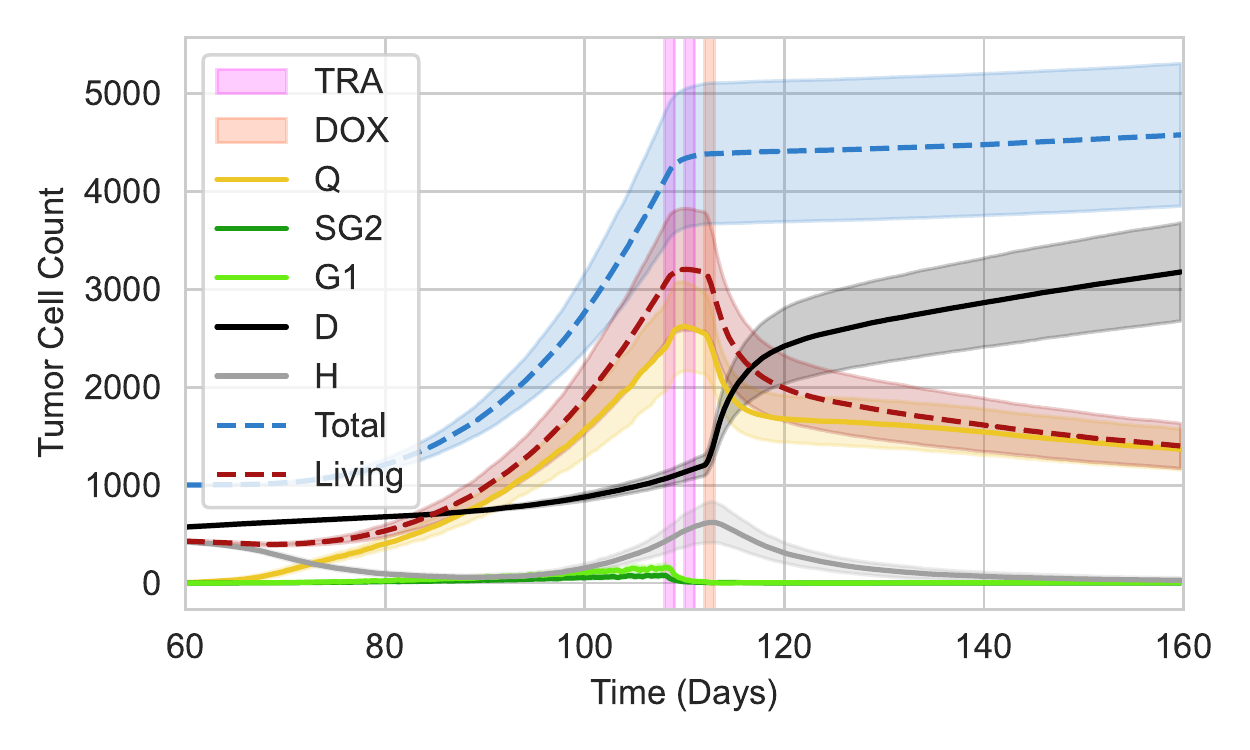}}
  } \qquad
  \subfloat[DOX-TRA-TRA]{
    {\includegraphics[width=.4\textwidth,valign=c]{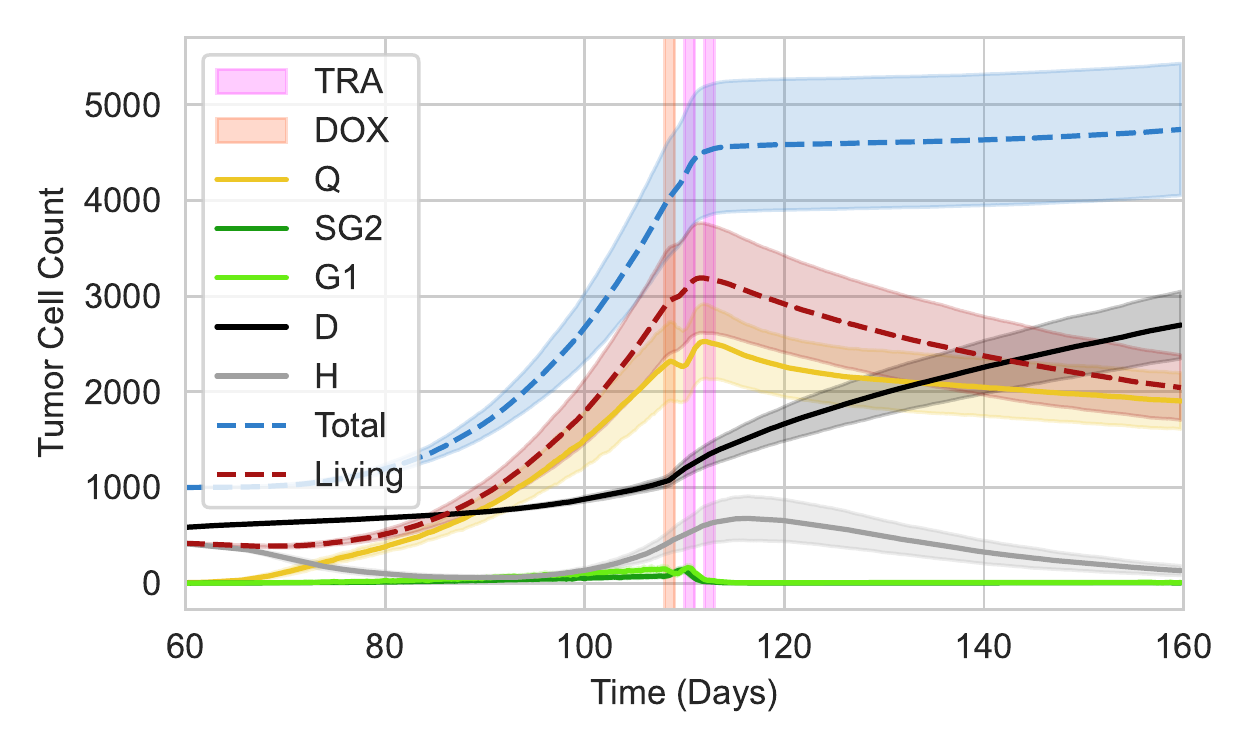}}
  }
  \caption{
    Evolution of the number of tumor cells in the different states for
    different treatment protocols. The line of the \textit{living} cells
    add up all states but the dead cells. The line of the \textit{total} number
    of cells further includes the dead cells.
  }
  \label{fig:result-treatment-comparison}
\end{figure}

Recall that a principal goal of this investigation is to
build a model to better understand the combination
of DOX and TRA for breast cancer treatment.
In this section, we initialize the simulation as in the previous two
experiments and Fig.~\ref{fig:result-full-concept} depicts the different
simulation stages. Here, we selected the treatment
parameters such that our model accurately describes Jain's
hypothesis~\cite{Jain2001} and matches the trends in the data~\cite{Sorace2016}.
For the treatment, we allocate three slots between the days
108-109, 110-111, and 112-112. During these intervals, the vasculature acts as
source terms for the cancer drugs. We consider four treatment scenarios
relating to the treatment groups 2, 3, 4, and 5 of Fig.~\ref{fig:growth-data},
i.e., DOX only, TRA only, TRA followed by DOX, and DOX followed by TRA. For each
treatment scenario, we run 10 simulation runs to account for the inherent
stochasticity and plot the mean and standard deviation of the number of cells
in different states over time. The results are depicted in
Fig.~\ref{fig:result-treatment-comparison}. All simulations use the identical
set of parameters and only differ in the treatment schedule.

Figure~\ref{fig:result-treatment-comparison} (a) shows the treatment effects
when we only apply DOX. In our simulations, this protocol is ineffective
and the tumor growth is barely disturbed. After the application, the quiescent
cells show a slight decline but they quickly recover. Recall that DOX has a
short half-life and, thus, long-term changes are not readily observed.
The unaffected tumor growth agrees qualitatively with the data in
Fig.~\ref{fig:growth-data}~(b).

TRA has a significantly longer half-life and we expect to see long-term effects.
In Fig.~\ref{fig:growth-data}~(c), the data shows that the TRA treatment stalls
the tumor growth. Our simulations show the same pattern, e.g., in
Fig.~\ref{fig:result-treatment-comparison}~(b), the tumor stops growing.
This effect is modeled with the $Q \rightarrow SG2$ suppression through TRA.

Next, we consider the scenario in which we first apply TRA and subsequently
supply DOX - the test case for Jain's hypothesis.
In Fig.~\ref{fig:result-treatment-comparison}~(c), right after DOX is applied,
we observe a sharp decline in quiescent cells and a strong increase in the
number of dead cells.
Here, the dose is significantly more effective than if only DOX is applied.
This is due to the improved supply properties of the
vasculature caused by the preceding TRA treatment allowing more DOX to
enter the system. Among all the simulated scenarios, the TRA-TRA-DOX
treatment shows the strongest treatment effect, i.e., the number of (living)
tumor cells at the end is the lowest. This result agrees with
Fig.~\ref{fig:growth-data}~(e) and Jain's hypothesis.

Lastly, we want to consider the inverse case, e.g., DOX treatment followed by
TRA (Fig.~\ref{fig:result-treatment-comparison}~(d)). Our simulation results
are, in this case, hard to distinguish from the case in which we solely
use TRA. Unfortunately, they disagree with the data
(Fig.~\ref{fig:growth-data}~(d)) which suggests
that the first dose of DOX prohibits TRA from being effective. The present model
does not capture this feature. Terms involving both drugs cannot explain the
observation because they do not differentiate between the order in which drugs
arrive. We hypothesize that DOX either affects the vessels and therefore the
supply, or effects of DOX in the cell's internals disturb the pathways used by
TRA. Both effects have not been considered in the model. While there are
some hints that DOX may in fact damage the microvasculature \cite{Hader2019},
this would also harm the nutrient supply and contradict the strong growth in
Fig.~\ref{fig:growth-data}~(d). Thus, we lean towards the latter and hypothesize
that DOX negatively influences the way TRA can work inside the cells.

\section{Towards Large Scale Simulations}\label{sec:large-scale}

Criticism of some ABMs has arisen
because of their high computational costs and lack of
scalability. Typical ABM simulations focus on small-scale systems and cannot
simulate medically relevant sizes.
However, recent advances in ABM software \cite{Breitwieser2021,Breitwieser2023}
address many of the computational bottlenecks and provide the foundation for
scaling up simulations.
Leveraging these optimizations through the BioDynaMo API, we here demonstrate
that our model and the associated C++ code can handle medically relevant system
sizes by reproducing the pre-treatment data of Sorace's pre-clinical
study~\cite{Sorace2016}.

We choose a $9 \times 9 \times 9 mm$ simulation volume.
For lack of data availability, we stochastically mimic
the vascular density across the simulation volume. For details, consider
\ref{appendix:stochastic-vessels}.
In the pre-clinical study \cite{Sorace2016},
the researchers injected $10$ million
tumor cells into rodents. To agree with the average tumor volume
observed on day seven, we initialize our simulation with $6$ million tumor
cells in a spheroid. We focus on the pretreatment stage and simulate from day
7 to 34 with a timestep of 10 minutes. In other words, we simulate 27 days with
3888 time steps. We discretize the continua with
$22.5 \times 22.5 \times 22.5 \mu m$ voxels (roughly the size of a tumor
cell).

\begin{figure}
  \centering
  \includegraphics[width=0.8\textwidth]{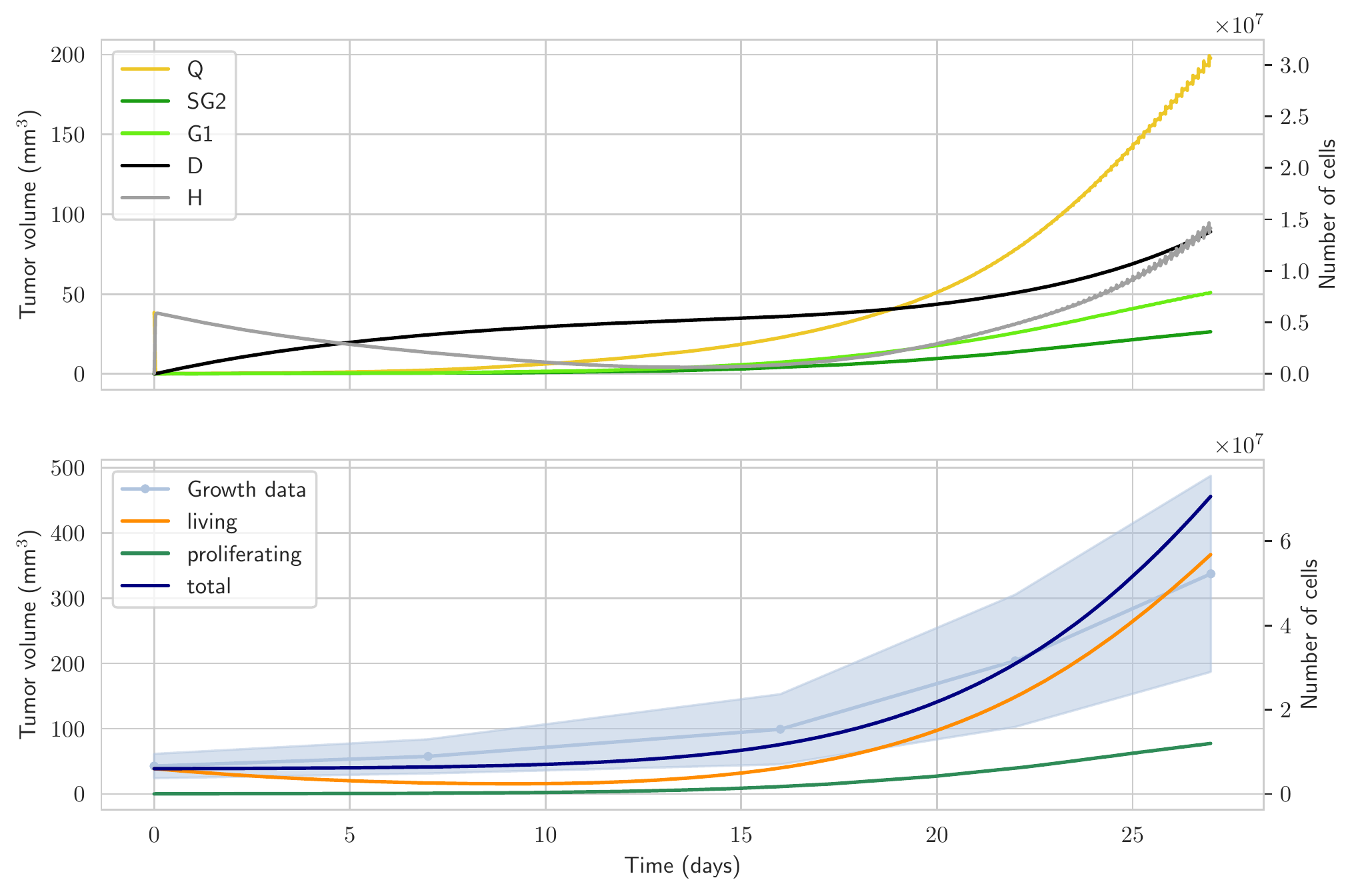}
  \caption{Large scale simulation: tumor volume and cell count over time.
    Top: volume and number of cells according to the five cell states.
    Bottom: Aggregated numbers proliferating ($SG2 + G1$), living ($SG2+G1+Q+H$),
    and total (all).
    Tumor volume computed as $V_{tumor} = N \cdot V_{cell} / 0.64$, with the
    measured number of cells $N$ and a correction factor accounting for
    sphere packing. The displayed \textit{growth data} combines the
    pre-treatment stage of all treatment groups in Fig.~\ref{fig:growth-data}.
  }
  \label{fig:large-scale-results}
\end{figure}

Figure~\ref{fig:large-scale-results} shows the tumor volume over time.
In the beginning, all cells are in
a hypoxic state. They then begin to stochastically transition into
the dead state and no longer consume any nutrients. At the same time, the
vasculature starts growing, and the available nutrients increase. From day
three, we observe cells transitioning into proliferative states, and the tumor
grows in some regions. Between days 10 and 15, we observe that
almost as many cells are in the necrotic state as we initialized, indicating that
most of the spheroid died, and the initial spheroid now forms a necrotic core.
From day ten on, parts of the tumor are well supplied with nutrients
after attracting the vasculature via VEGF, and we observe exponential growth.
Between days 15 and 20, the number of hypoxic cells increases again. This trend
suggests that the exponential growth of the tumor mass
depleted newly vascularized regions, and parts of the tumor begin to die.
The overall tumor dynamics seem reasonable and agree with the
pre-treatment data (Tab~\ref{tab:tumor-evolution-data-merged}) as can be seen
in the lower part of Fig.~\ref{fig:large-scale-results}.

While starting with 6 million tumor cells and roughly 4.5 million
vessel segments, the simulation concludes with 70.6 million tumor
cells and 21 million vessel agents, respectively. The vessel volume, and
therefore the vascular density increased by a factor of roughly 5.
Overall, the simulation took approximately 7.9 days on a
72-core server with 1 TB of RAM
and hyper-threading (4 x Intel(R) Xeon(R) E7-8890 v3 clocking at 2.50 GHz with
four NUMA domains) and the memory usage peaked around 50 GB. It appears that the
force computation is responsible for most of the computation. We expect that
the runtime can still be significantly reduced by optimizing the model code.

\section{Discussion}\label{sec:discussion}

The computational and mathematical models and algorithms described in this work
appear to be capable of simulating very complex growth of vascular structures
and variations in tumor volume in environments in which drug protocols are
designed and orchestrated to control, minimize, or eliminate tumor growth.

A limitation of the model is that the healthy tissue surrounding the tumor is
not considered. As long as the \textit{in silico} tumor floats in a vacuum, it is
difficult to mimic the physical traits \cite{Nia2020} such as stress, pressure,
and stiffness. Effects, such as vasculature being damaged by forces, can hardly
be modelled when the cells can escape into the empty space.
Ignoring the forces between tumor cells and the vasculature is another limiting
factor linked to the previous point. Without the healthy tissue surrounding
the vasculature and tumor, the vasculature would be pushed away rather than
forming a supply network.
Future work should also model the cell death in more detail
to free space for the proliferative tumor mass and the healthy
tissue (see \cite{Lejeune2020}).
Moreover, our model neglects cell migration which has proven to have a
significant impact on the tumor dynamics in theoretical studies
\cite{Gonzalez2018, Waclaw2015}.

While the generated vascular networks appear to be realistic and organic,
the growth does not entirely stop and vessels can begin growing in unexpected
directions. In fact, vasculature grows less structured in the presence of a
tumor; however, our model lacks pruning mechanisms for
the vessels growing in random directions.
Extending the model with a flow simulation seems a promising
direction for further research and would give additional information based on
which one may prune and
optimize the vasculature (see, e.g., \cite{Koeppl2020}).

Moving away from the vasculature, we note that the model has many parameters,
including some that have not been properly calibrated with data.
The number of parameters is, however, a by-product of the complexity that was
targeted in this work. Adding more mechanisms to the model inevitably adds
more parameters.
Nonetheless, by building on similar models and their (partially calibrated)
parameters \cite{Macklin2012,Rocha2018,Lima2021,Phillips2020}, we were able to
find reasonable parameter choices for the model which we demonstrated
through the model's ability to simulate vascular tumor growth and treatment.
We were able to reproduce the qualitative reaction of HER2+ breast cancer to
the combination treatment with DOX and TRA.
The parameter
choices and modeling approach is further justified by the fact that the model
produces realistic tumor volumes in agreement with the pre-clinical
study~\cite{Sorace2016}. However, our model ignores effects on the
tissue scale (e.g., cell death and migration) and is therefore not yet adequate
for predicting quantities of interest at this scale.

The models described in this work offer many interesting
opportunities. They have great potential for describing
small \textit{in vitro} experiments (see, for instance, \cite{Ehsan2014}).
Such data would further help calibrate our model.
For these small \textit{in vitro} scales, parameters may be inferred using
Bayesian frameworks because the model runs fast and shared-memory parallel such
that frequent model evaluation may be possible.
However, even though we emphasized computational efficiency during the
development, the calibration of stochastic models remains a challenging
subject because the repeated evaluation of the forward model may amount to
substantial run times.
Recent advances in Bayesian computation \cite{Rocha2022} suggest the
construction of surrogates based on Gaussian processes to reduce the number of
forward simulations. The method has proven efficient for other
stochastic cancer models and may help to calibrate ours.
Once those parameters have been calibrated, one may run large-scale simulations
and compare the results to macroscopic data, as demonstrated in
this work. The model and its implementation should enable researchers
to bridge scales, i.e., hypothesize phenomena on microscopic, cellular scales
and compare the simulation results to macroscopic data.

\section{Conclusion}\label{sec:conclusion}

In this work, a complex hybrid model is presented together with a
performant C++ implementation. The model is shown to capture
many characteristics of vascular tumor growth and, qualitatively,
describes the treatment effects of Doxorubicin and Trastuzumab on HER2+
breast cancer cells. Furthermore, the model and code can
scale to tissue-relevant sizes and may therefore help future research to
bridge scales, i.e, hypothesize cellular effects and test how they affect
macroscopic quantities.

\section*{Acknowledgements}
TD would like to thank Lukas Breitwieser, Fons Rademakers, and Ahmad Hesam for
the many technical discussions related to BioDynaMo that
greatly helped him progressing in the present work.
The authors would like to further thank Lukas Breitwieser for sharing his Docker
infrastructure, helping with its set up, and testing the reproducer.
The authors also thank Tobias Köppel for sharing code, data, and knowledge
that helped setting up the initial vessel structure.
The work of TD has been sponsored by the Wolfgang Gentner Programme of the German
Federal Ministry of Education and Research (grant no.~13E18CHA).
The work of EABFL is supported by the National Institute of Health via
grant~R01CA240589.
The work of JTO is supported by the U.S. Department of Energy,
Office of Science, Office of Advanced Scientific Computing Research,
Award~DE-960009286.
The work of BW was funded by the German Research Foundation by
grants~WO671/11-1.


\appendix

\section{Data}\label{appendix: data}

The data in \cite[Tab.~1 and 5]{Lima2022} contains six groups for the
tumor growth before treatment begin. Generally,
two sets $S_i = \{ x_1 , \dots , x_{n_i} \}$ of real numbers with their
respective cardinality ($n_1,n_2$), mean ($m_1,m_2$), and standard deviation
($\sigma_1, \sigma_2$) can be combined in one set $S$ with its characteristics
given by
\begin{align}
  n      & = n_1 + n_2 \ ,                                        \\
  m      & = \frac{n_1 m_1 + n_2 m_2}{n_1 + n_2} \ , \ \text{and} \\
  \sigma & = \sqrt{
    \frac{(n_1 - 1)\sigma_1^2 + (n_1 - 1)\sigma_1^2 +
      \frac{n_1 n_2}{n_1 + n_2} (m_1 - m_2)^2}
    {n_1 + n_2 - 1}
  } \ .
\end{align}
We apply these formulas iteratively to the dataset in
\cite[Tab.~5]{Lima2022} and obtain a dataset characterizing the
tumor growth before treatment in a statistically more sound way. The dataset is
given in Tab.~\ref{tab:tumor-evolution-data-merged}.

\begin{table}[h]
  \centering
  \begin{tabular}{@{}cc@{}}
    \toprule
    Days & Combined            \\ \midrule
    7    & 42.721 $\pm$ 18.69  \\
    14   & 57.53 $\pm$ 26.19   \\
    23   & 99.06 $\pm$ 53.76   \\
    29   & 203.95 $\pm$ 101.30 \\
    34   & 337.21 $\pm$ 150.22 \\ \bottomrule
  \end{tabular}
  \caption{Tumor volume in cubic millimeters over time until treatment begin.
    Data merged from \cite[Tab.~5]{Lima2022}.}
  \label{tab:tumor-evolution-data-merged}
\end{table}

\section{Model Parameter}\label{appendix:model_parameter}

\begin{table}[h]
  \centering
  \small
  \begin{tabular}{@{}lllllll@{}}
    \toprule
    \textbf{Parameter}    & \textbf{Notation}           & \textbf{Value}         & \textbf{Unit} & \textbf{Misc} \\ \midrule
    number of tumor cells & $N$                         & $10^3$, $6 \cdot 10^6$ & 1             & cube domain   \\
    lower bound           & $x_{min}, y_{min}, z_{min}$ & -1.3 , -4.5            & $mm$          & cube domain   \\
    upper bound           & $x_{max}, y_{max}, z_{max}$ & 1.3 , 4.5              & $mm$          & cube domain   \\
    resolution continua   & $N_n, N_v, N_d, N_t$        & 200, 400               & 1             & voxel per dim \\ \bottomrule
  \end{tabular}
  \caption{Overview of the parameters affecting the simulation.
    (regular simulation, large scale)
  }
\end{table}

\begin{table}[h]
  \centering
  \small
  \begin{tabular}{@{}llllll@{}}
    \toprule
    \textbf{Parameter}     & \textbf{Notation} & \textbf{Value} & \textbf{Unit} & \textbf{Source}     \\ \midrule
    Bifurcation distance   & $d_b$             & 80             & $\mu m$       & -                   \\
    Tip-cell distance      & $d_t$             & 150            & $\mu m$       & -                   \\
    Sprouting rate         & $p_s$             & 0.001          & $1/min$       & -                   \\
    Growth weight random   & $w_r$             & 0.2            & -             & -                   \\
    Growth weight old      & $w_o$             & 0.5            & -             & -                   \\
    Growth weight gradient & $w_\nabla$        & 0.3            & -             & -                   \\
    Growth speed           & $s$               & 0.033          & $\mu m / min$ & \cite{Phillips2023} \\
    Supply increase        & $\tau_\uparrow$   & 0.4            & days          & -                   \\
    Supply decrease        & $\tau_\downarrow$ & 10             & days          & -                   \\
    Supply maximum         & $\chi_{\max}$     & 9              & days          & -                   \\ \bottomrule
  \end{tabular}
  \caption{Overview of the parameters affecting the vessel.}
  \label{tab:ParameterAngiogenesis}
\end{table}

\begin{table}
  \centering
  \small
  \begin{tabular}{@{}llllll@{}}
    \toprule
    \textbf{Parameter}  & \textbf{Notation} & \textbf{Value} & \textbf{Unit} & \textbf{Source}                \\ \midrule
    cell radius         & $r_{c,\mu}$       & 9.953          & $\mu m$       & \cite{Macklin2012,Slooten1995} \\
    nuclear cell radius & $r_{c,n}$         & 5.296          & $\mu m$       & \cite{Macklin2012}             \\
    action cell radius  & $r_{c,a}$         & 12.083         & $\mu m$       & \cite{Macklin2012}             \\ \bottomrule
  \end{tabular}
  \caption{Overview of the parameters affecting the tumor cell.}
  \label{tab:ParameterCellRadii}
\end{table}

\begin{table}
  \centering
  \small
  \begin{tabular}{@{}lllllll@{}}
    \toprule
    \textbf{Transition}    & \textbf{Parameter}    & \textbf{Notation}                 & \textbf{Value}      & \textbf{Unit} & \textbf{Source}    \\ \midrule
    $ SG2 \rightarrow G1 $ & duration cell cycle   & $\tau_p = \tau_{G1} + \tau_{SG2}$ & 18                  & hour          & \cite{Macklin2012} \\ \midrule[.1pt]
    $ G1 \rightarrow Q $   & duration growth phase & $\tau_{G1}$                       & 9                   & hour          & \cite{Macklin2012} \\ \midrule[.1pt]
    $ D \rightarrow $ gone & duration apoptosis    & $\tau_D$                          & 8.6                 & hour          & \cite{Macklin2012} \\ \midrule[.1pt]
    $Q \rightarrow H$      & nutrient threshold    & $u^{Q \rightarrow H}_n$           & 0.09                & -             &                    \\ \midrule[.1pt]
    $Q \rightarrow D$      & nutrient threshold    & $u^{Q \rightarrow D}_n$           & 0.000538            & -             & -                  \\
                           & nutrient gamma        & $\gamma^{Q \rightarrow D}_n$      & 0.000408            & -             & -                  \\
                           & nutrient alpha        & $\alpha^{Q \rightarrow D}_n$      & $6.8 \cdot 10^{-6}$ & -             & -                  \\
                           & nutrient k            & $k^{Q \rightarrow D}_n$           & 50.0                & -             & -                  \\
                           & DOX zeta              & $\zeta^{Q \rightarrow D}_d$       & 100                 & -             & -                  \\
                           & TRA zeta              & $\zeta^{Q \rightarrow D}_t$       & 0                   & -             & -                  \\ \midrule[.1pt]
                           & Cross term            & $\zeta^{Q \rightarrow D}_{dt}$    & 0                   & -             & -                  \\ \midrule[.1pt]
    $Q \rightarrow SG2$    & nutrient threshold    & $u^{Q \rightarrow SG2}_n$         & 0.0538              & -             & \cite{Lima2021}    \\
                           & nutrient alpha        & $\alpha^{Q \rightarrow SG2}_n$    & 0.000821            & -             & \cite{Lima2021}    \\
                           & decay with TRA        & $a^{Q \rightarrow SG2}_t$         & 30                  & -             & -                  \\ \midrule[.1pt]
    $SG2 \rightarrow SG2$  & DOX threshold         & $u^{SG2 \rightarrow SG2}_d$       & 0.001               & -             & -                  \\
                           & DOX alpha             & $\alpha^{SG2 \rightarrow SG2}_d$  & 0.1                 & -             & -                  \\ \midrule[.1pt]
    $SG2 \rightarrow D$    & DOX threshold         & $u^{SG2 \rightarrow D}_d$         & 0.001               & -             & -                  \\
                           & DOX alpha             & $\alpha^{SG2 \rightarrow D}_d$    & 0.001               & -             & -                  \\ \midrule[.1pt]
    $H \rightarrow D$      & Base rate             & $r$                               & 0.00001             & -             & -                  \\
                           & DOX zeta              & $\zeta^{H \rightarrow D}_d$       & 100                 & -             & -                  \\
                           & TRA zeta              & $\zeta^{H \rightarrow D}_t$       & 0                   & -             & -                  \\
                           & Cross term            & $\zeta^{H \rightarrow D}_{dt}$    & 0                   & -             & -                  \\ \bottomrule
  \end{tabular}
  \caption{Overview of the parameters affecting the cell cycle.}
  \label{tab:ParameterCellCycle}
\end{table}

\begin{table}
  \centering
  \small
  \begin{tabular}{@{}llllllll@{}}
    \toprule
    \textbf{Agent} & \textbf{Parameter}       & \textbf{Notation}       & \textbf{Value} & \textbf{Unit}         & \textbf{Source} \\ \midrule
    Tumor Cell     & nutrient consumption     & $c_n$                   & 0.0483         & $h^{-1}$              & \cite{Lima2021} \\
    Tumor Cell     & VEGF supply              & $c_v$                   & 2.73           & $h^{-1}$              & -               \\
    Tumor Cell     & DOX consumption          & $c_d$                   & 0.0            & $h^{-1}$              & -               \\
    Tumor Cell     & TRA consumption          & $c_t$                   & 0.000983       & $h^{-1}$              & -               \\ \midrule[.1pt]
    Vessel         & Nutrient supply          & $v_n$                   & 0.0000819      & $1/(\mu m^2 \cdot h)$ & -               \\
    Vessel         & VEGF consumption         & $v_v$                   & 0.109239       & $1/(\mu m^2 \cdot h)$ & -               \\
    Vessel         & DOX supply               & $v_d$                   & 0.000136       & $1/(\mu m^2 \cdot h)$ & -               \\
    Vessel         & TRA supply               & $v_t$                   & 0.002730       & $1/(\mu m^2 \cdot h)$ & -               \\ \midrule[.1pt]
    Vessel         & VEGF  threshold          & $v_v^{sprout}$          & 0.001          & 1                     & -               \\
    Vessel         & VEGF  gradient threshold & $v_{\nabla v}^{sprout}$ & 0.00001        & 1                     & -               \\ \bottomrule
  \end{tabular}
  \caption{Overview of the parameters affecting the agent - continuum interaction.
    Estimation explained in the main text. Note that
    $c_v$ and $v_v$ may be chosen significantly smaller without affecting the
    results but we report the parameters used for the numerical experiments.
  }
  \label{tab:ParameterContinuumInteraction}
\end{table}

\begin{table}
  \centering
  \small
  \begin{tabular}{@{}llllll@{}}
    \toprule
    \textbf{Parameter}  & \textbf{Notation} & \textbf{Value} & \textbf{Unit} & \textbf{Source} \\ \midrule
    Cell-cell adhesion  & $f_a$             & 0.0489         & $\mu m / min$ & \cite{Lima2021} \\
    Cell-cell repulsion & $f_r$             & 10.0           & $\mu m / min$ & \cite{Lima2021} \\
    Maximal speed       & $v_{max}$         & 10.0           & $\mu m / min$ & \cite{Lima2021} \\
    Viscosity           & $\nu$             & 2.0            & -             & \cite{Lima2021} \\ \bottomrule
  \end{tabular}
  \caption{Overview of the parameters affecting the forces.}
  \label{tab:ParameterForces}
\end{table}

\begin{table}
  \centering
  \small
  \begin{tabular}{@{}lllllll@{}}
    \toprule
    \textbf{Continuum} & \textbf{Parameter} & \textbf{Notation} & \textbf{Value} & \textbf{Unit}     & \textbf{Source} \\ \midrule
    Nutrients          & Diffusion          & $D_n$             & 0.833          & $\mu m ^ 3 / min$ & \cite{Lima2021} \\
    Nutrients          & Decay              & $\lambda_n$       & 0.00001        & $min^{-1}$        & \cite{Lima2021} \\ \midrule[.1pt]
    VEGF               & Diffusion          & $D_v$             & 0.175          & $\mu m ^ 3 / min$ & -               \\
    VEGF               & Decay              & $\lambda_v$       & 0.00001        & $min^{-1}$        & -               \\ \midrule[.1pt]
    DOX                & Diffusion          & $D_d$             & 0.708          & $\mu m ^ 3 / min$ & -               \\
    DOX                & Decay              & $\lambda_d$       & 0.0002         & $min^{-1}$        & -               \\ \midrule[.1pt]
    TRA                & Diffusion          & $D_t$             & 0.09           & $\mu m ^ 3 / min$ & -               \\
    TRA                & Decay              & $\lambda_t$       & 0.0            & $min^{-1}$        & -               \\ \bottomrule
  \end{tabular}
  \caption{Overview of the parameters affecting the continua.
    Explanation of the estimation may be found in the main text.}
  \label{tab:ParameterContinua}
\end{table}
\FloatBarrier


\section{Large Scale Simulation - Mimicking the mirco-vasculature in
  the tumor environment}
\label{appendix:stochastic-vessels}

The initial structure of the micro-vasculature plays an important role. A denser
vasculature may supply more nutrients and accelerate tumor evolution. However,
data for the initialization is merely available. This section describes our
approach to statistically mimic a realistic tumor micro-vasculature.

We first note that we do not consider a flow model; thus, connections between
different segments of the vascular network do not affect the simulation outcome.
Here, we place different vessel segments in the simulation space without
explicitly considering their connections. Each segment consists of several
agents that connect a start and end point and is further characterized by its
diameter and length. Under these assumptions, we strive to find meaningful
probability distributions to sample the diameter and length of the segments, as
well as a reasonable number for the total number of segments.

In 1998, Secomb and coworkers \cite{Secomb1998} investigated the blood flow
through the vasculature in the tumor region. The region's volume is
$V_S = (550\times550\times230) \mu m^3$, and they describe the different vessel
segments by start and end point, diameter, and length. The data is available to
the public and is the base for our statistical arguments.

\begin{figure}
  \centering
  \subfloat[diameter]{
    {\includegraphics[width=.49\textwidth,valign=c]{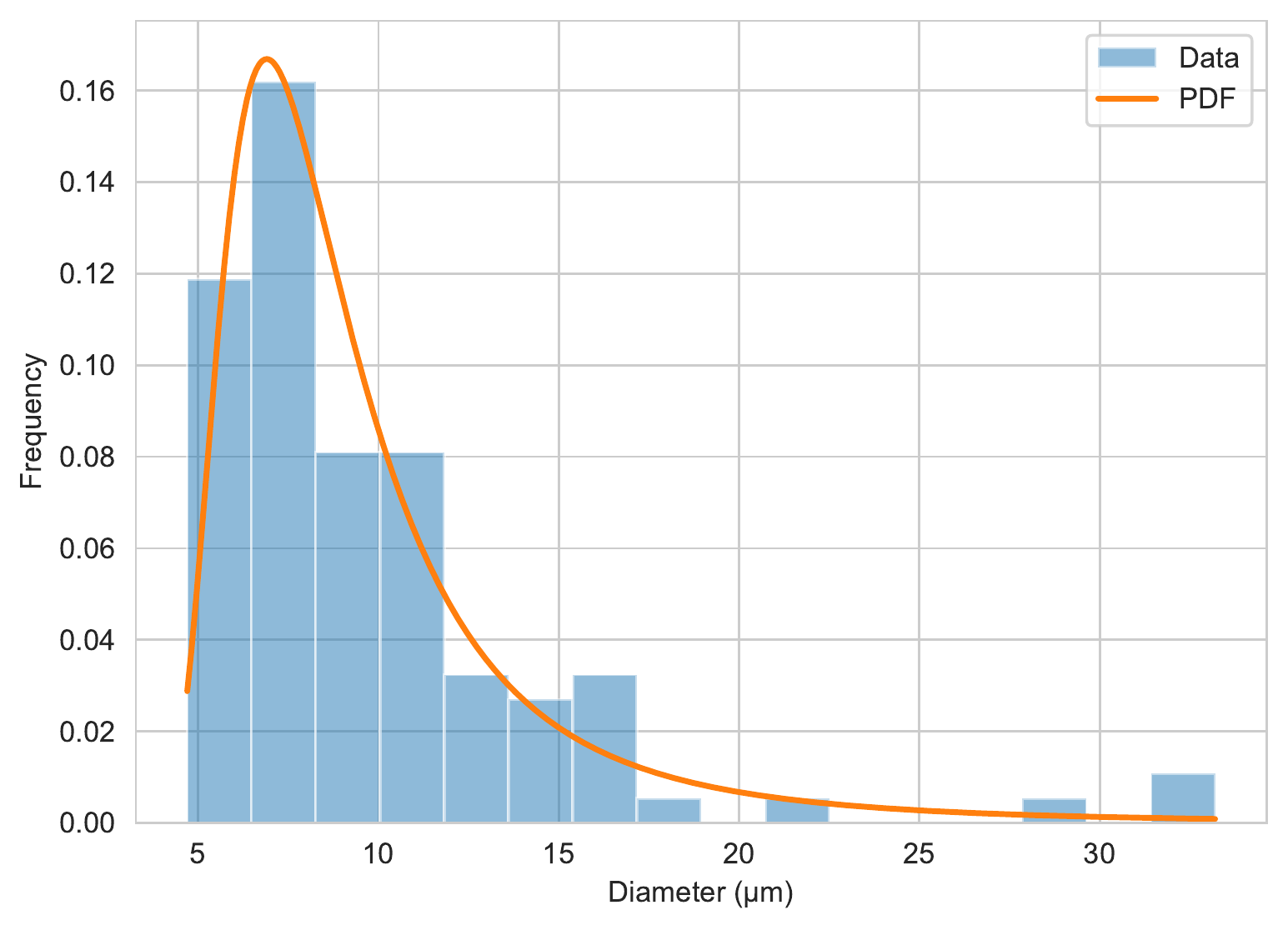}}
  }
  \subfloat[length]{
    {\includegraphics[width=.49\textwidth,valign=c]{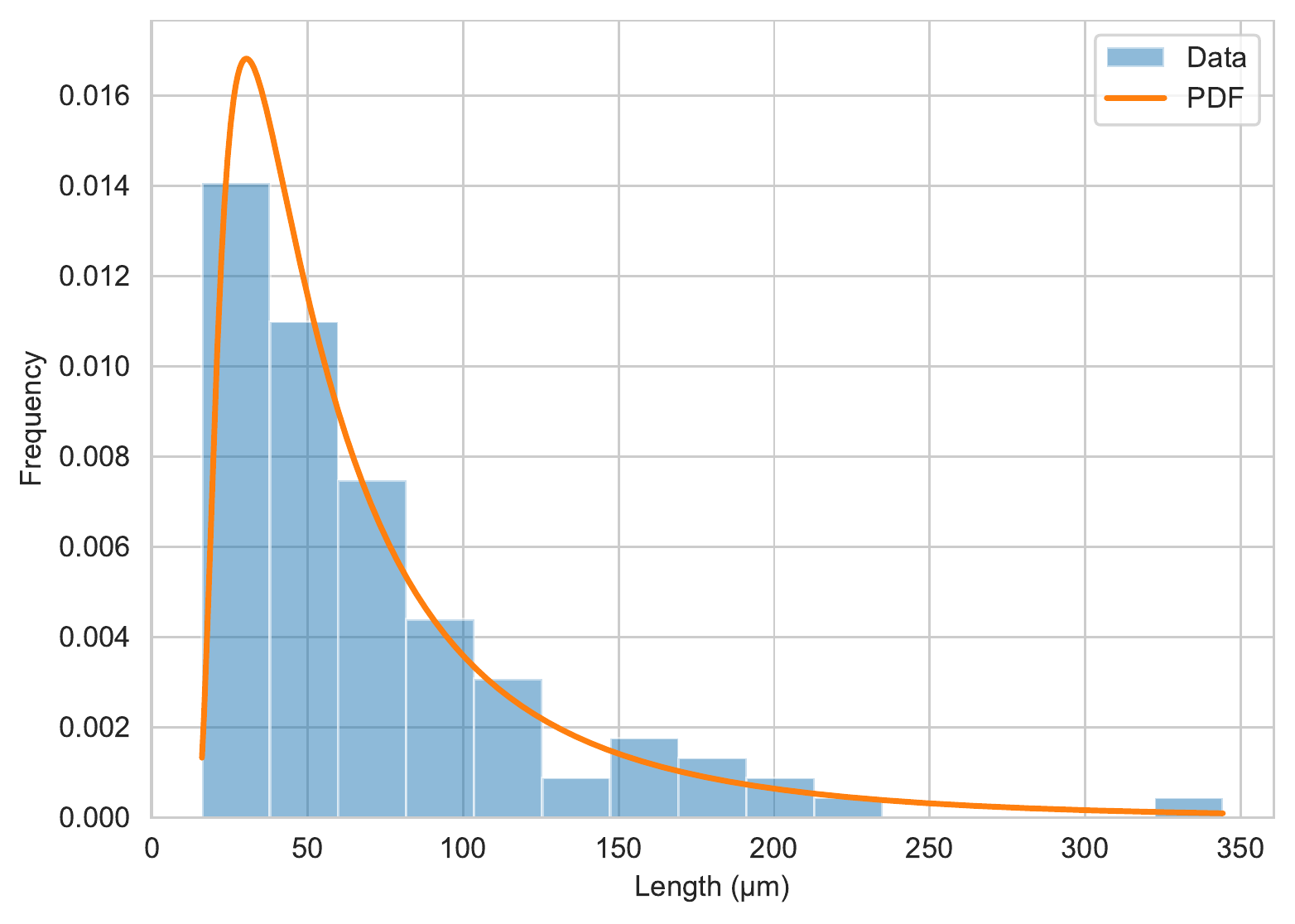}}
  }
  \caption{
    Histograms of the data \cite{Secomb1998} (blue) together with the best
    fitting PDF (orange). The PDF in (a) is a generalized extreme value
    distribution, and in (b) it is a Wald distribution.
  }
  \label{fig:VesselPDF}
\end{figure}

As we ignore the connections, we neglect the start and end point information and
focus on the data for diameter and length. We first compute the Pearson,
Spearman, and Kendall-Tau correlation coefficients and obtain the values
$0.35$, $0.10$, and $0.06$, respectively. We conclude that treating
the diameter and vessel length as independent random variables here is
justified. To find a suitable probability density function (PDF), we fitted 95
PDFs to the data and chose the one yielding the most significant p-value for the
Kolmogorov-Smirnov test.
We find that the diameter is well described by a generalized extreme value
continuous random variable, while a Wald continuous random variable better
describes the length. The data and the fitted PDFs are displayed in
Fig.~\ref{fig:VesselPDF}.

Before sampling from the PDFs, we truncate the Wald distribution, i.e. we
require a minimum segment length of $l_{\min} = 100 \mu m$. By computing the
integral
\begin{equation}\label{eq:IntegralOverWald}
  \zeta = \int_{- \infty }^{l_{\min}}  p_{w}\left(x,\Theta_{w}\right) dx \ ,
\end{equation}
where $\Theta_{w}$ denotes the fitted parameters of the distribution, we obtain
the ratio of samples that we ignore. Numerically, we evaluate the integral with
an adaptive integration rule (Gauss-Kronrod 21-point) and obtain
$\zeta \approx 0.78$. In total, Secomb's data contains
$N_S = 104$ vessel segments. According to our previous considerations, we
initialize our simulation volume $V$ with
\begin{equation}\label{eq:InitialNumberVessels}
  N = N_S \cdot (1 - \zeta) \cdot \frac{V}{V_S}
\end{equation}
vessel segments.

After evaluating Eq.~(\ref{eq:IntegralOverWald}) and
(\ref{eq:InitialNumberVessels}), we perform the following steps to initialize
the vessel segments. First, we sample the diameter $d_v$ from the
generalized extreme value distribution and the length $l_v$ from
the fitted, truncated Wald distribution. Second, we sample a random point
marking the segment's start in the simulation space. Third, we sample a random
point on a sphere of radius $l_v$ defining the segment's endpoint. If the
endpoint ends up outside the simulation volume, we resample. Once the start
point,
the endpoint, and the diameter are defined, we fill the line with cylindrical
agents of appropriate diameter. If the length $l_v$ is larger than $200 \mu m$,
we add smooth fluctuations from the center line, realizing that longer
micro-vessels are unlikely to be straight.

To verify the approach, we compute the volume fraction occupied by the
vasculature as
\begin{equation}
  \rho_S = \frac{1}{V_S} \cdot \sum_{i=1}^{N_S} V_i \ \text{and} \
  \rho = \frac{1}{V} \cdot \sum_{i=1}^{N} V_i \ ,
\end{equation}
where $V_i$ denotes the volume of a vessel. For Secomb's data, $V_i$ is computed
from the length and diameter; for the simulation, $V_i$ is the sum of the agent
volumes. We find the values $\rho_S = 0.014308$ and $\rho = 0.004559$ showing
that the simulated vascular density is roughly $30\%$ of the one given in the
data. Since the vasculature density increases during the simulation due to
angiogenesis, it is reasonable to begin with a lower vascular density.


\bibliographystyle{elsarticle-num}
\bibliography{bibliography}





\end{document}